\documentclass[11pt,a4paper,english,american]{article}
\usepackage[T1]{fontenc}
\usepackage[utf8]{inputenc}
\usepackage{float}
\usepackage{mathrsfs}
\usepackage{amsmath}
\usepackage{amssymb}
\usepackage{graphicx}

\makeatletter

\numberwithin{equation}{section}

\usepackage{jheppub}
\date{}
\usepackage{tikz}
\usepackage{mathrsfs}
\usetikzlibrary{
  arrows.meta,
  calc,
  positioning,
  angles,
  quotes,
  intersections,
  decorations.pathmorphing,
  decorations.markings
}
\definecolor{myblue}{RGB}{55,110,210}
\definecolor{handblue}{RGB}{55,110,210}
\definecolor{handgreen}{RGB}{0,128,0}
\tikzset{
  midarrow/.style={
    decoration={markings, mark=at position 0.55 with {\arrow{>}}},
    postaction={decorate}
  }
}

\makeatother

\usepackage{babel}
\begin{document}
\title{Soft momentum expansion about the Stokes topological limit of Bogoliubov coefficients}

\author{Daniel J. H. Chung,}
\author{Nidhi Sudhir}
\author{and Junqi Zhang}
\emailAdd{danchung@wisc.edu}
\emailAdd{kandathpatin@wisc.edu}
\emailAdd{jzhang2648@wisc.edu}

\affiliation{Department of Physics, University of Wisconsin-Madison,\\
Madison, WI 53706, USA}

\abstract{
It is known that particle production in time-dependent cosmological backgrounds can be computed using the Stokes phenomena occurring in Bogoliubov transformations with an analytically continued time coordinate.  We analytically compute  the soft momentum expansion of the particle production amplitude about the Stokes topological limit that was recently discovered in this context.  The computational technique relies on an exponentially improved asymptotic expansion that involves automorphisms in the space of functions defined partly by the angular step function intrinsic to the Stokes topological sectors.  We combine this with special discrete symmetries tied to the time dependence of the dispersion relation $\omega_k^2(z) = C(k) + A z^n$ for even $n$  to explicitly compute the leading momentum dependence of the amplitude which is nonperturbative in $A$.  We also use this result to give a semi-analytic estimate of the number density of particles in this class of scenarios.}

\maketitle
\flushbottom

\global\long\def\S{\bar{S}}%
\global\long\def\E{\bar{E}}%
\global\long\def\sol{\mathfrak{S}}%
\global\long\def\bf{\mathfrak{\beta}^{{\rm fic}}_{k}}%
\global\long\def\bp{\mathfrak{\beta}^{{\rm phy}}_{k}}%
\global\long\def\f{{\rm fic}}%
\global\long\def\p{{\rm phy}}%
\global\long\def\g{\Gamma}%
\global\long\def\l{N_{+}}%

\section{Introduction}

Because cosmological particle production in time-dependent backgrounds
is numerically expensive, almost any novel computational technique
is of interest. There has been some recent interest in studying particle
production using Stokes phenomena techniques \cite{Hashiba:2020rsi,Enomoto:2020xlf,Yamada:2021kqw,Hashiba:2021npn,Hashiba:2022bzi,Enomoto:2022nuj,Kolb:2023ydq,Racco:2024aac,Chung:2024pot}
that stems from propagation in an analytically continued time coordinate
in the functional space of any given choice of asymptotic expansion
basis. A key ingredient in this technique stems from the fact that
\begin{equation}
h(Re^{i\theta})e^{-R}\sim0\label{eq:firsteq}
\end{equation}
asymptotically as $R\rightarrow\infty$ to \emph{all }orders in $R^{-1}$
for almost any function $h(z)$ that is finite in that limit, which
allows $h(Re^{i\theta})$ to be discontinuous as a function of $\theta$
without changing the smoothness of the asymptotically expanded function
\foreignlanguage{english}{\cite{Stokes:1864,FF:1965,bender1969anharmonic,Voros:1983,Berry:1989zz}}.
At the end of the computation, multiplication of judiciously weighted
several of these discontinuous jumps of $h(z)$ will become the particle
production amplitude.

Many of the previous studies using the Stokes phenomena for particle
production focused on the large momenta region with a variety of spins,
which yields answers that are qualitatively similar to those derived
from saddle-point approximations\foreignlanguage{english}{ \cite{Enomoto:2020xlf,Dumlu:2010ua,Hashiba:2021npn}}.
Reference \cite{Chung:2024pot} considered the opposite limit in which
$C\rightarrow0$ in the dispersion relationship $\omega^{2}_{k}(\eta)\approx C(k)+A\eta^{n}$
(where $\eta$ is the conformal time and $n$ is even) and showed
that the probability amplitude $\beta$ governing particle production
was topological
\begin{equation}
\beta_{\sqrt{C}\rightarrow0^{+}}=-i\cot\left[\frac{\pi}{N_{\text{Stokes sectors}}}\right]\label{eq:topo}
\end{equation}
(where $N_{\text{Stokes sectors}}=n+2$) which for example in conformally
coupled scenarios corresponds to the small momenta limit. As this amplitude
showcases, one of the advantages of the Stokes phenomena techniques
compared to more traditional saddle-point approximations is that it
allows one to compute very large coefficients \cite{Chung:2024pot}.
In this paper, we will keep the leading $C$-dependent term in the
expansion of $\beta_{\sqrt{C}}$ as $C\rightarrow0$ about this topological
limit to obtain analytic results for the $C$ coefficient. The answer
turns out to be nonperturbative in $A$ and will be given for any
$n\in\text{even}$ in terms of an analytic formula summarized in eq.~(\ref{eq:finresult}).\footnote{We also show that the topological term can be derived using the novel
technique employed in this paper.} We use the leading $C$ dependence of $\beta_{\sqrt{C}}$ in conjunction
with numerical examples to obtain a semi-analytic estimate for the
number density of particles produced in eq.~(\ref{eq:quadraticresult}).

The key reasoning employed to obtain our result using the technique
of exponentially improved asymptotic expansion can be summarized as
follows. One can obtain $\beta_{\sqrt{C}}$ by considering $\theta$
propagation in analytically continued time $z\sim_{R\rightarrow\infty}Re^{i\theta}$
for certain discrete sectors of $\theta$. The jumps described in
eq.~(\ref{eq:firsteq}) can be thought of as introducing step functions:
i.e. non-analyticity in a coefficient multiplying a ``zero''. This
step function motivates a particular ansatz for the propagator across
a single sector involving a power series of $1/z$, exponential functions
of $z$, and a $\theta$ dependent step function. This turns the propagator
solution for one of these discrete sectors in $\theta$ into solving
a recursion relationship. Expanding in small $\sqrt{C}$ allows one
to find a solution to the recursion relationship explicitly. This
propagation solution across a single sector can be used to generate
$\beta_{\sqrt{C}}$ by employing a novel symmetry that analytically
continues $\sqrt{C}$ in addition to $\eta$ into the complex plane.

Mathematical ideas similar to this can be found in \cite{hinton1979stokesmultipliers,osherov2013starkproblem,zhu1992stokesconstants},
but the current work provides a novel perspective and a generalization
that is useful for the asymptotic basis in the present particle production
physics context. Furthermore, we have expressed the automorphism structure
in terms of operator maps that can be seen as a simplification of
the original recursion relation presentation given by ~\cite{zhu1992stokesconstants}.
In the process of computing the $C$ coefficient in the small $C$
expansion of $\beta_{\sqrt{C}}$, we show that despite the asymptotic
basis having non-analyticities in $\sqrt{C}\sim k$ coming from the
Lorentz group representational constraint $\omega=\sqrt{C+A\eta^{n}}$
and despite this non-analyticity manifesting as a type of Schroedinger
phase in intermediate expressions for $\beta_{\sqrt{C}}$, it disappears
consistently with what one would expect from time-reversal symmetry
in the system.

The order of presentation is as follows. In section \ref{sec:Particle-production},
we define the asymptotic basis and the analytic continuation for the
particle production computation. In section \ref{sec:Discrete-symmetries-1},
we discuss the discrete symmetries (one involving complexification
of both $\eta$ and $\sqrt{C}$) of the system that will help with
obtaining the particle production amplitude. In section \ref{sec:The-first-Stokes},
we compute the first Stokes constant for $n\geq4$. The key to the
computation that will be explained in this section is the exponentially
improved asymptotic expansions from the perspective of an automorphism
on the functional space induced by WKB amplitudes in the asymptotic
region. In section \ref{sec:Soft-corrections-to}, we combine the
discrete symmetry representation and the first Stokes constant computed
in section \ref{sec:The-first-Stokes} to derive the coefficient $c_{1}$
in $\beta_{\sqrt{C}}\approx\beta_{0}+c_{1}C/A^{\frac{2}{n+2}}$ which
is the main aim of this paper. There, we also use this to give a semi-analytic
formula for the total number density of particles produced. We will
then conclude with a brief summary. The Appendices contain many interesting
technical details related to the surprisingly intricate computation
of this paper.

\section{\label{sec:Particle-production}Particle production}

The objective of this section is to first set up a generic cosmological
particle production scenario and review the Bogoliubov computational
approach \cite{Birrell:1982ix,Ford1987,Kofman:1997yn,Parker:1969au,Parker:1971pt,Parker:1968mv,Parker:2009uva,Kaneta:2022gug,Winitzki:2005rw,Mukhanov:2007zz,Kolb:2023ydq,Chung:2024pot}.
We then specialize to particle mode equations
\begin{equation}
\left[\partial^{2}_{\eta}+\omega^{2}_{k}(\eta)\right]\chi_{k}(\eta)=0\label{eq:ModeEq}
\end{equation}
with dispersion relations of the form
\begin{equation}
\omega^{2}_{k}(\eta)=C(k)+A\eta^{n}\label{eq:dispersionRelation}
\end{equation}
where, as an extension to \cite{Chung:2024pot}, we will be interested
in understanding the produced number density in the soft regime ($C(k)/A\ll1$).
Here $\eta$ is the conformal time in the Friedmann--Lemaître--Robertson--Walker
(FLRW) spacetime, $n$ is a positive even integer, and $C\equiv C(k)$
and $A$ are some real constants. Such dispersion relations also arise
in many branches of physics (see e.g. \cite{bender1969anharmonic,Dumlu:2010ua,Enomoto:2020xlf,Hashiba:2020rsi,Hashiba:2021npn,hebenstreit2009momentum,Kofman:1997yn,Li:2019ves,Racco:2024aac,Voros:1983,osherov2013starkproblem,zhu1992thetwostate,matsuda2025exactwkb}).
In section \ref{subsec:Setup-and-Bogoliubov}, we briefly review particle
production on time-dependent backgrounds and the relation of the produced
number density to Bogoliubov coefficients. Following this, section
\ref{subsec:Analytic-continuation-and} reviews Stokes phenomena and
translates the problem of computing Bogoliubov coefficients (defined
for real $\eta$) to computing Stokes constants which characterize
the analytic continuation of the solutions to eq.\ (\ref{eq:ModeEq})
in the complexified (asymptotic) $\eta$ plane.

\subsection{Setup and Bogoliubov coefficients\label{subsec:Setup-and-Bogoliubov}}

This subsection provides a review of the well-known Bogoliubov coefficients
computation in cosmological particle production.

Consider a non-minimally coupled scalar field $\chi$ on flat FLRW
spacetime
\begin{equation}
S_{\chi}=\int d\eta d^{3}x\frac{\sqrt{-g}}{2}\left[\partial_{\mu}\chi\partial^{\mu}\chi-m^{2}_{\chi}(\eta)\chi^{2}+\xi R\chi^{2}\right]
\end{equation}
\begin{equation}
ds^{2}=a^{2}(\eta)\left(d\eta^{2}-\left|d\vec{x}\right|^{2}\right)
\end{equation}
where $\eta$ is the conformal time and $m_{\chi}(\eta)$ denotes
a time-dependent effective mass due to couplings to a background field.
After quantizing $\chi$ in the FLRW spacetime \cite{Birrell:1982ix},
we obtain a mode expansion
\begin{equation}
\chi(\eta, \vec{x})=\int\frac{d^{3}k}{(2\pi)^{3}a(\eta)}\left(a_{\vec{k}}\chi_{k}(\eta)e^{i\vec{k}\cdot\vec{x}}+a^{\dagger}_{\vec{k}}\chi^{*}_{k}(\eta)e^{-i\vec{k}\cdot\vec{x}}\right)
\end{equation}
where $a_{\vec{k}}$ and $a^{\dagger}_{\vec{k}}$ satisfy the usual
commutation relations and the Fourier modes $\chi_{k}(\eta)$ solve
the Heisenberg equation of motion
\begin{equation}
\left[\partial^{2}_{\eta}+\omega^{2}_{k}(\eta)\right]\chi_{k}(\eta)=0\label{eq:modeEqnChi}
\end{equation}
where
\begin{equation}
\omega^{2}_{k}(\eta)=k^{2}+a^{2}m^{2}_{\chi}(\eta)+(6\xi-1)\frac{a''}{a}\,.
\end{equation}
Here primes denote derivatives with respect to the conformal time,
and the Fourier modes $\chi_{k}(\eta)$ satisfy the Wronskian condition
imposed by canonical quantization
\begin{equation}
-iW(\chi_{k},\chi^{*}_{k})\equiv-i\left(\chi_{k}\partial_{\eta}\chi^{*}_{k}-\chi^{*}_{k}\partial_{\eta}\chi{}_{k}\right)=1\,.\label{eq:normChi}
\end{equation}

The vacuum state is then defined as the one annihilated by the annihilation
operators $a_{\vec{k}}$. Consequently, different choices of the mode
functions $\chi_{k}(\eta)$ corresponding to different boundary conditions
(BC) imply different notions of the vacuum. A natural choice is mode
functions which approximate to the positive-frequency WKB modes in
time intervals where $\omega_{k}(\eta)$ is adiabatic --- that is,
evolves slowly in $\eta$. Then, as the WKB modes limit to plane wave
solutions when $\omega_{k}(\eta)$ asymptotes to a constant, we may
associate an approximate notion of a single particle state with the
corresponding creation and annihilation operators. The corresponding
vacuum is referred to as the adiabatic vacuum \cite{Birrell:1982ix}.

A convenient approximate measure of nonadiabaticity (or time-translation
symmetry breaking) in a time region for our dispersion relationship
is
\begin{equation}
\delta_{k}(\eta)\equiv\frac{\omega_{k}'(\eta)}{4\omega^{2}_{k}(\eta)}
\end{equation}
corresponding to a linear dimensionless time derivative. The leading
approximate WKB mode functions are given by 
\begin{equation}
\chi^{\text{WKB}}_{k}(\eta)\approx\frac{e^{-i\int^{\eta}d\bar{\eta}\omega_{k}(\bar{\eta})}}{\sqrt{2\omega_{k}(\eta)}}\label{eq:WKB_mode}
\end{equation}
whose error is of order $\delta^{2}_{k}$. For dispersion relations
such as eq.\ (\ref{eq:dispersionRelation}) where the non-adiabaticities
in $\omega_{k}$ are approximately localized within a finite time
interval, vacua in the asymptotic past and future of this interval
may then be defined with respect to mode functions which asymptote
to eq.\ (\ref{eq:WKB_mode}) in these two regions \cite{bunch1978quantum,Birrell:1982ix}.

Let $\sol$ denote the space of solutions to the mode equation (\ref{eq:modeEqnChi}).
For the above reason, we introduce the two solutions $\chi_{k,1}(\eta),\ \chi_{k,2}(\eta)\in\sol$
satisfying the WKB BC in the asymptotic past and the asymptotic future,
respectively, as
\begin{equation}
\lim_{\eta\to\eta_{-\infty}}\chi_{k,1}(\eta)\sim\frac{e^{-i\int^{\eta}_{\eta_{*}}d\bar{\eta}\omega_{k}(\bar{\eta})}}{\sqrt{2\omega_{k}(\eta)}},\qquad\lim_{\eta\to\eta_{+\infty}}\chi_{k,2}(\eta)\sim\frac{e^{-i\int^{\eta}_{\eta_{*}}d\bar{\eta}\omega_{k}(\bar{\eta})}}{\sqrt{2\omega_{k}(\eta)}}\,,\label{eq:chi12}
\end{equation}
where $\eta_{-\infty}$ and $\eta_{+\infty}$ denote times in the
asymptotic past and future respectively, and $\eta_{*}$ is a common
reference point for the WKB phase.\footnote{We remark here that both $\omega_{k}(\eta)$ and $\sqrt{\omega_{k}(\eta)}$
in the WKB BCs (\ref{eq:chi12}) are non-negative to match the standard
Minkowski space field theory conventions. Further implications of
this non-negativity are discussed later in sections \ref{subsec:Analytic-continuation-and}
and \ref{subsec:Analytic-beta}.\label{fn:We-remark-here} The $\sim$
symbol here also means asymptotic matching as $\eta\rightarrow\eta_{\pm\infty}$.} The non-adiabaticities in $\omega_{k}$ imply that $\chi_{k,1}$
and $\chi_{k,2}$ are inequivalent functions such that $\{\chi_{k,1}(\eta),\,\chi^{*}_{k,1}(\eta)\}$
and $\{\chi_{k,2}(\eta),\,\chi^{*}_{k,2}(\eta)\}$ form two bases
of $\sol$ to quantize $\chi$ with. Basis independence of quantization
implies these are related to each other by a linear superposition
\begin{equation}
\chi_{k,1}(\eta)=\alpha_{k}\chi_{k,2}(\eta)+\beta_{k}\chi^{*}_{k,2}(\eta),\label{eq:bogoTransReal}
\end{equation}
where $\alpha_{k}$ and $\beta_{k}$ are constants, referred to as
the Bogoliubov coefficients.

What the above superposition implies in the context of the class of
dispersion relations given by eq.\ (\ref{eq:dispersionRelation})
is the following. The time dependence of $\omega_{k}$ implies that
the adiabatic condition $|\delta_{k}|\ll1$ breaks down in the region
$\vert\eta\vert\lesssim(C/A)^{1/n}$. Consequently, although the
solution $\chi_{k,1}(\eta)$ corresponds to a pure positive-frequency
mode in the asymptotic past, the violation of adiabaticity near $\vert\eta\vert\sim(C/A)^{1/n}$
causes $\chi_{k,1}(\eta)$ to evolve into a superposition of positive
and negative frequency modes in the asymptotic future. This in turn
implies that the Heisenberg vacuum state $|\text{past}\rangle$ fixed
in the asymptotic past is an excited state when measured with the
Heisenberg picture Hamiltonian in the asymptotic future. The consequently
produced particle number density given by
\begin{equation}
n_{\chi}\equiv\langle\text{past}|n^{(\text{future})}_{\chi}|\text{past}\rangle=\frac{1}{a^{3}}\int\frac{d^{3}k}{(2\pi)^{3}}\left|\beta_{k}\right|^{2}\label{eq:nchi}
\end{equation}
is obtained by computing the expectation value of the Heisenberg picture
number density operator $n^{(\text{future})}_{\chi}$ made of normal
ordered creation/annihilation operator that annihilates the future
vacuum state $|\text{future}\rangle$ (where $n^{(\text{future})}_{\chi}|\text{future}\rangle\approx0$).

In scenarios where the relevant mode functions can be computed ---
either analytically or numerically --- the Bogoliubov coefficient
$\beta_{k}$ can be computed by projecting eq.\ (\ref{eq:bogoTransReal})
onto $\chi^{*}_{k,2}(\eta)$ using the Wronskian. Recall that the
Wronskian provides the normalization (\ref{eq:normChi}) and vanishes
for linearly dependent solutions to eq.\ (\ref{eq:modeEqnChi}).
This projection leads to
\begin{equation}
\beta_{k}=iW(\chi_{k,1},\chi_{k,2})=i\left[\chi_{k,1}(\eta)\partial_{\eta}\chi_{k,2}(\eta)-\chi_{k,2}(\eta)\partial_{\eta}\chi_{k,1}(\eta)\right].\label{eq:betaReal}
\end{equation}
Because the nontrivial time dependences in the system often preclude
an analytic form of the mode functions, an accurate estimate of $\beta_{k}$
typically requires expensive numerical computations. Hence, analytic
treatments, such as saddle-point techniques, which may be used to
estimate $\beta_{k}$ in certain simplified scenarios are useful.
The remainder of this paper is devoted to answering one such question:
for the mode equation given by eqs.\ (\ref{eq:ModeEq}) and (\ref{eq:dispersionRelation}),
how can one derive an analytic expression for soft corrections to
the Bogoliubov coefficient $\beta_{k}$? As a preliminary step, we
review the formalism of the Stokes phenomena approach and establish
our conventions in the next section.

\subsection{Analytic continuation and Stokes phenomena\label{subsec:Analytic-continuation-and}}

Stokes phenomenon provides an approach to compute the Bogoliubov coefficients
based on the analyticity properties of the dispersion relation $\omega_{k}$.
We noted in section \ref{subsec:Setup-and-Bogoliubov} that particle
production may be understood in terms of the non-adiabatic evolution
of the positive frequency WKB mode (\ref{eq:WKB_mode}) in the asymptotic
past into a superposition of positive and negative frequency WKB modes
in the asymptotic future. The superposition coefficients, given by
the Bogoliubov coefficients, then determine the produced particle
number density. Therefore, the problem of computing particle production
is equivalent to that of tracking the evolution of the WKB modes across
non-adiabaticities in $\omega_{k}$ --- similar to scattering problems
in quantum mechanics. Stokes phenomenon describes exactly this evolution,
but in the analytically continued version of this problem.

Continuing the conformal time $\eta$ to $z\in\mathbb{C}$ results
in the complexified mode equation
\begin{equation}
\left[\partial^{2}_{z}+\omega^{2}_{k}(z)\right]\chi_{k}(z)=0\label{eq:ComplexMEQ}
\end{equation}
where $\omega_{k}(z)$ and $\chi_{k}(z)$ are the analytic continuations
of their real time versions. Similarly, we may define the complexified
WKB basis functions as 
\begin{equation}
f_{\pm}(k,z)\equiv\frac{e^{\pm i\theta(k,z)}}{\sqrt{2\omega_{k}(z)}}\,,\quad\theta(k,z)\equiv\int^{z}_{\eta_{*}}dz'\omega_{k}(z')\label{eq:WKBFic}
\end{equation}
where $\theta(k,z)$ is the WKB phase.\footnote{Depending on analyticity properties of the dispersion relation $\omega^{2}_{k}$,
this definition of $f_{\pm}(k,z)$ requires a choice of the branches
of $\omega_{k}(z)$ and $\sqrt{\omega_{k}(z)}$. This is further detailed
in the context of eq.\ (\ref{eq:dispersionRelation}) later in this
section.} For the WKB approximation\footnote{In this work we will refer to the leading order truncation of the
WKB series as the WKB approximation. That is, WKB approximations take
the form $\chi_{k}\sim\alpha f_{-}+\beta f_{+}$ where $\alpha$ and
$\beta$ are superposition coefficients. We further discuss the WKB
approximations from an asymptotic expansion perspective in section
\ref{sec:The-first-Stokes}.} of a given mode function $\chi_{k}(z)$ in the `adiabatic' regions
of the complex plane defined by
\begin{equation}
|\delta_{k}(z)|=\left|\frac{\omega_{k}'(z)}{4\omega^{2}_{k}(z)}\right|\ll1\,,\label{eq:dk-217}
\end{equation}
Stokes phenomenon implies that the superposition coefficients (defined
explicitly in eq.\ (\ref{eq:modeDecomposition})) will shift discretely
across certain contours in the complex plane called the Stokes lines.
These are contours along which the WKB phase is such that $i\theta(k,z)$
is real. Hence, along these, the two WKB basis functions $f_{\pm}(k,z)$
which either diverge or converge exponentially as $\vert z\vert$
increases, are therefore maximally dominant or subdominant with respect
to each other. These contours are further classified as ``$+$''(``$-$'')
Stokes lines if $f_{+}(k,z)$ ($f_{-}(k,z)$) is the dominant WKB
function.

Given this, the statement of Stokes phenomenon may be understood as
a consequence of the mismatch between the basis functions $f_{\pm}(k,z)$
and the exact mode $\chi_{k}(z)$ as well as the peculiar property
of the zero asymptotic expansion representation of exponentially suppressed
functions as $|z|\rightarrow\infty$. In adiabatic regions where $f_{\pm}(k,z)$
present good approximate solutions to eq.\ (\ref{eq:ComplexMEQ}),
this mismatch between a given $\chi_{k}(z)$ and $f_{\pm}(k,z)$ arises
as shifts in the WKB superposition coefficients as $\chi_{k}(z)$
is analytically continued across the Stokes lines. These shifts occur
only in the coefficient of the exponentially suppressed WKB function
which in the limit $\vert z\vert\rightarrow\infty,\ \vert\omega_{k}\vert\rightarrow\infty$
makes zero contribution to the WKB approximation truncated at $O(z^{-1})$.
This preserves the continuity of $\chi_{k}(z)$ in its WKB approximation.

The constants which determine the above shifts are called Stokes constants.
In the following, we determine the relation between the Bogoliubov
coefficients and the relevant Stokes constants in the context of dispersion
relations of the form eq.\ (\ref{eq:dispersionRelation}), where
we have set $C=k^{2}$ for physical clarity. This replaces the problem
of computing $\beta_{k}$ with that of computing the Stokes constants
which we will turn to in sections \ref{sec:Discrete-symmetries} and
\ref{sec:The-first-Stokes}.

To formalize the idea of tracking the evolution of the analytically
continued WKB modes, we begin by first rewriting $\chi_{k}(\eta)$
on the \textit{real} line as
\begin{equation}
\chi_{k}(\eta)=\alpha_{k}(\eta)f^{\p}_{-}(k,\eta)+\beta_{k}(\eta)f^{\p}_{+}(k,\eta),\quad\eta\in\mathbb{R}\label{eq:modeDecompositionReal}
\end{equation}
where $f^{\p}_{\pm}(k,\eta)$ are the physical WKB basis functions defined as
\begin{equation}
f^{\p}_{\pm}(k,\eta)\equiv\frac{e^{\pm i\theta_{k}(\eta)}}{\sqrt{2\omega_{k}(\eta)}},\quad\theta_{k}(\eta)\equiv\int^{\eta}_{0}d\eta'\omega_{k}(\eta')
\end{equation}
and the functions $\alpha_{k}(\eta)$ and $\beta_{k}(\eta)$ will
be called the WKB amplitudes. Here both $\omega_{k}(\eta)$ and $\sqrt{\omega_{k}(\eta)}$
are non-negative for all $\eta\in\mathbb{R}$ as remarked in footnote
\ref{fn:We-remark-here}. Consequently, the physical WKB basis functions
asymptote to the WKB BCs defined in eq.\ (\ref{eq:chi12}), therefore
the Bogoliubov coefficients from eq.\ (\ref{eq:bogoTransReal}) are
related to the WKB amplitudes $\alpha_{k}(\eta)$ and $\beta_{k}(\eta)$
via
\begin{equation}
\left(\alpha_{k},\beta_{k}\right)=\lim_{\eta\rightarrow\eta_{+\infty}}\left(\alpha_{k}(\eta),\beta_{k}(\eta)\right)\,.
\end{equation}
Henceforth, we set the common reference point $\eta_{*}$ of the WKB
phase (see eq.~(\ref{eq:WKBFic})) to $0$ for the ease of describing
symmetry representations in section \ref{sec:Discrete-symmetries}.

Note that in rewriting the mode function $\chi_{k}(\eta)$ in terms
of the WKB amplitudes we have expressed one complex degree of freedom
in terms of two, allowing multiple functions $\alpha_{k}(\eta)$ and
$\beta_{k}(\eta)$ to represent the same $\chi_{k}(\eta)$. This additional
freedom can be fixed by specifying $\chi_{k}'(\eta)$ in terms of
the WKB functions and amplitudes. A convenient choice is
\begin{equation}
\partial_{\eta}\chi_{k}(\eta)=i\omega_{k}(\eta)\left(\beta_{k}(\eta)f^{\p}_{+}(k,\eta)-\alpha_{k}(\eta)f^{\p}_{-}(k,\eta)\right)\label{eq:DervCond}
\end{equation}
which allows the second order ODE (\ref{eq:ComplexMEQ}) of $\chi_{k}(\eta)$
to be rewritten as the first order vector ODE of the WKB amplitudes
(see eq.\ (37) in \cite{Chung:2024pot})
\begin{equation}
\partial_{\eta}V_{k}(\eta)=M_{0}(k,\eta)V_{k}(\eta),\quad V_{k}(\eta)\equiv\begin{pmatrix}\alpha_{k}(\eta)\\
\beta_{k}(\eta)
\end{pmatrix}\,,\label{eq:amplitudeODE0}
\end{equation}
\begin{equation}
M_{0}(k,\eta)=\frac{\omega'_{k}(\eta)}{2\omega_{k}(\eta)}\begin{pmatrix}0 & e^{+2i\int^{\eta}_{\eta_{*}}d\eta'\omega_{k}(\eta')}\\
e^{-2i\int^{\eta}_{\eta_{*}}d\eta'\omega_{k}(\eta')} & 0
\end{pmatrix}\,.\label{eq:M0ZeroGauge}
\end{equation}
Integrating the above we see that the propagator which relates the
WKB amplitudes at two times $\eta_{0}$ and $\eta$ is given by
\begin{equation}
V_{k}(\eta)=U_{0}(k,\eta,\eta_{0})V_{k}(\eta_{0})
\end{equation}
\begin{equation}
U_{0}(k,\eta,\eta_{0})=\mathbb{T}\left[\mathbb{I}_{2\times2}\exp\left(\int^{\eta}_{\eta_{0}}dwM_{0}(k,w)\right)\right]\label{eq:propagatorZeroGauge}
\end{equation}
where $\mathbb{T}$ denotes time-ordering. Note, since $M_{0}\left(k,\eta\right)$
belongs to the Lie algebra of ${\rm SU}(1,1)$ (see eq.\ (22) in
\cite{Chung:2024pot}), the WKB amplitudes on the real line satisfy
\begin{equation}
|\alpha_{k}(\eta)|^{2}-|\beta_{k}(\eta)|^{2}=1,\quad\forall\eta\in\mathbb{R}.\label{eq:alphaBetaSU11}
\end{equation}
This naturally extends the normalization condition on the Bogoliubov
coefficients arising from eq.\ (\ref{eq:normChi}) to the WKB amplitudes
at all real times.\footnote{As noted in \cite{Chung:2024pot} we may choose from a family of conditions
on $\chi'_{k}(\eta)$ (similar to eq.\ (\ref{eq:DervCond})) which
preserve the BCs at asymptotic past and future and also, for convenience
lead to WKB amplitudes satisfying eq.\ (\ref{eq:alphaBetaSU11}).} We will see in sections \ref{sec:Discrete-symmetries} and \ref{sec:The-first-Stokes}
that the choice of eq.\ (\ref{eq:DervCond}) and eq.\ (\ref{eq:M0ZeroGauge})
will be convenient for analyzing discrete symmetry representations
and asymptotic behaviors of the analytically continued WKB amplitudes.

Now, to extend the problem to the complex $z$-plane, we may analytically
continue the first order ODE in eq.\ (\ref{eq:amplitudeODE0}) and
the propagator of eq.\ (\ref{eq:propagatorZeroGauge}). The propagators
$U_{0}\left(k,z,z_{0}\right)$ then relate the analytically continued
WKB amplitudes between two complex points $z$ and $z_{0}$. Specializing
to dispersion relations of the form (\ref{eq:dispersionRelation}),
the mode equation becomes
\begin{equation}
\left[\partial^{2}_{z}+\underline{\omega}^{2}_{\bar{k}}(z)\right]\underline{\chi}_{\bar{k}}(z)=0,\quad\underline{\omega}^{2}_{\bar{k}}(z)=\bar{k}^{2}+z^{n},\quad z\in\mathbb{C}\label{eq:modeEqnChiReduced}
\end{equation}
where $n$ is a positive even integer and the complexified conformal
time has been rescaled $z=A^{1/(n+2)}\eta$ to absorb the constant
$A$ resulting in $\bar{k}=C^{1/2}/A^{1/(n+2)}$.\footnote{Existence of the analytic continuation $\chi_{k}(z)$ is discussed
in \cite{shibuya1975globaltheory}.} If we define $\underline{\chi}_{\bar{k},j}(z)$ to be the mode equation
solution satisfying eq.~(\ref{eq:chi12}) with $\eta\rightarrow z$,
one can easily check eq.~(\ref{eq:betaReal}) becomes
\begin{equation}
\underline{\chi}_{\bar{k},1}(z)\partial_{z}\underline{\chi}_{\bar{k},2}(z)-\underline{\chi}_{\bar{k},2}(z)\partial_{z}\underline{\chi}_{\bar{k},1}(z)=\chi_{k,1}(\eta)\partial_{\eta}\chi_{k,2}(\eta)-\chi_{k,2}(\eta)\partial_{\eta}\chi_{k,1}(\eta)
\end{equation}
leading to $\underline{\beta}_{\bar{k}}=\beta_{k}.$ To reduce visual
clutter in notation, we will from now until a later section eq.~(\ref{eq:kbarisntk})
drop the underscore on $\underline{\beta}_{\bar{k}}$ and the bar
on $\bar{k}$ 
\begin{equation}
\boxed{\underline{\beta}_{\bar{k}}\rightarrow\beta_{k}}\,.\label{eq:rescalek}
\end{equation}

The analog of eq.\ (\ref{eq:modeDecompositionReal}) is
\begin{equation}
\chi_{k}(z)=\alpha_{k}(z)f_{-}(k,z)+\beta_{k}(z)f_{+}(k,z)\label{eq:modeDecomposition}
\end{equation}
where $\alpha_{k}(z)$ and $\beta_{k}(z)$ are the complexified WKB
amplitudes and $f_{\pm}(k,z)$ are analytically continued WKB basis
functions defined as
\begin{equation}
f_{\pm}(k,z)\equiv\frac{e^{\pm i\theta(k,z)}}{\sqrt{2\omega_{k}(z)}}\,,\quad\theta(k,z)\equiv\int^{z}_{0}dz'\omega_{k}(z')\,.
\end{equation}
We note that $f_{\pm}(k,z)$ inherit multivaluedness from the fractional
powers of $\omega^{2}_{k}$, such as $\omega_{k}$ and $\sqrt{\omega_{k}}$.
Therefore, the decomposition (\ref{eq:modeDecomposition}) of the
complexified WKB amplitudes $\alpha_{k}(z)$ and $\beta_{k}(z)$ over
the fictitious WKB basis requires a choice of branch of $\omega_{k}(z)$
and $\sqrt{\omega_{k}(z)}$. Since on the real line, $\omega_{k}(\eta)$
corresponds to the energy of approximate single particle states, we
require it to be non-negative and continuous for all $\eta\in\mathbb{R}$.
This implies that the branch cuts defining $\omega_{k}(z)$ and $\sqrt{\omega_{k}(z)}$,
arising from the zeros of $\omega^{2}_{k}$, have to be taken away
from the real line.\footnote{Note $\omega_{k}(z)$ and $\sqrt{\omega_{k}(z)}$ share the same set
of branch cuts.} In section \ref{sec:Discrete-symmetries-1} and appendix \ref{subsec:analytically-continue-time},
we detail this branch choice and its consequences, one of which is
that the fictitious WKB basis matches the physical WKB basis on the
real line, ensuring that the complexified WKB amplitudes are equal
to their real versions on the real line.

With this branch choice, the propagation of the analytically continued
WKB amplitudes between two complex points $z$ and $z_{0}$ is
\begin{equation}
V_{k}(z)=U_{0}(k,z,z_{0})V_{k}(z_{0}),\quad V_{k}(z)=\begin{pmatrix}\alpha_{k}(z)\\
\beta_{k}(z)
\end{pmatrix}\,,\label{eq:propagation-complex}
\end{equation}
where the propagator is given by
\begin{equation}
U_{0}(k,z,z_{0})=\mathbb{P}\left[\mathbb{I}_{2\times2}\exp\left(\int_{\Gamma(z,z_{0})}dwM_{0}(k,w)\right)\right]\label{eq:propagatorComplex}
\end{equation}
\begin{equation}
M_{0}(k,z)=\frac{\omega'_{k}(z)}{2\omega_{k}(z)}\begin{pmatrix}0 & e^{+2i\int^{z}_{0}dz'\omega_{k}(z')}\\
e^{-2i\int^{z}_{0}dz'\omega_{k}(z')} & 0
\end{pmatrix}\,,\label{eq:M0Complex}
\end{equation}
and $\mathbb{P}$ denotes path-ordering and $\Gamma(z,z_{0})$ denotes
the integration path connecting $z_{0}$ and $z$. In terms of these,
the Bogoliubov coefficients of eq.\ (\ref{eq:bogoTransReal}) are
given as the $11$ and $21$ entries of the propagator from the asymptotic
past to the asymptotic future, that is
\begin{equation}
\alpha_{k}\equiv\left[\lim_{z\to\eta_{+\infty}}\lim_{z_{0}\to\eta_{-\infty}}U_{0}(k,z,z_{0})\right]_{11}\,,\label{eq:alphaFicPropagator11}
\end{equation}
\begin{equation}
\beta_{k}\equiv\left[\lim_{z\to\eta_{+\infty}}\lim_{z_{0}\to\eta_{-\infty}}U_{0}(k,z,z_{0})\right]_{21}.\label{eq:betaFicPropagator21}
\end{equation}

\begin{figure}[h]
\begin{centering}
\includegraphics[scale=0.5]{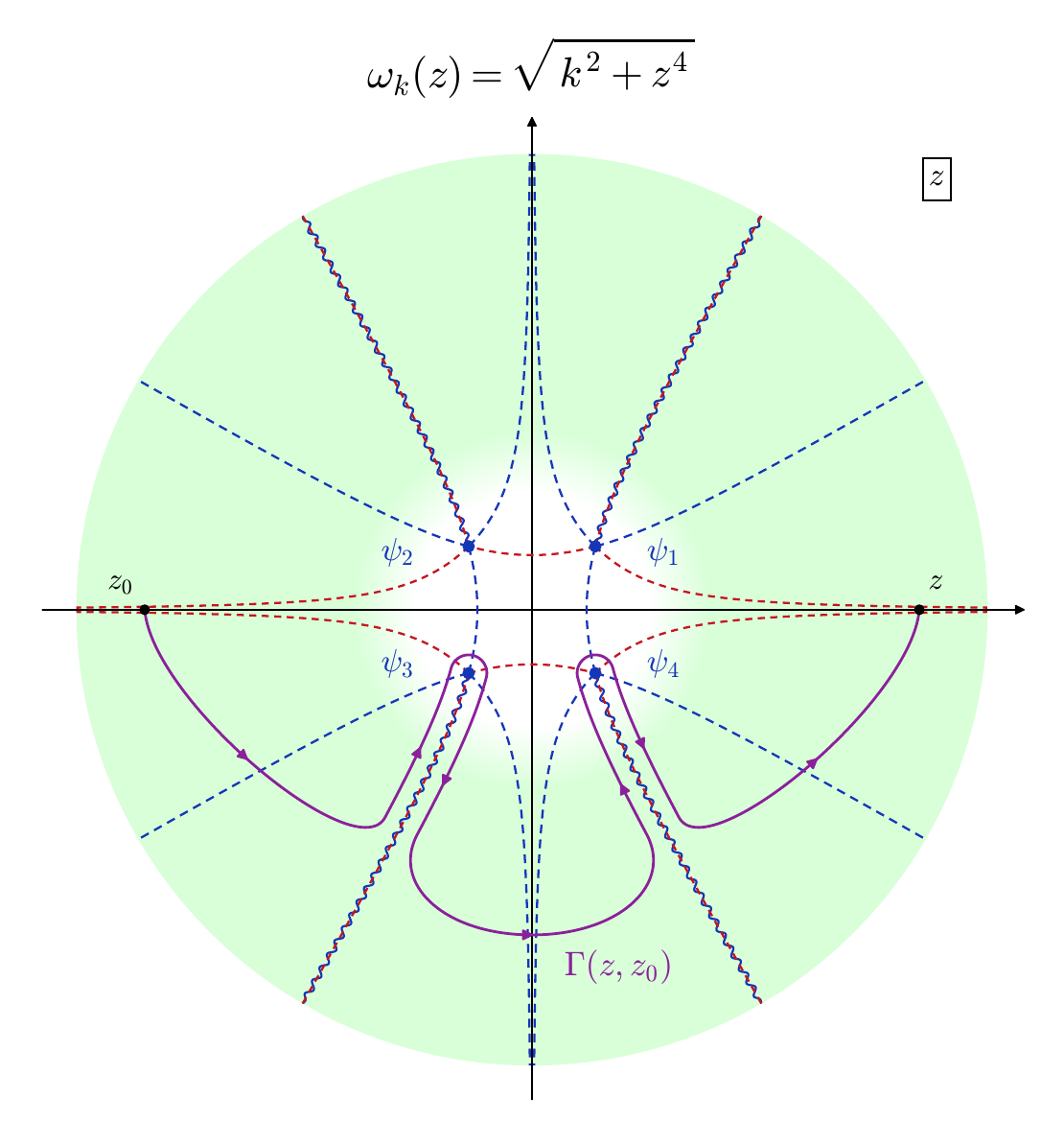}
\par\end{centering}
\caption{The figure depicts the analytic structure associated with the WKB
basis functions $f_{\pm}(k,z)$ on the complex plane, for the dispersion
relation $\omega_{k}(z)=\sqrt{k^{2}+z^{4}}$. The zeros of the dispersion
relation denoted as $\psi_{i}$ have been shown as blue dots from
which wavy lines representing branch cuts arise. The green annulus
and the white disk at the center correspond to the regions where the
dispersion relation becomes adiabatic and non-adiabatic, respectively;
and the blue and red dashed lines to Stokes lines (SLs) and anti-Stokes
lines (ASLs). Note, the branch cuts have been taken to lie along the
ASLs directed away from the real line. The adiabatic region between
any two ASLs corresponds to a Stokes sector --- denoted $\mathscr{S}_{j}$,
$j=1,2,\dots,n+2$. The first Stokes sector $\mathscr{S}_{1}$ is
taken to be the sector in the first quadrant with the real line as
one of its boundaries; higher $j$s are labelled counterclockwise.
Also shown is a representative integration contour $\Gamma(z,z_{0})$
connecting a point $z_{0}$ on the negative real axis to a point $z$
on the positive real axis. The contour is deformed near the branch
cuts with keyhole shaped detours.\label{fig:gamma-contour-n4}}
\end{figure}

The above expressions provide a precise formulation of tracking the
evolution of the complexified WKB modes we mentioned in the beginning
of this subsection. The advantage of these expressions, for dispersion
relations like eq.\ (\ref{eq:modeEqnChiReduced}), is that the propagation
contour $\Gamma(z,z_{0})$ can be chosen as depicted in figure \ref{fig:gamma-contour-n4}
such that the propagator along it becomes easier to evaluate. In particular,
away from the branch cuts, $\Gamma(z,z_{0})$ is chosen to lie in
the asymptotic annulus (AA) where $|z|\gg k^{2/n}$, whereas near
the branch cuts the contour is deformed with keyhole shaped detours.\footnote{The outer radius of the annulus can be arbitrarily large and is simply
a reminder that we do not compactify to a point at infinity where
the differential equation solution has an essential singularity. This
facilitates a clean separation of the Stokes sectors.} In the following, we quantify the propagators along contour segments
in the AA and segments near the branch cuts.

\paragraph{Across the Stokes lines (Stokes phenomenon)}

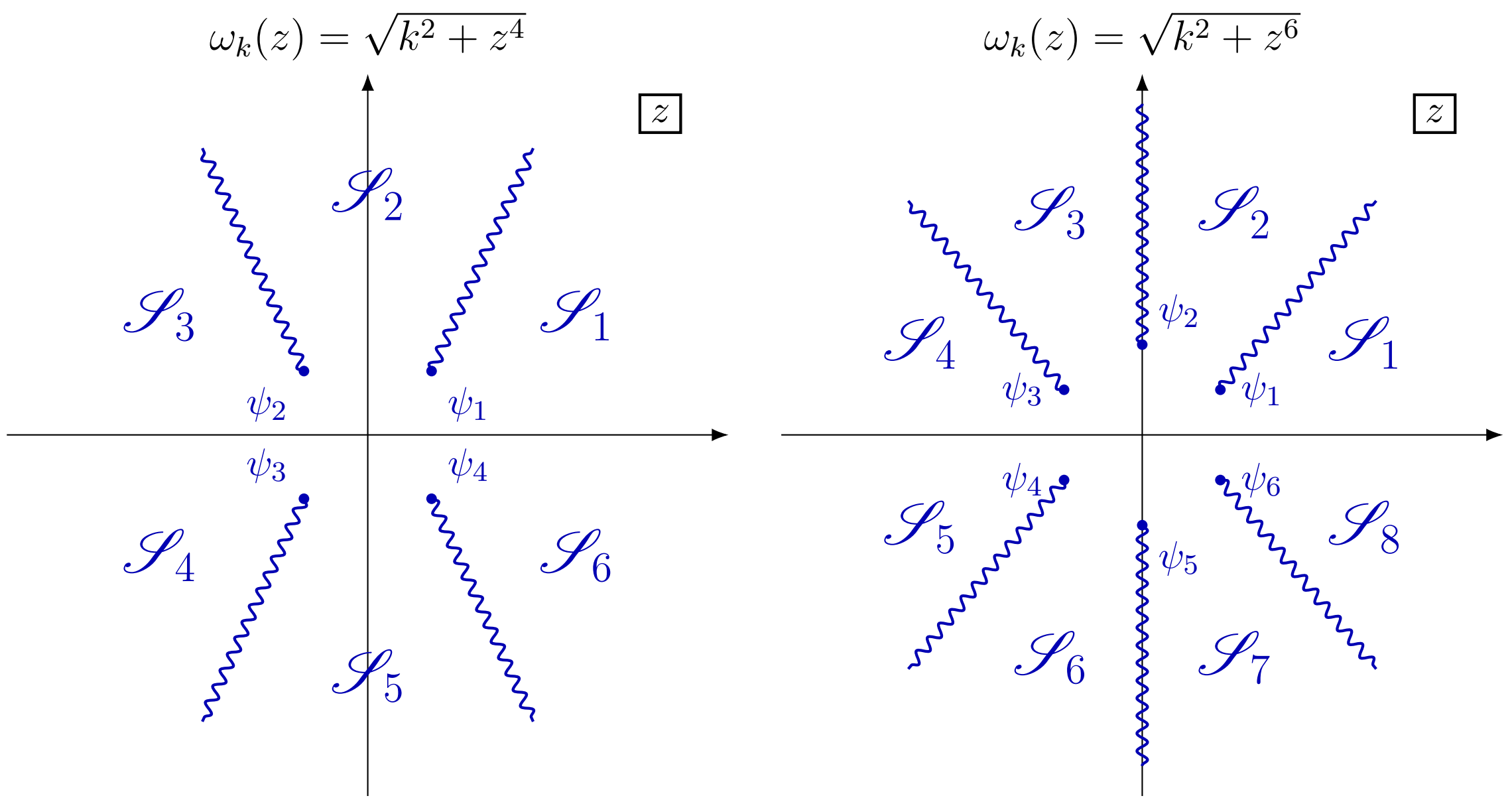
\begin{figure}[H]
\begin{center}

\begin{tikzpicture}[>=Latex, scale=0.6]
    \def\Rpsi{1.5} 
    \def\Rout{5.5} 

    \draw[->] (-\Rout-0.5, 0) -- (\Rout+0.5, 0);
    \draw[->] (0, -\Rout-0.5) -- (0, \Rout+0.5);

    \node[anchor=south, font=\large, align=center] at (0, \Rout+0.5) {$\omega_k(z) = \sqrt{k^2 + z^4}$};
    \node[draw, thick, anchor=south west, font=\normalsize, inner sep=3pt] at (\Rout-1, \Rout-0.5) {$z$};

    \coordinate (psi1) at (45:\Rpsi);
    \coordinate (psi2) at (135:\Rpsi);
    \coordinate (psi3) at (225:\Rpsi);
    \coordinate (psi4) at (315:\Rpsi);

    \coordinate (end1) at (60:\Rout);
    \coordinate (end2) at (120:\Rout);
    \coordinate (end3) at (240:\Rout);
    \coordinate (end4) at (300:\Rout);

    \begin{scope}[decoration={snake, amplitude=1.5pt, segment length=5pt}]
        \draw[blue!70!black, thick, decorate] (psi1) -- (end1);
        \draw[blue!70!black, thick, decorate] (psi2) -- (end2);
        \draw[blue!70!black, thick, decorate] (psi3) -- (end3);
        \draw[blue!70!black, thick, decorate] (psi4) -- (end4);
    \end{scope}

    \fill[blue!70!black] (psi1) circle (2.5pt) node[below right=1pt, font=\normalsize] {$\psi_{1}$};
    \fill[blue!70!black] (psi2) circle (2.5pt) node[below left=1pt, font=\normalsize] {$\psi_{2}$};
    \fill[blue!70!black] (psi3) circle (2.5pt) node[above left=1pt, font=\normalsize] {$\psi_{3}$};
    \fill[blue!70!black] (psi4) circle (2.5pt) node[above right=1pt, font=\normalsize] {$\psi_{4}$};

	\def\Rsec{4};
    \node[blue!70!black, font=\LARGE] at (30:\Rsec) {$\mathscr{S}_1$};
    \node[blue!70!black, font=\LARGE] at (90:\Rsec) {$\mathscr{S}_2$};
    \node[blue!70!black, font=\LARGE] at (150:\Rsec) {$\mathscr{S}_3$};
    \node[blue!70!black, font=\LARGE] at (210:\Rsec) {$\mathscr{S}_4$};
    \node[blue!70!black, font=\LARGE] at (270:\Rsec) {$\mathscr{S}_5$};
    \node[blue!70!black, font=\LARGE] at (330:\Rsec) {$\mathscr{S}_6$};

\end{tikzpicture}
\quad
\begin{tikzpicture}[>=Latex, scale=0.6]
    \def\Rpsi{1.5} 
    \def\Rout{5.5} 

    \draw[->] (-\Rout-0.5, 0) -- (\Rout+0.5, 0);
    \draw[->] (0, -\Rout-0.5) -- (0, \Rout+0.5);

    \node[anchor=south, font=\large, align=center] at (0, \Rout+0.5) {$\omega_k(z) = \sqrt{k^2 + z^6}$};
    \node[draw, thick, anchor=south west, font=\normalsize, inner sep=3pt] at (\Rout-1, \Rout-0.5) {$z$};

    \coordinate (psi1) at (30:\Rpsi);
    \coordinate (psi2) at (90:\Rpsi);
    \coordinate (psi3) at (150:\Rpsi);
    \coordinate (psi4) at (210:\Rpsi);
    \coordinate (psi5) at (270:\Rpsi);
    \coordinate (psi6) at (330:\Rpsi);

    \coordinate (end1) at (45:\Rout);
    \coordinate (end2) at (90:\Rout);
    \coordinate (end3) at (135:\Rout);
    \coordinate (end4) at (225:\Rout);
    \coordinate (end5) at (270:\Rout);
    \coordinate (end6) at (315:\Rout);

    \begin{scope}[decoration={snake, amplitude=1.5pt, segment length=5pt}]
        \draw[blue!70!black, thick, decorate] (psi1) -- (end1);
        \draw[blue!70!black, thick, decorate] (psi2) -- (end2);
        \draw[blue!70!black, thick, decorate] (psi3) -- (end3);
        \draw[blue!70!black, thick, decorate] (psi4) -- (end4);
        \draw[blue!70!black, thick, decorate] (psi5) -- (end5);
        \draw[blue!70!black, thick, decorate] (psi6) -- (end6);
    \end{scope}

    \fill[blue!70!black] (psi1) circle (2.5pt) node[right=2pt, font=\normalsize] {$\psi_{1}$};
    \fill[blue!70!black] (psi2) circle (2.5pt) node[above right=1pt, font=\normalsize] {$\psi_{2}$};
    \fill[blue!70!black] (psi3) circle (2.5pt) node[left=2pt, font=\normalsize] {$\psi_{3}$};
    \fill[blue!70!black] (psi4) circle (2.5pt) node[left=2pt, font=\normalsize] {$\psi_{4}$};
    \fill[blue!70!black] (psi5) circle (2.5pt) node[below right=1pt, font=\normalsize] {$\psi_{5}$};
    \fill[blue!70!black] (psi6) circle (2.5pt) node[right=2pt, font=\normalsize] {$\psi_{6}$};

    \def\Rsec{4}
    \node[blue!70!black, font=\LARGE] at (22.5:\Rsec) {$\mathscr{S}_1$};
    \node[blue!70!black, font=\LARGE] at (67.5:\Rsec) {$\mathscr{S}_2$};
    \node[blue!70!black, font=\LARGE] at (112.5:\Rsec) {$\mathscr{S}_3$};
    \node[blue!70!black, font=\LARGE] at (157.5:\Rsec) {$\mathscr{S}_4$};
    \node[blue!70!black, font=\LARGE] at (202.5:\Rsec) {$\mathscr{S}_5$};
    \node[blue!70!black, font=\LARGE] at (247.5:\Rsec) {$\mathscr{S}_6$};
    \node[blue!70!black, font=\LARGE] at (292.5:\Rsec) {$\mathscr{S}_7$};
    \node[blue!70!black, font=\LARGE] at (337.5:\Rsec) {$\mathscr{S}_8$};

\end{tikzpicture}
\end{center}\caption{Shown is a schematic for branch cuts and Stokes sectors in the complex
$z$-plane for the dispersion relation $\omega_{k}(z)=\sqrt{k^{2}+z^{n}}$.
The left and right panels correspond to $n=4$ and $n=6$, respectively.
The wavy blue lines represent the chosen branch cuts emanating from
the zeros $\psi_{i}$ and the calligraphic $\mathscr{S}$ labels the
Stokes sectors.\label{fig:Branch-cut-structures}}
\end{figure}

As noted in the introduction to this section, Stokes phenomenon implies
that the WKB amplitude of the subdominant mode undergoes a discrete
shift by an $O(1)$ constant (to $O(z^{-1})$) across the Stokes lines.
In terms of the propagator in the AA, this shift across one Stokes
line (SL) may be represented as \cite{Froman:2002,Chung:2024pot}
\begin{equation}
U_{+}(S(k))\sim\begin{pmatrix}1 & S(k)\\
0 & 1
\end{pmatrix},\quad U_{-}(S(k))\sim\begin{pmatrix}1 & 0\\
S(k) & 1
\end{pmatrix}\,,\label{eq:Upm-S}
\end{equation}
where $U_{\pm}$ refer to the propagators across $\pm$ SL (defined
below eq.\ (\ref{eq:dk-217})), respectively, and $S(k)$ are the
Stokes constants.\footnote{In section \ref{sec:The-first-Stokes} eq.\ (\ref{eq:betaZ}), we
provide another perspective on eq.\ (\ref{eq:Upm-S}) by recognizing
a step function in the asymptotic behavior of the WKB amplitudes.} Between any two Stokes lines, there exists an anti-Stokes line (ASL)
defined as a contour along which 
\begin{equation}
i\theta(k,z)\in\text{imaginary}.
\end{equation}
These ASLs separate the complex plane into different Stokes sectors.
In the AA, for the given $\omega_{k}(z)=\sqrt{k^{2}+z^{n}}$, the
anti-Stokes lines asymptote towards radial lines at the arguments
$2\pi(j-1)/(n+2)$ and the Stokes lines to those at $\pi(2j-1)/(n+2)$,
where $j=1,2,\cdots,n+2$. We define the \textit{asymptotic region}
between the $j$-th and the $(j+1)$-th anti-Stokes lines the $j$-th
Stokes sector $\mathscr{S}_{j}$, as illustrated in figure \ref{fig:Branch-cut-structures}.

\paragraph{Across branch cuts}

To work out the propagator across branch cuts with keyhole shaped
contours as shown in figure \ref{fig:gamma-contour-n4}, we begin
by describing the transformation of the WKB mode functions across
a branch cut arising from the $j$-th zero $\psi_{j}=k^{2/n}e^{(2j-1)i\pi/n}$
of the dispersion relation $\omega_{k}(z)$ (\ref{eq:modeEqnChiReduced}).
The transformation of the WKB amplitudes across the complementary
branch cuts is then taken to be inverses of these. Given two points
$z^{\pm}_{j}$ infinitesimally close and padding the branch cut from
the $j$-th zero of $\omega_{k}(z)$, the values taken by the dispersion
relation at these two points are related as 
\begin{equation}
\omega_{k}(z^{+}_{j})=-\omega_{k}(z^{-}_{j}).\label{eq:omega-z-pm}
\end{equation}
 This in turn implies that the WKB phases $\theta(k,z^{\pm}_{j})$
are related as
\begin{equation}
\theta(k,z^{+}_{j})=\int^{\psi_{j}}_{0}dz\omega_{k}(z)+\int^{z^{+}_{j}}_{\psi_{j}}dz\omega_{k}(z)=2\Delta_{j}(k)-\theta(k,z^{-}_{j})
\end{equation}
where
\begin{equation}
\Delta_{j}(k)\equiv\int^{\psi_{j}}_{0}dz\omega_{k}(z)\,.\label{eq:Deltaj}
\end{equation}
Using this and taking into account the transformation of $\omega^{-1/2}_{k}(z)$
(see eq.\ (\ref{eq:sqrt-omega-continued})) in the denominator of
the WKB basis $f_{\pm}(k,z)$, we have 
\begin{equation}
F^{T}_{k}(z^{+}_{j})=F^{T}_{k}(z^{-}_{j})\begin{pmatrix}0 & ie^{2i\Delta_{j}(k)}\\
ie^{-2i\Delta_{j}(k)} & 0
\end{pmatrix}\,\label{eq:Fpm}
\end{equation}
where $F_{k}(z)\equiv\begin{pmatrix}f_{-}(k,z) & f_{+}(k,z)\end{pmatrix}^{T}$
denotes the WKB basis vector. Writing the exact mode decomposition
(\ref{eq:modeDecomposition}) in vector form $\chi_{k}(z)=F^{T}_{k}(z)V_{k}(z)$
and noting the smoothness of the exact mode $\chi_{k}(z^{-}_{j})=\chi_{k}(z^{+}_{j})$,
we obtain the WKB amplitudes propagation across the $j$-th branch
cut:
\begin{equation}
V_{k}(z^{+}_{j})=\mathscr{B}_{j}(k)V_{k}(z^{-}_{j}),\qquad\mathscr{B}_{j}(k)\equiv\begin{pmatrix}0 & -ie^{2i\Delta_{j}(k)}\\
-ie^{-2i\Delta_{j}(k)} & 0
\end{pmatrix}\label{eq:branchj}
\end{equation}
where $\mathscr{B}_{j}(k)$ denotes the propagator matrix across the
branch cut from the $j$-th zero.

With the above, we can now rewrite $U_{0}(k,\eta_{+\infty},\eta_{-\infty})$
appearing in the Bogoliubov $\beta_{k}$ computation (\ref{eq:betaFicPropagator21})
as a product of propagator matrices across the relevant SLs and branch
cuts. For convenience, the branch cuts are further taken to be such
that in the adiabatic region these asymptote to the ASLs directed
away from the real line. With this choice of the branch cuts, the
SLs in the upper half plane are of the ``$-$'' type and those of
the lower half plane of the ``$+$'' type. To see this note, for
this branch choice the dispersion relation asymptotes to the form
$\omega_{k}(z)\sim s_{j}z^{n/2}$ for $z\in\mathscr{S}_{j}$.\footnote{A more detailed derivation of this result is in eqs.\ (\ref{eq:omega-continued-sj})
and (\ref{eq:sj}).} Therefore, the propagator matrix across each Stokes sector takes
the form
\begin{equation}
U_{j}(k)\sim\begin{cases}
\begin{pmatrix}1 & 0\\
S_{j}(k) & 1
\end{pmatrix}+O(z^{-1}) & 1\leq j\leq n/2+1\\
\begin{pmatrix}1 & S_{j}(k)\\
0 & 1
\end{pmatrix}+O(z^{-1}) & n/2+2\leq j\leq n+2
\end{cases}\,.\label{eq:Uj}
\end{equation}

Combining eqs.\ (\ref{eq:Uj}) and (\ref{eq:branchj}) for the contour
shown in figure \ref{fig:gamma-contour-n4}, we obtain the asymptotic
expression for the propagator from the asymptotic past to the asymptotic
future 
\begin{equation}
\lim_{z\to\eta_{+\infty}}\lim_{z_{0}\to\eta_{-\infty}}U_{0}(k,z,z_{0})=U_{2N+2}\mathscr{B}_{2N}U_{2N+1}\cdots\mathscr{B}_{N+1}U_{N+2}\,,\label{eq:matrixforbeta}
\end{equation}
where we have suppressed the $k$ dependence in the right hand side
and introduced $N\equiv n/2$. From eq.\ (\ref{eq:betaFicPropagator21}),
we then get
\begin{equation}
\boxed{\beta_{k}=\left[U_{2N+2}\mathscr{B}_{2N}U_{2N+1}\cdots\mathscr{B}_{N+1}U_{N+2}\right]_{21}}\label{eq:betaPropagatorExp}
\end{equation}
with the subscript $21$ denoting the $21$ entry of the matrix product.
We see that $\beta_{k}$ is now expressed in terms of the Stokes constants
$S_{j}(k)$ and the non-analytic phase factors $\Delta_{j}(k)$.

To obtain the soft corrections to the Bogoliubov $\beta_{k}$, what
we need are the expressions for Stokes constants $S_{j}(k)$ for $j=1,\cdots,2N+2$,
which are typically very complicated for nonzero momentum. To overcome
this complication, in section \ref{sec:Discrete-symmetries}, we will
first establish representations of a discrete rotational symmetry
of the mode equation (\ref{eq:modeEqnChiReduced}) on the propagators,
which allows us to express all Stokes constants in terms of the first
Stokes constant $S_{1}$. We also show that a time-reversal symmetry
for the real variable mode equation (\ref{eq:modeEqnChi}) implies
that the Bogoliubov $\beta_{k}$ must be purely imaginary. 

\section{\label{sec:Discrete-symmetries-1}Discrete symmetries\label{sec:Discrete-symmetries}}

In the topological limit case ($k=0$), as noted in \cite{Chung:2024pot},
the mode equation (\ref{eq:modeEqnChiReduced}) is invariant under
the rotation $z\rightarrow\gamma z$ where $\gamma\equiv\exp(2\pi i/(n+2))$.
This invariance induces a symmetry representation on the set of propagators
across SLs (see eq.\ (\ref{eq:Uj})), equating all Stokes constants
in the system. The single Stokes constant can then be determined using
the single-valuedness of the exact mode $\chi_{k}(z)$ upon propagation
by $2\pi$ on a circle.

For $k\neq0$, we hope to extend the above argument to relate the
different Stokes constants of the mode equation. Note, under $z\rightarrow\gamma z$
the mode equation (\ref{eq:modeEqnChiReduced}) transforms as
\begin{equation}
\left[\partial^{2}_{z}+(\gamma^{-n/2}k)^{2}+z^{n}\right]\chi_{k}(\gamma z)=0\label{eq:RotModeEq}
\end{equation}
where $\chi_{k}(z)$ is a solution to the unrotated mode equation
(\ref{eq:modeEqnChiReduced}). This implies a relation between the
Stokes constants of eq.~(\ref{eq:modeEqnChiReduced}), which are
functions of $k$, and the Stokes constants of eq.~(\ref{eq:RotModeEq}),
which are functions of $\gamma^{-n/2}k$. We derive this relation
in section \ref{subsec:Rotational-symmetry} which relates the different
Stokes constants of eq.\ (\ref{eq:modeEqnChiReduced}) to analytic
continuations in $k$ of a single function $S_{1}(k)$. This function
is further constrained by the single-valuedness of the mode function,
but as we will see in appendix \ref{sec:Remnant-symmetry}, this is
insufficient to constrain the Stokes constants due to a redundancy
in the type of functions which can solve this constraint. Nevertheless,
the relation among all Stokes constants reduces the problem to computing
just $S_{1}(k)$, which we turn to in section \ref{sec:The-first-Stokes}.

In addition to the above, restricting to real conformal time, the
mode equation (\ref{eq:modeEqnChiReduced}) is symmetric under the
time-reversal transformation $\{\eta\to-\eta\}$. This induces another
symmetry representation on the solution space $\sol$  of eq.\ (\ref{eq:modeEqnChiReduced}),
yielding a constraint on the Bogoliubov $\beta_{k}$. In section \ref{subsec:Time-reversal-symmetry},
we construct this representation and use the Wronskian projection
(\ref{eq:betaReal}) to prove that the Bogoliubov $\beta_{k}$ must
be imaginary, providing a simple consistency check for our analysis
in section \ref{sec:The-first-Stokes}.

\subsection{Rotational symmetry\label{subsec:Rotational-symmetry}}

\selectlanguage{english}%
Here, we show that the propagators across the different SLs corresponding
to eq.~(\foreignlanguage{american}{\ref{eq:modeEqnChiReduced}}),
as a consequence of the latter's symmetry under $\{z\to\gamma z,\ k\to\gamma^{-1}k\}$,
can be related to analytic continuations of the single function $S_{1}(k)$.

We begin by working out a more transparent representation of $\omega_{k}(z)$
given the branch cut choice depicted in figure \ref{fig:gamma-contour-n4}
(see appendix \ref{subsec:analytically-continue-time} for a more
detailed version). As depicted in figure \ref{fig:Branch-cut-structure.-1},
we define branch sectors $\mathfrak{B}_{l}$ as the region bounded
between the branch cuts arising from $\psi_{(l-1)\mod n}$ and $\psi_{l\mod n}$
for $\vert z\vert>k^{2/n}$.\footnote{\selectlanguage{american}%
Note $\mathfrak{B}_{l=1}$ is the region bounded by the branch cuts
emanating from $\psi_{n}$ and $\psi_{1}$, which includes the positive
real axis.\selectlanguage{english}%
} Across these regions, the analytic expression for $\omega_{k}(z)$
changes discretely by overall factors of $-1$ corresponding to the
two roots $\pm\sqrt{\omega^{2}_{k}(z)}$.\footnote{\selectlanguage{american}%
The factors of $-1$ can be understood more precisely by comparing
the values of the dispersion relation at two infinitesimally close
points $z_{+}$ and $z_{-}$ padding a branch cut --- these are related
as $\omega_{k}(z_{+})=-\omega_{k}(z_{-})$.\selectlanguage{english}%
} More precisely, in terms of the following Taylor series $\underbar{\ensuremath{\omega}}_{k}(z)$
defined to be analytic in the annulus $\vert z\vert>k^{2/n}$
\begin{equation}
\underline{\omega}_{k}(z)\equiv z^{n/2}\sum^{\infty}_{m=0}\begin{pmatrix}1/2\\
m
\end{pmatrix}\left(\frac{k^{2}}{z^{n}}\right)^{m}\,,\label{eq:underomega-1}
\end{equation}
$\omega_{k}(z)$ in $\mathfrak{B}_{l}$ for the above branch choice
is given by
\begin{equation}
\omega_{k}(z)=(-1)^{l-1}\underline{\omega}_{k}(z),\quad z\in\mathfrak{B}_{l}\label{eq:BidefOmega}
\end{equation}
The choice of eq. (\ref{eq:BidefOmega}) is such that for real time
$\eta$ and nonzero momentum, for any even integer $n$, $\omega_{k}(\eta)>0$.
Correspondingly, in the inner disk $D$ defined as $|z|\leq k^{2/n}$,
$\omega_{k}(z)$ is given by the root whose restriction on the real
line is positive definite, denoted $+\sqrt{k^{2}+z^{n}}$. Combining
these we have

\begin{equation}
\omega_{k}(z)=\begin{cases}
+\sqrt{k^{2}+z^{n}} & z\in D\\
(-1)^{l-1}\underline{\omega}_{k}(z) & z\in\mathfrak{B}_{l}
\end{cases}\,.\label{eq:omega-continued-1}
\end{equation}
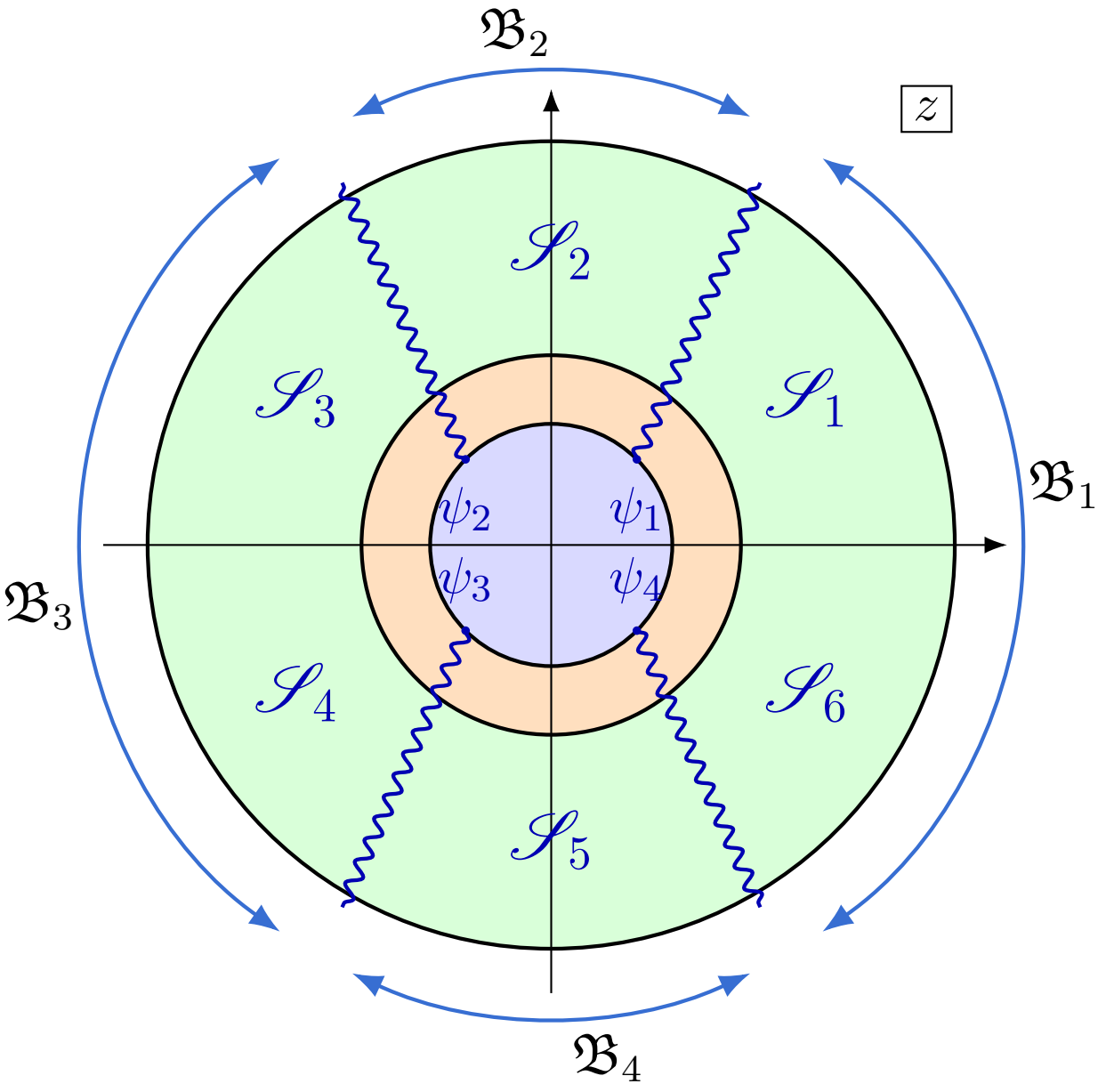
\begin{figure}[t]
\begin{centering}
\centering
\begin{tikzpicture}[>=Latex,scale=0.6]

\def\Rpsi{1.5}
\def\Rmid{2.35}
\def\Rout{5}
\def\Rcut{5.18}
\def\Rarr{5.85}

\fill[green!15] (0,0) circle (\Rout);
\fill[orange!25] (0,0) circle (\Rmid);
\fill[blue!15!white] (0,0) circle (\Rpsi);

\draw[->] (-5.55,0)--(5.65,0);
\draw[->] (0,-5.55)--(0,5.65);

\draw[thick] (0,0) circle (\Rpsi);
\draw[thick] (0,0) circle (\Rmid);
\draw[thick] (0,0) circle (\Rout);

\node[draw,inner sep=2.5pt,font=\normalsize] at (4.65,5.40) {$z$};

\foreach \i/\ang in {1/45,2/135,3/225,4/315}
    \coordinate (psi\i) at (\ang:\Rpsi);

\begin{scope}[
    blue!70!black,
    thick,
    decoration={snake,amplitude=1.5pt,segment length=5pt}
]
    \draw[decorate] (psi1)--(60:\Rcut);
    \draw[decorate] (psi2)--(120:\Rcut);
    \draw[decorate] (psi3)--(240:\Rcut);
    \draw[decorate] (psi4)--(300:\Rcut);
\end{scope}

\fill[blue!70!black] (psi1) circle (1.5pt)
    node[right=4pt,below=2pt,text=blue!70!black,font=\normalsize] {$\psi_1$};

\fill[blue!70!black] (psi2) circle (1.5pt)
    node[left=4pt,below=2pt,text=blue!70!black,font=\normalsize] {$\psi_2$};

\fill[blue!70!black] (psi3) circle (1.5pt)
    node[left=4pt,above=2pt,text=blue!70!black,font=\normalsize] {$\psi_3$};

\fill[blue!70!black] (psi4) circle (1.5pt)
    node[right=4pt,above=2pt,text=blue!70!black,font=\normalsize] {$\psi_4$};

\foreach \i/\ang in {1/30,2/90,3/150,4/210,5/270,6/330}
    \node[blue!70!black,font=\Large] at (\ang:3.65) {$\mathscr{S}_{\i}$};

\begin{scope}[myblue,thick,<->]

    \draw (55:\Rarr) arc[start angle=55,end angle=-55,radius=\Rarr];

    \draw (65:\Rarr) arc[start angle=65,end angle=115,radius=\Rarr];

    \draw (125:\Rarr) arc[start angle=125,end angle=235,radius=\Rarr];

    \draw (245:\Rarr) arc[start angle=245,end angle=295,radius=\Rarr];

\end{scope}

\node[font=\large] at (6.35,0.75) {$\mathfrak{B}_1$};
\node[font=\large] at (-0.45,6.35) {$\mathfrak{B}_2$};
\node[font=\large] at (-6.35,-0.75) {$\mathfrak{B}_3$};
\node[font=\large] at (0.70,-6.35) {$\mathfrak{B}_4$};

\end{tikzpicture}
\par\end{centering}
\caption{\label{fig:Branch-cut-structure.-1}Branch cut structure of $\omega_{k}(z)$
is shown schematically for $\omega_{k}(z)=\sqrt{k^{2}+z^{4}}$. The
branch cuts are chosen to be coincident with anti-Stokes boundaries
that are radially emanating in the AA shown in green. In contrast,
the basis function $\underline{\omega}_{k}(z)$ of eq.~(\ref{eq:underomega-1})
has no branch cuts and is analytic for $|z|>|\psi_{j}|$ where $\psi_{j}$
are the branch points of $\omega_{k}(z)$. We define the branch sectors
$\mathscr{\mathfrak{B}}_{j}$ as the region bounded by two branch
cuts exterior to the disk $|z|\protect\leq|\psi_{j}|$. The inner
annular sector (orange) is not an asymptotic expansion region, but
$\underline{\omega}_{k}(z)$ is analytic there as well. Also, the
anti-Stokes lines and the attendant branch cuts are not as straight
as schematically shown here.}
\end{figure}

Note the Stokes sectors $\mathscr{S}_{j}$ are subsets of corresponding
branch sectors. With our branch choice, eq.\ (\ref{eq:omega-continued-1})
can be expressed in terms of Stokes sector labels
\begin{equation}
\omega_{k}(z)=\begin{cases}
+\sqrt{k^{2}+z^{n}} & z\in D\\
s_{j}\underline{\omega}_{k}(z) & z\in\mathscr{S}^{E}_{j}
\end{cases}\label{eq:omega-k-sj}
\end{equation}
where
\begin{equation}
s_{j}\equiv\begin{cases}
(-1)^{j-1} & 1\leq j\leq n/2+1\\
(-1)^{j} & n/2+2\leq j\leq n+2
\end{cases}
\end{equation}
and where the superscript ``E'' in $\mathscr{S}^{E}_{j}$ corresponds
to the extension of the Stokes sector to cover the region between
the Stokes sector and the inner disk.\footnote{More formally, we define
\[
\mathscr{S}^{E}_{j}=\{|z|>|\psi_{j}||z\text{ restricted to the region bounded by anti-Stokes boundaries belonging to }\mathscr{S}_{j}\}
\]
where we remind the reader that the branch cut has been chosen to
be coincident with the outward emanating anti-Stokes boundaries. Each
real axis sliver between each set of two anti-Stokes boundaries surrounding
the real axis can be evenly included in the surrounding $\mathscr{S}^{E}_{j}$
to not leave out any $|z|>|\psi_{j}|$.}\foreignlanguage{american}{ }

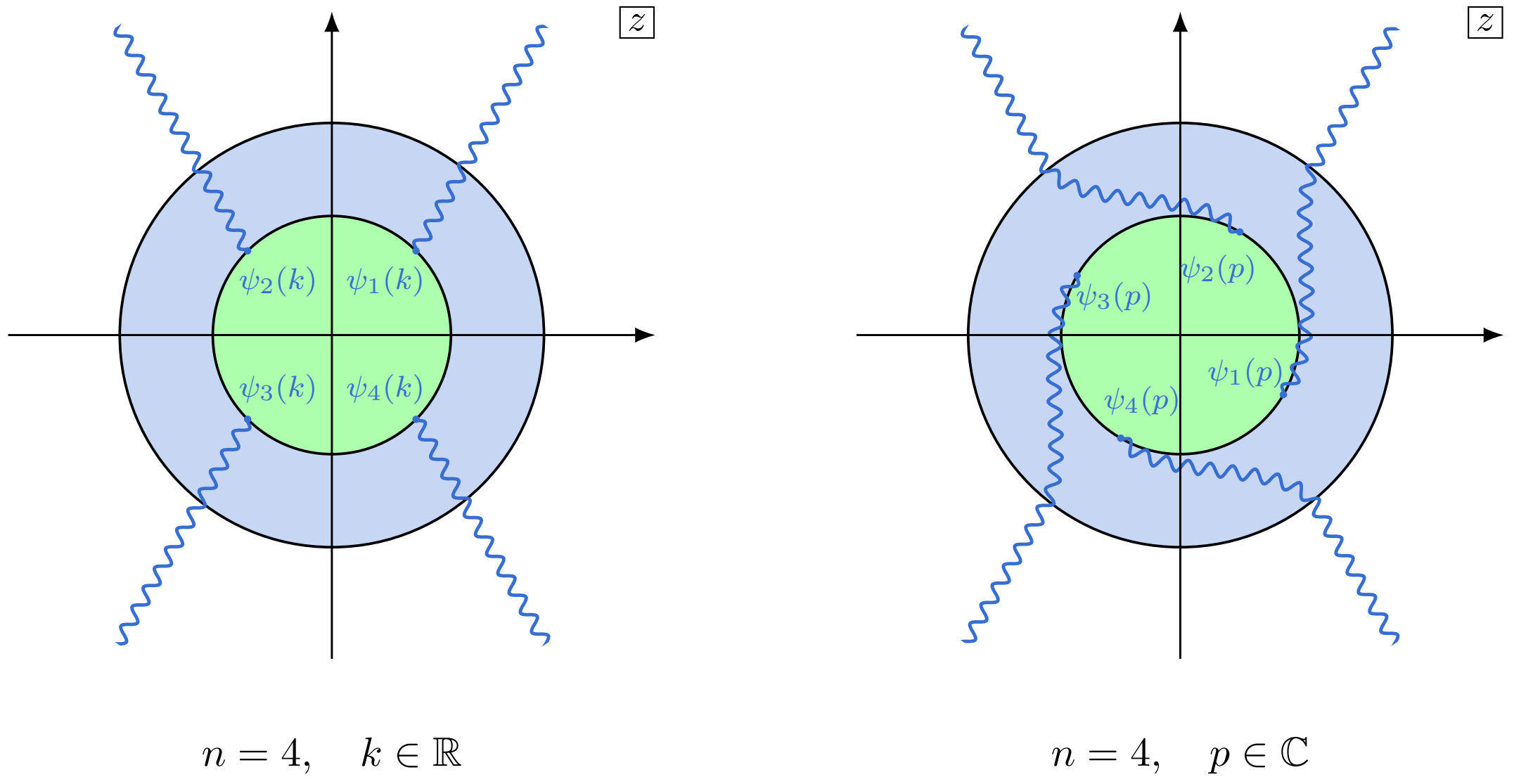
\begin{figure}[h]
\centering
\begin{tikzpicture}[>=Latex,scale=0.7,font=\small]

\def\Rgreen{1.60}
\def\Rpsi{1.60}
\def\Rout{2.85}
\def\Rcut{5.05}
\def\Xaxis{4.35}
\def\Yaxis{4.35}

\tikzset{
  branchcut/.style={
    myblue,
    line width=0.9pt,
    decoration={
      snake,
      amplitude=1.7pt,
      segment length=6pt
    }
  },
  circleboundary/.style={
    black,
    line width=0.7pt,
  }
}

\begin{scope}[shift={(-5.7,0)}]

  \fill[myblue!28!white,even odd rule]
      (0,0) circle (\Rout)
      (0,0) circle (\Rpsi);

  \fill[green!32!white]
      (0,0) circle (\Rgreen);

  \draw[circleboundary]
      (0,0) circle (\Rpsi);

  \draw[circleboundary]
      (0,0) circle (\Rout);

  \foreach \i/\ang in {
      1/45,
      2/135,
      3/225,
      4/315
  }{
      \coordinate (Lpsi\i) at (\ang:\Rpsi);
  }

  \begin{scope}
    \clip (0,0) circle (\Rcut);

    \begin{scope}[branchcut]

      \draw[decorate]
          (Lpsi1)
          -- ($(Lpsi1)+(60:8)$);

      \draw[decorate]
          (Lpsi2)
          -- ($(Lpsi2)+(120:8)$);

      \draw[decorate]
          (Lpsi3)
          -- ($(Lpsi3)+(240:8)$);

      \draw[decorate]
          (Lpsi4)
          -- ($(Lpsi4)+(300:8)$);

    \end{scope}
  \end{scope}

  \draw[->,line width=0.55pt]
      (-\Xaxis,0)--(\Xaxis,0);

  \draw[->,line width=0.55pt]
      (0,-\Yaxis)--(0,\Yaxis);

  \foreach \i in {1,2,3,4}{
      \fill[myblue]
          (Lpsi\i) circle (1.4pt);
  }

  \node[myblue,font=\scriptsize]
      at (45:1.02) {$\psi_1(k)$};

  \node[myblue,font=\scriptsize]
      at (135:1.02) {$\psi_2(k)$};

  \node[myblue,font=\scriptsize]
      at (225:1.02) {$\psi_3(k)$};

  \node[myblue,font=\scriptsize]
      at (315:1.02) {$\psi_4(k)$};

  \node[
      draw,
      inner sep=2pt,
      font=\small
  ] at (4.10,4.20) {$z$};

  \node[font=\normalsize]
      at (0,-5.65)
      {$n=4,\quad k\in\mathbb{R}$};

\end{scope}

\begin{scope}[shift={(5.7,0)}]

  \fill[myblue!28!white,even odd rule]
      (0,0) circle (\Rout)
      (0,0) circle (\Rpsi);

  \fill[green!32!white]
      (0,0) circle (\Rgreen);

  \draw[circleboundary]
      (0,0) circle (\Rpsi);

  \draw[circleboundary]
      (0,0) circle (\Rout);

  \foreach \i/\ang in {
      1/-30,
      2/60,
      3/150,
      4/240
  }{
      \coordinate (Rpsi\i) at (\ang:\Rpsi);
  }

  \foreach \i/\ang in {
      1/45,
      2/135,
      3/225,
      4/315
  }{
      \coordinate (Rref\i) at (\ang:\Rpsi);
  }

  \path[name path=Routercircle]
      (0,0) circle (\Rout);

  \path[name path=Rrayone]
      (Rref1)
      -- ($(Rref1)+(60:8)$);

  \path[name path=Rraytwo]
      (Rref2)
      -- ($(Rref2)+(120:8)$);

  \path[name path=Rraythree]
      (Rref3)
      -- ($(Rref3)+(240:8)$);

  \path[name path=Rrayfour]
      (Rref4)
      -- ($(Rref4)+(300:8)$);

  \path[
      name intersections={
        of=Routercircle and Rrayone,
        by={Rjoin1}
      }
  ];

  \path[
      name intersections={
        of=Routercircle and Rraytwo,
        by={Rjoin2}
      }
  ];

  \path[
      name intersections={
        of=Routercircle and Rraythree,
        by={Rjoin3}
      }
  ];

  \path[
      name intersections={
        of=Routercircle and Rrayfour,
        by={Rjoin4}
      }
  ];

  \begin{scope}
    \clip (0,0) circle (\Rcut);

    \begin{scope}[branchcut]

      \draw[decorate]
          (Rpsi1)
          .. controls
              (0:1.95)
              and ($(Rjoin1)-(60:0.75)$)
          ..
          (Rjoin1)
          -- ($(Rjoin1)+(60:8)$);

      \draw[decorate]
          (Rpsi2)
          .. controls
              (82:1.95)
              and ($(Rjoin2)-(120:0.75)$)
          ..
          (Rjoin2)
          -- ($(Rjoin2)+(120:8)$);

      \draw[decorate]
          (Rpsi3)
          .. controls
              (180:1.95)
              and ($(Rjoin3)-(240:0.75)$)
          ..
          (Rjoin3)
          -- ($(Rjoin3)+(240:8)$);

      \draw[decorate]
          (Rpsi4)
          .. controls
              (260:1.95)
              and ($(Rjoin4)-(300:0.75)$)
          ..
          (Rjoin4)
          -- ($(Rjoin4)+(300:8)$);

    \end{scope}
  \end{scope}

  \draw[->,line width=0.55pt]
      (-\Xaxis,0)--(\Xaxis,0);

  \draw[->,line width=0.55pt]
      (0,-\Yaxis)--(0,\Yaxis);

  \foreach \i in {1,2,3,4}{
      \fill[myblue]
          (Rpsi\i) circle (1.4pt);
  }

  \node[myblue,font=\scriptsize]
      at (-30:1.02) {$\psi_1(p)$};

  \node[myblue,font=\scriptsize]
      at (60:1.02) {$\psi_2(p)$};

  \node[myblue,font=\scriptsize]
      at (150:1.02) {$\psi_3(p)$};

  \node[myblue,font=\scriptsize]
      at (240:1.02) {$\psi_4(p)$};

  \node[
      draw,
      inner sep=2pt,
      font=\small
  ] at (4.10,4.20) {$z$};

  \node[font=\normalsize]
      at (0,-5.65)
      {$n=4,\quad p\in\mathbb{C}$};

\end{scope}

\end{tikzpicture}\caption{Shown is our convention for branch cut deformation under the complexification
$k\to p$ for $n=4$. As $\arg p$ varies, the branch points $\psi_{j}(p)$
rotate and drag the branch cuts emanating from them. We choose to
deform the cuts only in the annulus while keeping their outer segments
fixed. The corresponding branch sectors, and hence the factors $s_{j}$,
may therefore be kept fixed. This guarantees that $\omega_{p}(z)$
and $\omega_{k}(z)$ have the same leading asymptotic term $s_{j}z^{n/2}$
for $z\in\mathscr{S}_{j}$.\label{fig:Shown-is-our-1}}
\end{figure}

\selectlanguage{american}%
With the above, consider the $j$-th Stokes constant given by
\begin{equation}
S_{j}(k)=\lim_{|z_{j0}|,|z_{j1}|\to\infty}[U_{j}(k)]_{21(12)}
\end{equation}
\foreignlanguage{english}{where $z_{j1}$ and $z_{j0}$ are two points
on the ASLs bounding the $j$-th Stokes sector $\mathscr{S}_{j}$
--- avoiding the branch cuts,}\footnote{An explicit representation of $z_{j0}$ and $z_{j1}$ is 
\[
z_{j0}\equiv R\exp\left[i\left(\frac{2\pi(j-1)}{n+2}+\epsilon\right)\right]\,,\quad z_{j1}\equiv R\exp\left[i\left(\frac{2\pi j}{n+2}-\epsilon\right)\right]
\]
where $R$ is a large real number and $\epsilon\to0^{+}$ a positive
infinitesimal regulator.} and where
\begin{equation}
U_{j}(k)\equiv U_{0}(k,z_{j1},z_{j0})=\mathbb{P}\left[\mathbb{I}_{2\times2}\exp\left(\int_{\Gamma(z_{j1},z_{j0})}dwM_{0}(k,w)\right)\right]
\end{equation}
\begin{equation}
M_{0}(k,z)=\frac{\omega'_{k}(z)}{2\omega_{k}(z)}\begin{pmatrix}0 & e^{+2i\theta(k,z)}\\
e^{-2i\theta(k,z)} & 0
\end{pmatrix},\quad\theta(k,z)\equiv\int^{z}_{0}dz'\omega_{k}(z')\,.\label{eq:M0-thetakz}
\end{equation}
To derive the symmetry representation on the Stokes constants as functions
of $k$, we need to consider the rotation $z\to\gamma z$, as mentioned
above eq.\ (\ref{eq:RotModeEq}), in the WKB phase $\theta(k,z)$
in the propagators. However, this rotation is obstructed when there
is a branch cut between $z$ and $\gamma z$, as is the case when
rotating between two Stokes sectors $\mathscr{S}_{j}$ and $\mathscr{S}_{j+1}$
separated by the $j$-th branch cut. Nevertheless, for this case,
we may introduce an auxiliary dispersion relation $\omega^{{\rm II}[j]}_{k}(z)$
that lands on the second Riemann sheet in $\mathscr{S}^{E}_{j+1}$
which equals $\omega_{k}(z)$ otherwise. This construction implies
that $\omega^{{\rm II}[j]}_{k}(z)$ is continuous across the $j$-th
branch cut of $\omega_{k}(z)$, which makes rotating the WKB phase
of $\omega^{{\rm II}[j]}_{k}(z)$ more transparent. Formally, we define
\begin{equation}
\omega^{{\rm II}[j]}_{k}(z)\equiv\begin{cases}
-\omega_{k}(z) & z\in\mathscr{S}^{E}_{j+1}\\
+\omega_{k}(z) & \text{otherwise}
\end{cases}\,,\quad b=1\,,\label{eq:auxiliary-omega}
\end{equation}
where $b\in\{0,1\}$ denotes the number of branch cuts between $\mathscr{S}_{j}$
and $\mathscr{S}_{j+1}$. Compared to the original WKB phase $\theta(k,z)$,
the phase integral of the auxiliary dispersion relation in $\mathscr{S}_{j}$
and $\mathscr{S}_{j+1}$ satisfies
\begin{equation}
\theta^{{\rm II}[j]}(k,z)\equiv\int^{z}_{0}dz'\omega^{{\rm II}[j]}_{k}(z')=\begin{cases}
\theta(k,z) & z\in\mathscr{S}_{j}\\
2\Delta_{j}(k)-\theta(k,z) & z\in\mathscr{S}_{j+1}
\end{cases}\,,\quad b=1\,,\label{eq:auxiliary-WKB-phase}
\end{equation}
where the phase factor $\Delta_{j}(k)$ is defined in eq.\ (\ref{eq:Deltaj}).
\foreignlanguage{english}{Consequently, in the following where we
define analytic continuations of the functions $S_{j}(k)$ to complex
momenta $p=\gamma^{-nm/2}k$ where $m\in\mathbb{Z}_{\geq0}$, the
corresponding dispersion relation $\omega_{p}(z)$ is taken to have
branch cuts such that for $\vert z\vert\gg\vert p\vert^{2/n}$, $\omega_{p}(z)/\omega_{k}(z)\sim1$.
Therefore, for this momentum complexification we have
\begin{equation}
\omega_{p}(z)=s_{j}\underline{\omega}_{p}(z),\quad z\in\mathscr{S}_{j}\,,
\end{equation}
where $\underline{\omega}_{p}(z)$ denotes the Taylor series (\ref{eq:underomega-1})
with $k$ replaced by $p$. Such a branch choice, shown in figure
\ref{fig:Shown-is-our-1}, can be obtained by replacing $k$ with
$p$ in eq.\ (\ref{eq:underomega-1}) and analytically continuing
to $z\in D$. This complexifies the momentum in the WKB phase as $\theta(p,z)\equiv\int^{z}_{0}dz'\omega_{p}(z').$}\footnote{We provide in appendix \ref{subsec:analytically-continue-momentum}
another perspective on complexifying $\theta(k,z)$.}\foreignlanguage{english}{ Note this preserves the Stokes sectors
$\mathscr{S}_{j}$ for the case of continued momenta. }

\selectlanguage{english}%
We now turn to deriving the symmetry representation of the propagators.
Consider two sets of points belonging to adjacent Stokes sectors,
given by $\{\gamma z_{j1},\gamma z_{j0}\}\subset\mathscr{S}_{j+1}$ and
$\{z_{j1},z_{j0}\}\subset\mathscr{S}_{j}$, where these points lie close
to the ASLs in each sector while avoiding branch cuts. Given this,
the propagator in $\mathscr{S}_{j+1}$ is given by \foreignlanguage{american}{
\begin{align}
U_{j+1}(k) & =\mathbb{P}\left[\mathbb{I}_{2\times2}\exp\left(\int_{\Gamma(\gamma z_{j1},\gamma z_{j0})}dzM_{0}(k,z)\right)\right]\nonumber \\
 & =\mathbb{P}\left[\mathbb{I}_{2\times2}\exp\left(\int_{\Gamma(\gamma z_{j1},\gamma z_{j0})}dz\frac{\omega'_{k}(z)}{2\omega_{k}(z)}\begin{pmatrix}0 & e^{+2i\theta(k,z)}\\
e^{-2i\theta(k,z)} & 0
\end{pmatrix}\right)\right]\,
\end{align}
where $\Gamma(\gamma z_{j1},\gamma z_{j0})$ denotes the propagation
contour.}

\selectlanguage{american}%
Under a variable change $z=\gamma y$, the exponent of the above path
ordered exponential integral goes to
\begin{align}
 & \int_{\Gamma(\gamma z_{j1},\gamma z_{j0})}dz\frac{\omega'_{k}(z)}{2\omega_{k}(z)}\begin{pmatrix}0 & e^{+2i\theta(k,z)}\\
e^{-2i\theta(k,z)} & 0
\end{pmatrix}\nonumber \\
= & \int_{\Gamma(z_{j1},z_{j0})}dy\frac{ny^{n-1}}{4\left(\gamma^{-n}k^{2}+y^{n}\right)}\begin{pmatrix}0 & e^{+2i\theta(k,\gamma y)}\\
e^{-2i\theta(k,\gamma y)} & 0
\end{pmatrix}
\end{align}
where $\Gamma(z_{j1},z_{j0})$ now lies in $\mathscr{S}_{j}$ and
note the pre-factor $\omega_{k}'/\omega_{k}$ is single valued. Now,
since $\gamma y$ still lies in $\mathscr{S}_{j+1}$, we use another
variable change $z'=\gamma u$ in $\theta(k,\gamma y)=\int^{\gamma y}_{0}dz'\omega_{k}(z')$
to relate the phase integral $\theta(k,\gamma y)$ to the corresponding
integrals for $\omega_{\gamma^{-n/2}k}$ in $\mathscr{S}_{j}$. If
the Stokes sectors are not separated by branch cuts, then this simple
variable change leads to the relation
\begin{align}
\theta(k,\gamma y) & =\int^{\gamma y}_{0}dz'\omega_{k}(z')\nonumber \\
 & =\int^{y}_{0}\gamma du\sqrt{k^{2}+\gamma^{n}u^{n}}\nonumber \\
 & =\int^{y}_{0}\gamma^{1+n/2}du\sqrt{\gamma^{-n}k^{2}+u^{n}}\nonumber \\
 & =-\theta(\gamma^{-n/2}k,y)\,,\label{eq:theta-rot-1}
\end{align}
where we have used $\gamma^{1+n/2}=-1$. If there is a branch cut
of $\omega_{k}(z')$ between $\mathscr{S}_{j}$ and $\mathscr{S}_{j+1}$,
then we may first use the auxiliary dispersion $\omega^{{\rm II}[j]}_{k}(z')$
which is continuous across this branch cut to get the analogue of
eq.\ (\ref{eq:theta-rot-1}):
\begin{equation}
\theta^{{\rm II}[j]}(k,\gamma y)=-\theta^{{\rm II}[j]}(\gamma^{-n/2}k,y)\,.
\end{equation}
Then, we map the above auxiliary WKB phase to $\theta(k,z)$ using
eq.\ (\ref{eq:auxiliary-WKB-phase}), yielding
\begin{equation}
\theta(k,\gamma y)=2\Delta_{j}(k)+\theta(\gamma^{-n/2}k,y)\,.\label{eq:theta-rot2}
\end{equation}
Combining eqs.\ (\ref{eq:theta-rot-1}) and (\ref{eq:theta-rot2})
leads to
\begin{equation}
\theta(k,\gamma y)=\begin{cases}
2\Delta_{j}(k)+\theta(\gamma^{-n/2}k,y) & b=1\\
-\theta(\gamma^{-n/2}k,y) & b=0
\end{cases}\,,\quad y\in\mathscr{S}_{j}\,.\label{eq:theta-rotation}
\end{equation}
This with eq.\ (\ref{eq:M0-thetakz}) leads to
\begin{equation}
\int_{\Gamma(z_{j1},z_{j0})}\gamma dyM_{0}(k,\gamma y)=\begin{cases}
\mathscr{R}_{j}(k)\left[\int_{\Gamma(z_{j1},z_{j0})}dyM_{0}(\gamma^{-n/2}k,y)\right]\mathscr{R}^{-1}_{j}(k) & b=1\\
B\left[\int_{\Gamma(z_{j1},z_{j0})}dyM_{0}(\gamma^{-n/2}k,y)\right]B & b=0
\end{cases}\,,
\end{equation}
where we have used eq.\ (\ref{eq:theta-rotation}) and introduced
\begin{equation}
B\equiv\begin{pmatrix}0 & 1\\
1 & 0
\end{pmatrix},\quad\mathscr{R}_{j}(k)\equiv\begin{pmatrix}e^{2i\Delta_{j}(k)} & 0\\
0 & e^{-2i\Delta_{j}(k)}
\end{pmatrix}\,.\label{eq:BRj}
\end{equation}
This yields a rotational symmetry representation on the propagator
in the zero gauge
\begin{equation}
U_{j+1}(k)=\begin{cases}
\mathscr{R}_{j}(k)U_{j}(\gamma^{-n/2}k)\mathscr{R}^{-1}_{j}(k) & 1\leq j\leq n/2\\
BU_{j}(\gamma^{-n/2}k)B^{-1} & j=n/2+1\\
\mathscr{R}_{j-1}(k)U_{j}(\gamma^{-n/2}k)\mathscr{R}^{-1}_{j-1}(k) & n/2+2\leq j\leq n+1
\end{cases}\,.\label{eq:propagatorRotation}
\end{equation}

Composing the above $j$ times, it then follows from eqs.\ (\ref{eq:propagatorComplex}),
(\ref{eq:propagatorRotation}) and (\ref{eq:DeltaRotation}) that
(see appendix \ref{sec:Propagator-rotation-details})
\begin{align}
S_{j+1}(k) & =e^{-4i\tilde{\Delta}_{j}(k)}S_{1}(\gamma^{-Nj}k),\quad1\leq j\leq N,\label{eq:SRot1}\\
S_{N+2}(k) & =e^{4i\tilde{\Delta}_{N}(k)}S_{1}(\gamma^{-N(N+1)}k),\label{eq:SRot2}\\
S_{N+2+j}(k) & =e^{4i\tilde{\Phi}_{j}(k)}S_{1}(\gamma^{-N(N+1+j)}k),\quad1\leq j\leq N,\label{eq:SRot3}
\end{align}
where $\tilde{\Delta}_{j}$ and $\tilde{\Phi}_{j}$ correspond to
the following alternating sums
\begin{equation}
\tilde{\Delta}_{j}(k)\equiv\sum^{j}_{r=1}(-1)^{j-r}\Delta_{r}(k),\quad\tilde{\Phi}_{j}(k)\equiv(-1)^{j}\tilde{\Delta}_{N}(k)-\tilde{\Delta}_{j}(k).\label{eq:alternatingPhase}
\end{equation}
To arrive at the above we have used 
\begin{equation}
\Delta_{j}(\gamma^{-N}k)=-\Delta_{j}(k),\quad\Delta_{N+j}(k)=-\Delta_{j}(k)\label{eq:DeltaRotation}
\end{equation}
which follows from $\Delta_{j}(k)\propto k^{\frac{n+2}{n}}$ as may
be derived from the definition (\ref{eq:Deltaj}) of $\Delta_{j}(k)$
with the variable change $z=uk^{2/n}$. The three relations (\ref{eq:SRot1})-(\ref{eq:SRot3})
express all the Stokes constants in terms of the first Stokes constant
and certain phase factors. As a result, to compute the Bogoliubov
$\beta_{k}$ using propagator with eq.\ (\ref{eq:betaPropagatorExp}),
we only need to compute the first Stokes constant $S_{1}(k)$, which
will be the task for section \ref{sec:The-first-Stokes}.

\subsection{Time-reversal symmetry\label{subsec:Time-reversal-symmetry}}

The even parity of $n$ guarantees that the dispersion relation $\omega_{k}(\eta)=\sqrt{k^{2}+\eta^{n}}$
is invariant under time-reversal $\eta\to-\eta$ with $k,\,\eta\in\mathbb{R}$.
We will show in this subsection that this symmetry forces the Bogoliubov
 $\beta_{k}$ to be purely imaginary. Throughout this subsection,
we will assume $k,\,\eta\in\mathbb{R}$.

We start with representing the time-reversal symmetry on the solution
space $\sol$ (defined above eq.\ (\ref{eq:bogoTransReal})), which
has two bases $\{\chi_{k,1}(\eta)\,,\chi^{*}_{k,1}(\eta)\}$ and $\{\chi_{k,2}(\eta),\,\chi^{*}_{k,2}(\eta)\}$
defined in eq.\ (\ref{eq:chi12}). As $\chi_{k,1}(\eta)\in\sol$,
under $\eta\to-\eta$ the mode equation (\ref{eq:modeEqnChiReduced})
for $\chi_{k,1}(\eta)$ goes to 
\begin{equation}
\left[\partial^{2}_{\eta}+\omega^{2}_{k}(\eta)\right]\chi_{k,1}(-\eta)=0\,,
\end{equation}
where we have used $\omega_{k}(\eta)=\omega_{k}(-\eta)$, which means
$\chi_{k,1}(-\eta)\in\sol$. We then expand $\chi_{k,1}(-\eta)$ over
the basis $\{\chi_{k,2}(\eta),\,\chi^{*}_{k,2}(\eta)\}$ as
\begin{equation}
\chi_{k,1}(-\eta)=c\,\chi_{k,2}(\eta)+d\,\chi^{*}_{k,2}(\eta)\label{eq:chik1Expansion}
\end{equation}
with $c$ and $d$ constants. These constants can be obtained by investigating
the asymptotic behavior of $\chi_{k,1}(-\eta)$. Note as $\eta$ approaches
the asymptotic future $\eta_{+\infty}$, $-\eta$ approaches the far
past $\eta_{-\infty}$, meaning that
\begin{equation}
\lim_{\eta\to\eta_{+\infty}}\chi_{k,1}(-\eta)=\lim_{-\eta\to\eta_{-\infty}}\frac{e^{-i\int^{-\eta}_{0}d\bar{\eta}\omega_{k}(\bar{\eta})}}{\sqrt{2\omega_{k}(-\eta)}}=\lim_{\eta\to\eta_{+\infty}}\frac{e^{i\int^{\eta}_{0}d\bar{\eta}\omega_{k}(\bar{\eta})}}{\sqrt{2\omega_{k}(\eta)}}\,,
\end{equation}
where we have used $\omega_{k}(\eta)=\omega_{k}(-\eta)$ again and
$\eta_{+\infty}=-\eta_{-\infty}$. Note the right hand side of the
above equation is identical to the boundary condition of $\chi^{*}_{k,2}(\eta)$,
giving 
\begin{equation}
\lim_{\eta\to\eta_{+\infty}}\chi_{k,1}(-\eta)=\lim_{\eta\to\eta_{+\infty}}\chi^{*}_{k,2}(\eta)\,,
\end{equation}
then from the expansion (\ref{eq:chik1Expansion}) we obtain $c=0$
and $d=1$, yielding 
\begin{equation}
\chi_{k,1}(-\eta)=\chi^{*}_{k,2}(\eta)\,.\label{eq:chi1mEta}
\end{equation}
This leads to 
\begin{equation}
\chi_{k,2}(-\eta)=\chi^{*}_{k,1}(\eta)\,.\label{eq:chi2mEta}
\end{equation}

Next, the Wronskian projection (\ref{eq:betaReal}) implies
\begin{align}
\beta_{k} & =i\left[\chi_{k,1}(\eta)\partial_{\eta}\chi_{k,2}(\eta)-\chi_{k,2}(\eta)\partial_{\eta}\chi_{k,1}(\eta)\right]\label{eq:betakEta}\\
 & =i\left[\chi_{k,1}(-\eta)\partial_{-\eta}\chi_{k,2}(-\eta)-\chi_{k,2}(-\eta)\partial_{-\eta}\chi_{k,1}(-\eta)\right]\,,\label{eq:betakmEta}
\end{align}
where in the second line we have changed the time variable to $-\eta$
since $\beta_{k}$ is time-independent. On one hand, substituting
eqs.\ (\ref{eq:chi1mEta}) and (\ref{eq:chi2mEta}) into eq.\ (\ref{eq:betakmEta})
yields
\begin{equation}
\beta_{k}=i\left[\chi^{*}_{k,1}(\eta)\partial_{\eta}\chi^{*}_{k,2}(\eta)-\chi^{*}_{k,2}(\eta)\partial_{\eta}\chi^{*}_{k,1}(\eta)\right]\,.
\end{equation}
On the other hand, conjugating eq.\ (\ref{eq:betakEta}) gives
\begin{equation}
\beta^{*}_{k}=-i\left[\chi^{*}_{k,1}(\eta)\partial_{\eta}\chi^{*}_{k,2}(\eta)-\chi^{*}_{k,2}(\eta)\partial_{\eta}\chi^{*}_{k,1}(\eta)\right]\,.
\end{equation}
Hence,
\begin{equation}
\boxed{\beta^{*}_{k}=-\beta_{k}}\,,\label{eq:betaImaginary}
\end{equation}
which means that the Bogoliubov $\beta_{k}$ must be purely imaginary
as a result of the time-reversal symmetry.

\section{The first Stokes constant\label{sec:The-first-Stokes}}

In section \ref{subsec:Rotational-symmetry} we noted that all the
Stokes constants are expressible in terms of the first Stokes constant
function $S_{1}(k)$ with rotated argument and certain phase factors.
As may be seen from the complexified version of the amplitude ODE
(\ref{eq:amplitudeODE0}) or its integrated version
\begin{align}
\alpha_{k}(z) & =\alpha_{k}(z_{0})+\int^{z}_{z_{0}}dz'\,\frac{\omega'_{k}(z')}{2\omega_{k}(z')}e^{+2i\theta(k,z')}\beta_{k}(z')\label{eq:IntModeEq1}\\
\beta_{k}(z) & =\beta_{k}(z_{0})+\int^{z}_{z_{0}}dz'\,\frac{\omega'_{k}(z')}{2\omega_{k}(z')}e^{-2i\theta(k,z')}\alpha_{k}(z')\label{eq:IntMode2}
\end{align}
an explicit computation of the WKB amplitudes, thereby the Stokes
constants, requires integrals of oscillatory exponentials along contours
in the asymptotic Stokes sectors. To mitigate this obstacle, one may
naively consider evaluating the asymptotic expansions of eqs.\ (\ref{eq:IntModeEq1})
and (\ref{eq:IntMode2}) instead, which seems to simplify the computation
of the Stokes constants. However, ordinary asymptotic expansions are
not accurate enough to capture the Stokes constants. To see this,
consider the exact mode decomposition over the fictitious WKB basis
$\chi_{k}(z)=\alpha_{k}(z)f_{-}(k,z)+\beta_{k}(z)f_{+}(k,z)$. In
the first Stokes sector $\mathscr{S}_{1}$ and in our branch convention,
$f_{-}$ and $f_{+}$ are exponentially amplified and suppressed by
$|z|$, respectively. This means the first Stokes constant $S_{1}(k)$,
an $O(1)$ shift in $\beta_{k}(z)$, is exponentially suppressed in
the mode decomposition and is invisible to all algebraic order asymptotic
contributions. Therefore, to compute $S_{1}(k)$ by evaluating the
asymptotic behavior of eqs.\ (\ref{eq:IntModeEq1}) and (\ref{eq:IntMode2}),
one must go beyond ordinary asymptotic expansions to take the exponentially
suppressed contributions into account. We therefore consider the exponentially
improved asymptotic expansion of the WKB amplitudes as $|z|\rightarrow\infty$,
which can be schematically expressed as
\begin{equation}
V_{k}(z)\,\dot{\sim}\,\text{constant}+\text{power suppressed}+\text{exponentially suppressed}\label{eq:WKB-eae-sloppy}
\end{equation}
where $\dot{\sim}$ denotes exponentially improved asymptotic expansion
(which we will often abbreviate as ``e.a.e.'').\footnote{Exponentially improved asymptotic expansions are defined by truncating
an ordinary asymptotic series near its least term and re-expanding
the resulting remainder, which improves the series accuracy to exponentially
small terms beyond all algebraic orders. It can be viewed as making
the Stokes phenomena manifest.}

After plugging the exponentially improved asymptotic expansions of
the WKB amplitudes into eqs.\ (\ref{eq:IntModeEq1}) and (\ref{eq:IntMode2}),
we encounter integrals of the form
\begin{equation}
E_{p}(z)\equiv z^{p-1}\int^{\infty}_{z}dt\frac{e^{-t}}{t^{p}},\quad z\neq0\,.
\end{equation}
These are called generalized exponential integrals and have known
exponentially improved asymptotic expansions given by
\begin{equation}
E_{p}(z)\dot{\sim}\frac{e^{-z}}{z}\sum^{\infty}_{s=0}\frac{(-1)^{s}(p)_{s}}{z^{s}}+\Theta_{\mathbb{R}}\left(\arg z-\pi\right)2\pi i\frac{e^{-ip\pi}}{\Gamma(p)}z^{p-1},\quad-\frac{3\pi}{2}<\arg z<\frac{7\pi}{2}\label{eq:ExpIntAsymp}
\end{equation}
where $\Theta_{\mathbb{R}}(x)$ denotes a step-function that jumps
at $x=0$ with $\Theta_{\mathbb{R}}(0)=1/2$ (see appendix \ref{sec:The-generalized-exponential}).
As seen in the above, these expansions undergo Stokes jumps, with
known Stokes constants, as $z$ is analytically continued across the
negative real line at $\arg z=\pi$.\footnote{The analogs of the WKB basis functions in the expansion eq.~(\ref{eq:ExpIntAsymp})
are $e^{-z}/z$ and $z^{p-1}$, corresponding to the leading exponentially
improved asymptotic expansions expressible as a superposition of these.
For $p=1$, the basis functions can be easily seen from the differential
equation satisfied by $E_{1}(z)$, which is $zE_{1}''(z)+(z+1)E_{1}'(z)=0.$
Stokes lines in these scenarios are defined as contours along which
the two basis functions $e^{-z}/z$ and $z^{p-1}$ are maximally dominant/subdominant
with respect to each other --- here this occurs along the negative
real line.}

Given the expansions (\ref{eq:WKB-eae-sloppy}) and (\ref{eq:ExpIntAsymp}),
we note that the coupled integrated equations (\ref{eq:IntModeEq1})
and (\ref{eq:IntMode2}) could be understood as an automorphism on
$\mathcal{V}_{k}$ --- the functional space of exponentially improved
asymptotic expansions of the WKB amplitudes, denoted as $[V_{k}(z)]_{{\rm e.a.e}.}$.
In particular, plugging $[V_{k}(z)]_{{\rm e.a.e}.}$ given by eq.\ (\ref{eq:WKB-eae-sloppy})
into the integrals on the right hand side of eqs.\ (\ref{eq:IntModeEq1})
and (\ref{eq:IntMode2}) and computing them term by term, we find
both 
\begin{equation}
\int^{z}_{z_{0}}dz'\,\frac{\omega'_{k}(z')}{2\omega_{k}(z')}e^{+2i\theta(k,z')}[\beta_{k}(z')]_{{\rm e.a.e.}}\quad\text{and}\quad\int^{z}_{z_{0}}dz'\,\frac{\omega'_{k}(z')}{2\omega_{k}(z')}e^{-2i\theta(k,z')}[\alpha_{k}(z')]_{{\rm e.a.e.}}\label{eq:int-eae-sloppy}
\end{equation}
have exponentially improved asymptotic expansions belonging to $\mathcal{V}_{k}$.
Therefore, we may view
\begin{equation}
\int^{z}_{z_{0}}dz'\,\frac{\omega'_{k}(z')}{2\omega_{k}(z')}e^{\pm2i\theta(k,z')}
\end{equation}
acting on $[\beta_{k}(z')]_{{\rm e.a.e.}}$ and $[\alpha_{k}(z')]_{{\rm e.a.e.}}$
as an automorphism on $\mathcal{V}_{k}$. We will see that this automorphism
yields a set of recursion relations on the expansion coefficients
of $[\alpha_{k}(z)]_{{\rm e.a.e.}}$ and $[\beta_{k}(z)]_{{\rm e.a.e.}}$,
whose solution leads to a power series expression of the first Stokes
constant.

The rest of this section is organized as follows. In section \ref{subsec:Non-analyticity-isolation},
we decompose the WKB phase $\theta(k,z)$ in the first Stokes sector
into three components: the leading term $z^{n/2+1}/(n/2+1)$, a power
series of $1/z$, and a constant factor non-analytic in $k$. Then,
in section \ref{subsec:Automorphism-on-WKB}, we show that this decomposition
allows us to compute the integrals in eq.~(\ref{eq:int-eae-sloppy})
using the expression (\ref{eq:ExpIntAsymp}) for the generalized exponential
integrals. This computation leads to a detailed representation of
the automorphism on $\mathcal{V}_{k}$. In section \ref{subsec:Solvable-technique},
we show how to translate this automorphism to a set of recursion relations
of the expansion coefficients of the WKB amplitudes, which is solved
in section \ref{subsec:Explicit-recursion-relation}, yielding a closed
form expression for the first Stokes constant up to quadratic order.

The above method follows \cite{hinton1979stokesmultipliers,zhu1992stokesconstants}
up to the automorphism perspective and the choice of the WKB basis
functions. The WKB basis functions in \cite{hinton1979stokesmultipliers,zhu1992stokesconstants}
were chosen to capture only the exponential asymptotics implied by
the dominant essential singularity at $\vert z\vert\rightarrow\infty$.
In our formalism, we have instead chosen a WKB basis more familiar
to physicists and suited to the rotational symmetry arguments of section
\ref{sec:Discrete-symmetries-1}. As we will show in section \ref{subsec:Non-analyticity-isolation},
this choice primarily differs from \cite{hinton1979stokesmultipliers,zhu1992stokesconstants}
by a constant factor non-analytic in $k^{2}$ --- the non-analyticity
arising due to contributions to the phase $dz$ integral between $0$
and $\vert z\vert\leq k^{2/n}$. Up to this factor, our WKB basis
(and that in \cite{hinton1979stokesmultipliers,zhu1992stokesconstants})
is entire in $k^{2}$.\footnote{A theorem in \cite{shibuya1975globaltheory} states that  asymptotic
boundary conditions that are entire in $k^{2}$, combined with the
fact that the mode equation is also entire in $k^{2}$, imply that
mode functions are entire in $k^{2}$. This in turn implies Stokes
constants  are entire functions of $k^{2}$. For our choice of basis
functions, this theorem implies that the Stokes constants will be
a product of a non-analytic factor and an entire function of $k^{2}$
--- which we derive in (\ref{eq:betaSTrans}).}

\subsection{Non-analyticity isolation\label{subsec:Non-analyticity-isolation}}

In this subsection, we note that the WKB basis functions can be written
as a product of a $z$-independent factor non-analytic in $k^{2}$
and a $z$-dependent entire function of $k^{2}$. The latter captures
the exponential asymptotics implied by the irregular singularity at
$\vert z\vert\rightarrow\infty$. Consider the WKB phase
\begin{equation}
\theta(k,z)\equiv\int^{z}_{0}dz'\omega_{k}(z'),\quad\omega_{k}(z')=\sqrt{k^{2}+z'^{n}}
\end{equation}
where the non-analyticity in $k^{2}$ arises from integration between
$0$ and $\vert z\vert\leq k^{2/n}$. Consequently, we may separate
the two contributions as 
\begin{equation}
\theta(k,z)=(\text{analytic in }k)+(\text{non-analytic in }k)\,.\label{eq:k-non-ana-isolation}
\end{equation}
by splitting the integral at $\psi_{1}$ (since we are interested
in the first Stokes sector), that is
\begin{equation}
\theta(k,z)=\int^{\psi_{1}}_{0}dz'\omega_{k}(z')+\int^{z}_{\psi_{1}}dz'\omega_{k}(z'),\quad z\in\mathscr{S}_{1}\,.
\end{equation}
The first integral is the phase factor $\Delta_{1}(k)$ and is non-analytic
in $k^{2}$. The second integral can be evaluated term by term using
the Taylor series representations (\ref{eq:underomega}) and (\ref{eq:omega-continued-sj}).
Evaluating the resulting primitive of this Taylor series at $\psi_{1}$
then yields another term non-analytic in $k^{2}$. Combining these
two terms we obtain for $n\geq4$ in the first Stokes sector
\begin{equation}
\theta(k,z)=\frac{z^{n/2+1}}{n/2+1}-\underset{\text{analytic in }k^{2}}{\underbrace{A(k,z)}}+\underset{\text{non-analytic in }k}{\underbrace{\bar{\theta}(k)}}\,,\quad k\in\mathbb{R}_{\geq0},\quad z\in\mathscr{S}_{1}\label{eq:WKBPhaseAE}
\end{equation}
where 
\begin{equation}
\bar{\theta}(k)\equiv\Delta_{1}(k)+\Psi_{1}(k)
\end{equation}
denotes the term non-analytic in $k$ and where explicit expressions
for $A(k,z)$ and $\Psi_{1}(k)$ are given by eqs.\ (\ref{eq:Akz})
and (\ref{eq:Psij-0}), respectively, as
\begin{equation}
A\left(k,z\right)=\sum^{\infty}_{m=1}\begin{pmatrix}1/2\\
m
\end{pmatrix}\frac{k^{2m}}{2mN-\left(N+1\right)}\frac{1}{z^{2mN-\left(N+1\right)}}
\end{equation}
\begin{equation}
\Psi_{1}(k)=\sum^{\infty}_{m=0}\begin{pmatrix}1/2\\
m
\end{pmatrix}\frac{1}{mn-(n/2+1)}\frac{k^{2m}}{\psi^{mn-(n/2+1)}_{1}}\,,
\end{equation}
where this expansion breaks down for $n=2$ ($N=n/2=1$) for example
because of the denominator $mn-\left(n/2+1\right)$ appearing in the
expression for $\Psi_{1}$.\footnote{We give an explicit solution to the $n=2$ case in eq.~(\ref{eq:betan2}).}

In appendix \ref{sec:Non-analytic--factor}, we show that the non-analytic
residue $\bar{\theta}(k)$ is proportional to the phase factor $\Delta_{1}(k)$:
\begin{equation}
\bar{\theta}(k)=\frac{e^{-\frac{i\pi}{n}}}{\cos\left(\frac{\pi}{n}\right)}\Delta_{1}(k)\,.\label{eq:thetaBar}
\end{equation}
We will also see in section \ref{sec:Soft-corrections-to} that this
cancels the non-analytic phase factors arising from branch cuts, making
the Bogoliubov $\beta_{k}$ analytic in $k^{2}$ in the soft limit.

\subsection{Automorphism\label{subsec:Automorphism-on-WKB}}

Here we show in detail  how using the WKB amplitude expansion (\ref{eq:WKB-eae-sloppy})
in eqs.~(\ref{eq:IntModeEq1}) and (\ref{eq:IntMode2}) leads to
an automorphism on $\mathcal{V}_{k}$, the functional space of the
exponentially improved asymptotic expansion of the WKB amplitudes.
The main result is eq.\ (\ref{eq:Vk}), the explicit form of $\mathcal{V}_{k}$.

We start with simplifying eqs.\ (\ref{eq:IntModeEq1}) and (\ref{eq:IntMode2})
by taking $z_{0}\to\infty$ on the real axis and the boundary condition\footnote{We choose this initial condition to isolate the first Stokes constant
term in $[\beta_{k}(z)]_{{\rm e.a.e.}}$.}
\begin{equation}
\lim_{z\to z_{0}}V_{k}(z)=\lim_{z\to z_{0}}\begin{pmatrix}\alpha_{k}(z)\\
\beta_{k}(z)
\end{pmatrix}=\begin{pmatrix}1\\
0
\end{pmatrix}\,.
\end{equation}
Using the decomposition (\ref{eq:WKBPhaseAE}) for the WKB phase in
the first Stokes sector, we obtain
\begin{align}
\alpha_{k}(z) & =1+\int^{z}_{\infty}dw\,e^{\zeta(w)}B_{-}(w)e^{2i\bar{\theta}(k)}\beta_{k}(w)\,,\label{eq:alpha-int-int1}\\
\beta_{k}(z) & =\int^{z}_{\infty}dw\,e^{-\zeta(w)}B_{+}(w)e^{-2i\bar{\theta}(k)}\alpha_{k}(w)\,,\label{eq:beta-int-int1}
\end{align}
where $\zeta(w)\equiv2iw^{N+1}/(N+1)$ and

\begin{equation}
B_{\pm}(z)\equiv\frac{\omega'_{k}(z)}{2\omega_{k}(z)}e^{\pm2iA(k,z)}=\sum^{\infty}_{m=1}B^{(\pm)}_{m}z^{-m},\quad z\in\mathscr{S}_{1}\label{eq:Bpm}
\end{equation}
with $B^{(\pm)}_{m}$ denoting expansion coefficients of $B_{\pm}(z)$.

To make the expansion (\ref{eq:WKB-eae-sloppy}) concrete, we define
the functional space
\begin{equation}
\mathcal{P}(z)\equiv\left\{ \sum^{\infty}_{m=1}a_{m}z^{-m}|a_{m}\in\mathbb{C}\right\} \label{eq:Pz-series}
\end{equation}
allowing us to classify $B_{\pm}(z)\in\mathcal{P}(z)$. Also note
in the asymptotic regions, the exponentially improved asymptotic expansions
of the WKB amplitudes $[\alpha_{k}(z)]_{{\rm e.a.e.}}$ and $[\beta_{k}(z)]_{{\rm e.a.e.}}$
must contain $z^{-p}$ terms because of the structure of eqs.~(\ref{eq:alpha-int-int1})-(\ref{eq:Bpm}).
Plugging such terms  into eqs.\ (\ref{eq:alpha-int-int1}) and (\ref{eq:beta-int-int1})
yields terms of the form
\begin{equation}
I^{\pm}_{p}(z)\equiv\int^{z}_{\infty}dw\frac{e^{\pm\zeta(w)}}{w^{p}}\,.
\end{equation}
A change of variable $w\to\pm\zeta=\pm2iw^{N+1}/(N+1)$ gives
\begin{equation}
I^{\pm}_{p}(z)\propto\zeta^{1-\frac{N+p}{N+1}}E_{\frac{N+p}{N+1}}(\mp\zeta)\,.
\end{equation}
Since we are interested in the first Stokes sector where $\arg\zeta\in(\pi/2,3\pi/2)$
as $\arg w\in(0,\pi/\left(N+1\right))$, the asymptotic expansion
of $I^{-}_{p}(z)$ sees a Stokes jump as $\arg\zeta$ crosses $\pi$,
or equivalently, as $\arg w$ crosses $\pi/(2N+2)$. Consequently,
as may be seen from eq.\ (\ref{eq:ExpIntAsymp}), the exponentially
improved asymptotic expansions of these integrals in $\mathscr{S}_{1}$
take the form 
\begin{align}
\left[\int^{z}_{\infty}dw\frac{e^{\zeta(w)}}{w^{p}}\right]_{{\rm e.a.e.}} & \in e^{\zeta(z)}\mathcal{P}(z)\label{eq:EAEplus}\\
\left[\int^{z}_{\infty}dw\frac{e^{-\zeta(w)}}{w^{p}}\right]_{{\rm e.a.e}.} & \in e^{-\zeta(z)}\mathcal{P}(z)+\Theta(z)\mathbb{C}\label{eq:EAEminus}
\end{align}
where $\Theta(z)$ is a step-function that jumps at the bisector of
the first Stokes sector
\begin{equation}
\Theta(z)\equiv\begin{cases}
0 & 0\leq\arg z<\frac{\pi}{2(N+1)}\\
1/2 & \arg z=\frac{\pi}{2(N+1)}\\
1 & \frac{\pi}{2(N+1)}<\arg z\leq\frac{\pi}{N+1}
\end{cases}\,.\label{eq:thetaz-step}
\end{equation}

Now since $\alpha_{k}(z)$ is the coefficient of the dominant mode
$f_{-}$ in $\mathscr{S}_{1}$, which doesn't see a non-exponentially-suppressed
$\Theta(z)$  and satisfies the boundary condition $\alpha_{k}(z_{0})=1$,
we have the ordinary asymptotic expansion (without exponential improvements)
of $\alpha_{k}(z)$ having the form $1+\mathcal{P}(z)$. Plugging
this into eqs.\ (\ref{eq:alpha-int-int1}) and (\ref{eq:beta-int-int1})
and using the generalization of eq.~(\ref{eq:EAEplus}) and eq.~(\ref{eq:EAEminus})
where $w^{-p}$ is replaced by an element of $\mathcal{P}(w)$, that
is 
\begin{align}
\int^{z}_{\infty}dwe^{\zeta(w)}\mathcal{P}(w) & \dot{\sim} e^{\zeta(z)}\mathcal{P}(z)\label{eq:automorPlus}\\
\int^{z}_{\infty}dwe^{-\zeta(w)}\mathcal{P}(w) & \dot{\sim} e^{-\zeta(z)}\mathcal{P}(z)+\Theta(z)\mathbb{C}\label{eq:automorMinus}
\end{align}
we note that the exponentially improved asymptotic expansions of the
WKB amplitudes should take the form (for a more rigorous justification
of the following, see appendix \ref{sec:Detailed-derivation-of})
\begin{align}
[\alpha_{k}(z)]_{{\rm e.a.e.}} & \in1+\mathcal{P}(z)+e^{\zeta(z)}\Theta(z)\mathcal{P}(z)\\{}
[\beta_{k}(z)]_{{\rm e.a.e.}} & \in e^{-\zeta(z)}\mathcal{P}(z)+\Theta(z)\left[\mathcal{P}(z)+\mathbb{C}\right].
\end{align}
This is equivalent to
\begin{equation}
\mathcal{V}_{k}=\begin{pmatrix}1+\mathcal{P}(z)+e^{\zeta(z)}\Theta(z)\mathcal{P}(z)\\
e^{-\zeta(z)}\mathcal{P}(z)+\Theta(z)\left[\mathcal{P}(z)+\mathbb{C}\right]
\end{pmatrix}\label{eq:Vk}
\end{equation}
where recall that $\mathcal{V}_{k}$ denotes functional space of the
exponentially improved asymptotic expansions of the WKB amplitudes.
Consequently, the integrals on the right hand side of eqs.\ (\ref{eq:alpha-int-int1})
and (\ref{eq:beta-int-int1}) may be viewed as automorphisms on $\mathcal{V}_{k}$
via series multiplications by $B_{\pm}(z)$, generalized exponential
integrals (\ref{eq:automorPlus}) and (\ref{eq:automorMinus}) and
ordinary integrals, which we will further detail in section \ref{subsec:Solvable-technique}.

For the purposes of the exact computations in the following section
we define the power series
\begin{equation}
\lambda_{1,2}(z)\equiv\sum^{\infty}_{m=1}\lambda^{(1,2)}_{m}z^{-m},\quad\mu_{1,2}(z)\equiv\sum^{\infty}_{m=1}\mu^{(1,2)}_{m}z^{-m}
\end{equation}
such that
\begin{align}
[\alpha_{k}(z)]_{{\rm e.a.e.}} & =1+\lambda_{1}(z)+S_{1}\Theta(z)e^{\zeta(z)}\mu_{1}(z)\label{eq:alphaZ}\\{}
[\beta_{k}(z)]_{{\rm e.a.e.}} & =e^{-\zeta(z)}\mu_{2}(z)+S_{1}\Theta(z)+S_{1}\Theta(z)\lambda_{2}(z).\label{eq:betaZ}
\end{align}
Here $S_{1}$ corresponds to the Stokes constant, and evidently $\lambda_{1,2}(z)$
and $\mu_{1,2}(z)$ belong to $\mathcal{P}(z)$.

\subsection{Solvable technique\label{subsec:Solvable-technique}}

In this section, we work out the explicit expressions for the automorphism
on $\mathcal{V}_{k}$ given the series representation (\ref{eq:alphaZ})
and (\ref{eq:betaZ}) for the exponentially improved asymptotic expansion
of the WKB amplitudes. This will determine $\lambda^{(1,2)}_{m}$
and $\mu^{(1,2)}_{m}$ as functions of $B^{(\pm)}_{m}$, and an expression
for $S_{1}(k)$ as a function of these coefficients.

Rewriting eqs.\ (\ref{eq:alpha-int-int1}) and (\ref{eq:beta-int-int1})
in terms of 
\begin{equation}
\bar{\beta}(z)\equiv e^{2i\bar{\theta}(k)}\beta(z),\quad T_{1}\equiv e^{2i\bar{\theta}(k)}S_{1},\label{eq:betaSTrans}
\end{equation}
we have
\begin{align}
\alpha(z) & =1+\int^{z}_{\infty}dw\,e^{\zeta(w)}B_{-}(w)\bar{\beta}(w),\label{eq:alphaInted}\\
\bar{\beta}(z) & =\int^{z}_{\infty}dw\,e^{-\zeta(w)}B_{+}(w)\alpha(w).\label{eq:betaInted}
\end{align}
Similarly, eqs.\ (\ref{eq:alphaZ}) and (\ref{eq:betaZ}) can be
rewritten as
\begin{align}
[\alpha_{k}(z)]_{{\rm e.a.e.}} & =1+\lambda_{1}(z)+T_{1}\Theta(z)e^{\zeta(z)}\bar{\mu}_{1}(z)\label{eq:alphaZ1}\\{}
[\bar{\beta}_{k}(z)]_{{\rm e.a.e.}} & =e^{-\zeta(z)}\bar{\mu}_{2}(z)+T_{1}\Theta(z)+T_{1}\Theta(z)\lambda_{2}(z)\label{eq:betaZ1}
\end{align}
where
\begin{equation}
\bar{\mu}_{1}(z)\equiv e^{-2i\bar{\theta}(k)}\mu_{1}(z),\quad\bar{\mu}_{2}(z)\equiv e^{2i\bar{\theta}(k)}\mu_{2}(z)\,.
\end{equation}
To carry out the integrals on the right hand side, we use eq.~(\ref{eq:ExpIntAsymp})
which with the relevant change of variables implies in the first Stokes
sector (see appendix \ref{sec:The-generalized-exponential})
\begin{equation}
\int^{z}_{\infty}dw\frac{e^{\zeta(w)}}{w^{p}}\dot{\sim} e^{\zeta(z)}\sum^{\infty}_{s=0}\frac{\E_{s,p}}{z^{(N+1)s+N+p}}\label{eq:geiPlus}
\end{equation}
\begin{equation}
\int^{z}_{\infty}dw\frac{e^{-\zeta(w)}}{w^{p}}\dot{\sim} e^{-\zeta(z)}\sum^{\infty}_{s=0}\frac{(-1)^{s+1}\E_{s,p}}{z^{(N+1)s+N+p}}-\Theta(z)\S_{\frac{N+p}{N+1}}\label{eq:geiMinus}
\end{equation}
  where we defined
\begin{equation}
\E_{s,p}\equiv\frac{(N+1)^{s}}{(2i)^{s+1}}\frac{\Gamma\left(\frac{N+p}{N+1}+s\right)}{\Gamma\left(\frac{N+p}{N+1}\right)},\qquad\S_{\frac{N+p}{N+1}}\equiv\frac{\pi e^{-\frac{i\pi}{2}\frac{N+p}{N+1}}}{\Gamma\left(\frac{N+p}{N+1}\right)\left(\frac{N+1}{2}\right)^{\frac{N+p}{N+1}}}
\end{equation}
and $\Theta(z)$  in eq.\ (\ref{eq:thetaz-step}).

With the above, the simplification of the right hand sides of eqs.~(\ref{eq:alphaInted})
and (\ref{eq:betaInted}) may be understood in terms of the following
linear maps on $\mathcal{P}(z)$:
\begin{itemize}
\item $\mathcal{B}_{\pm}:\mathcal{P}(z)\to\mathcal{P}_{\geq2}(z)$ defined
as multiplication by the series $B_{\pm}(z)\in\mathcal{P}(z)$, that
is 
\begin{equation}
\mathcal{B}_{\pm}\left[f(z)\right]\equiv B_{\pm}(z)f(z),\label{eq:polyTimes}
\end{equation}
where $\mathcal{P}_{\geq2}(z)$ is the subspace of $\mathcal{P}(z)$
with vanishing terms of order $z^{-1}$.
\item $\mathcal{E}_{\pm}:\mathcal{P}(z)\to\mathcal{P}(z)$ and the Stokes
functional $\mathcal{S}:\mathcal{P}(z)\to\mathbb{C}$, defined as
\begin{align}
\int^{z}_{\infty}dw\,e^{\zeta(w)}f(w) & \equiv e^{\zeta(z)}\mathcal{E}_{+}\left[f(z)\right]\label{eq:geiPlus-1}\\
\int^{z}_{\infty}dw\,e^{-\zeta(w)}f(w) & \equiv e^{-\zeta(z)}\mathcal{E}_{-}\left[f(z)\right]+\Theta(z)\mathcal{S}\left[f\right]\label{eq:geiMinus-1}
\end{align}
where the integrals on $f(z)\in\mathcal{P}(z)$ are to be understood
as evaluated term-wise using eq.~(\ref{eq:geiPlus}) and (\ref{eq:geiMinus}).\footnote{Even though eq.~(\ref{eq:geiMinus-1}) looks as if it cannot define
two functions with one equation, it can since $\mathcal{E}_{\pm}$
is restricted to a particular series functional space $\mathcal{P}(z)$
defined in eq.~(\ref{eq:Pz-series}). For an explicit expression
of $\mathcal{S}[f(z)]$, see eq.~(\ref{eq:repStokes}).}
\item For $f(z)\in\mathcal{P}_{\geq2}(z)$, the ordinary integral induces
the map $\mathcal{I}:\mathcal{P}_{\geq2}(z)\to\mathcal{P}(z)$: 
\begin{equation}
\int^{z}_{\infty}dw\,f(w)\equiv\mathcal{I}\left[f(z)\right].\label{eq:ordinaryInt}
\end{equation}
\end{itemize}
Explicit expressions for the above are provided in appendix \ref{sec:Linear-maps-from}.

In terms of the above, the integrals of eq.~(\ref{eq:alphaInted})
and (\ref{eq:betaInted}) can be expressed as 
\begin{equation}
\int^{z}_{\infty}dwe^{\zeta(w)}B_{-}(w)\bar{\beta}(w)\dot{\sim} T_{1}\Theta(z)e^{\zeta(z)}\mathcal{E}_{+}\left[B_{-}(z)\right]+T_{1}\Theta(z)e^{\zeta(z)}\mathcal{E}_{+}\mathcal{B}_{-}\left[\lambda_{2}(z)\right]+\mathcal{I}\mathcal{B}_{-}\left[\bar{\mu}_{2}(z)\right]
\end{equation}
\begin{align}
\int^{z}_{\infty}dwe^{-\zeta(w)}B_{+}(w)\alpha(w)\dot{\sim} & e^{-\zeta(z)}\mathcal{E}_{-}\left[B_{+}(z)+\mathcal{B}_{+}\left[\lambda_{1}(z)\right]\right]+\Theta(z)\mathcal{S}\left[B_{+}(z)+\mathcal{B}_{+}\left[\lambda_{1}(z)\right]\right]\nonumber \\
 & +T_{1}\Theta(z)\mathcal{I}\left[\mathcal{B}_{+}\left[\bar{\mu}_{1}(z)\right]\right].
\end{align}
Comparing with the left hand sides which are of the form eq. (\ref{eq:alphaZ1})
and eq. (\ref{eq:betaZ1}) we find
\begin{align}
\lambda_{1}(z) & =\mathcal{I}\mathcal{B}_{-}\left[\bar{\mu}_{2}(z)\right]\label{eq:c1}\\
\lambda_{2}(z) & =\mathcal{I}\mathcal{B}_{+}\left[\bar{\mu}_{1}(z)\right]\\
\bar{\mu}_{1}(z) & =\mathcal{E}_{+}\left[B_{-}(z)+\mathcal{B}_{-}\left[\lambda_{2}(z)\right]\right]\\
\bar{\mu}_{2}(z) & =\mathcal{E}_{-}\left[D(z)\right]\label{eq:d2}
\end{align}
\begin{equation}
T_{1}=\mathcal{S}\left[D(z)\right]\label{eq:T1D}
\end{equation}
where 
\begin{equation}
D(z)\equiv B_{+}(z)+\mathcal{B}_{+}\left[\lambda_{1}(z)\right].
\end{equation}
We note that the above relations recursively solve for $\lambda^{(1,2)}_{m}$
and $\mu^{(1,2)}_{m}$ in terms of $B^{(\pm)}_{m}$, and allow $T_{1}$
to be computed as a known function of these. In particular, note composing
(\ref{eq:c1}) and (\ref{eq:d2}) we obtain 
\begin{equation}
D(z)=B_{+}(z)+\mathcal{B}_{+}\mathcal{I}\mathcal{B}_{-}\mathcal{E}_{-}\left[D(z)\right]\label{eq:Dz}
\end{equation}
which gives $T_{1}$ through eq.\ (\ref{eq:T1D}) and in turn the
Stokes constant through eq.~(\ref{eq:betaSTrans}). We will discuss
the detailed structure of the power series representation for $T_{1}(k^{2})$
in the next subsection.

\subsection{Explicit recursion relation and its solution\label{subsec:Explicit-recursion-relation}}

The goal of this subsection is to derive a closed form expression
for $T_{1}(k^{2})$ to leading order of $k^{2}$. Specifically, we
will derive eqs.\ (\ref{eq:T0-closed-form}) and (\ref{eq:T2-closed-form}),
the $T^{(0)}_{1}$ and $T^{(2)}_{1}$ coefficients in the soft momentum
expansion of $T_{1}(k^{2})$ about the topological limit $k=0$:
\begin{equation}
T_{1}(k^{2})\equiv T^{(0)}_{1}+T^{(2)}_{1}k^{2}+O(k^{4})\,.\label{eq:T-split}
\end{equation}
We have seen in eq.\ (\ref{eq:T1D}) that $T_{1}$ is given by the
Stokes functional $\mathcal{S}$ acting on the $D(z)$ function with
$D(z)$ determined by the recursion relation (\ref{eq:Dz}). In this
subsection, we will solve a rescaled $D(z)$ recursion relation (see
eq.\ (\ref{eq:G-recursion})) in the $\{z^{-m}\}$ basis of the functional
space $\mathcal{P}(z)$. Based on this solution, we will derive closed
form expressions for $T^{(0)}_{1}$ and $T^{(2)}_{1}$ as functions
of the dispersion index $N=n/2$. They will be used in section \ref{subsec:Analytic-beta}
for deriving soft corrections for the Bogoliubov coefficient $\beta_{k}$.

To start, we expand the functions $D(z)$ and $B_{\pm}(z)$ in the
$\left\{ z^{-m}\right\} $ basis of the functional space $\mathcal{P}(z)$
as
\begin{equation}
D(z)\equiv\sum^{\infty}_{m=1}D_{m}z^{-m},\quad B_{+}(z)\equiv\sum^{\infty}_{m=1}B^{(+)}_{m}z^{-m},\quad B_{-}(z)\equiv\sum^{\infty}_{m=1}B^{(-)}_{m}z^{-m}\,.
\end{equation}
Next, we reduce the linear integral equation (\ref{eq:Dz}) for $D(z)$
by working out the explicit representation of the object $\mathcal{B}_{+}\mathcal{I}\mathcal{B}_{-}\mathcal{E}_{-}\left[D(z)\right]$.
Using the explicit representations of the linear operators given by
eqs.\ (\ref{eq:repEp})-(\ref{eq:repI}), we obtain
\begin{align}
\mathcal{E}_{-}\left[D(z)\right] & =\sum^{\infty}_{m=1}\frac{1}{z^{m+N}}\sum^{m-1}_{p=0}D_{p+1}(-1)^{\frac{m-p-1}{N+1}+1}\E_{\frac{m-p-1}{N+1},p+1}\delta\left(\frac{m-p-1}{N+1}\right)\nonumber \\
 & =\sum^{\infty}_{l=N+1}\frac{1}{z^{l}}\sum^{l-N-1}_{s=0}F_{l-N,s}D_{s+1}
\end{align}
where
\begin{equation}
F_{m,s}\equiv(-1)^{\frac{m-s-1}{N+1}+1}\E_{\frac{m-s-1}{N+1},s+1}\delta\left(\frac{m-s-1}{N+1}\right)\,.\label{eq:Fms}
\end{equation}
It follows that
\begin{equation}
\mathcal{B}_{+}\mathcal{I}\mathcal{B}_{-}\mathcal{E}_{-}\left[D(z)\right]=-\sum^{\infty}_{p=2}\frac{1}{z^{p}}\sum_{m+l=p}B^{(+)}_{m}\frac{1}{l}\left(\sum_{q+r=l+1}B^{(-)}_{q}\sum^{r-N-1}_{s=0}F_{r-N,s}D_{s+1}\right).
\end{equation}
With the above results, comparing the coefficients of the $\left\{ z^{-m}\right\} $
basis in eq.\ (\ref{eq:Dz}), we get the explicit recursion relation
for $D(z)$:
\begin{equation}
D_{p+1}=\begin{cases}
B^{(+)}_{p+1} & 0\leq p\leq N\\
B^{(+)}_{p+1}-\sum^{p}_{m=N+1}\sum^{m}_{l=N+1}\sum^{l-N-1}_{s=0}\frac{B^{(+)}_{p-m+1}B^{(-)}_{m-l+1}}{m}F_{l-N,s}D_{s+1} & p\geq N+1
\end{cases}\,,\label{eq:explicitRecursion}
\end{equation}
where $\E_{s,p}$ and $\delta$ are defined by eq.\ (\ref{eq:EandS})
and eq.\ (\ref{eq:delta}), respectively.

Now, it is convenient to define a rescaled variable for $D(z)$ encoded
by $D_{p}$. Use eqs.\ (\ref{eq:T1D}) and (\ref{eq:repStokes})
to find a series expression for $T_{1}$:
\begin{equation}
T_{1}=-\sum^{\infty}_{m=1}D_{m}\S_{\frac{N+m}{N+1}}=-\sum^{\infty}_{m=1}G_{m}\label{eq:TG}
\end{equation}
with $\S$ defined by eq.\ (\ref{eq:EandS}) coming from $\mathcal{S}$
operator acting on $z^{-m}$ and
\begin{equation}
G_{m}\equiv D_{m}\S_{\frac{N+m}{N+1}},\,\,\,\,G(z)\equiv\sum^{\infty}_{m=1}G_{m}z^{-m}\label{eq:Gz}
\end{equation}
with the $G(z)$ and $G_{m}$ definitions identified as naturally
rescaled $D(z)$ and $D_{p}$. The explicit recursion relation for
$G(z)$ is then obtained by rescaling each $D_{p}$ term in eq.~(\ref{eq:explicitRecursion})
by $\S_{(N+p)/(N+1)}$ according to eq.\ (\ref{eq:Gz}), after which
\begin{equation}
G_{p+1}=\begin{cases}
B^{(+)}_{p+1}\S_{\frac{N+p+1}{N+1}} & 0\leq p\leq N\\
B^{(+)}_{p+1}\S_{\frac{N+p+1}{N+1}}-\sum^{p}_{m=N+1}\sum^{m}_{l=N+1}\sum^{l-N-1}_{s=0}K_{p;m,l,s}G_{s+1} & p\geq N+1
\end{cases}\label{eq:G-recursion}
\end{equation}
\begin{equation}
K_{p;m,l,s}=\frac{B^{(+)}_{p-m+1}B^{(-)}_{m-l+1}}{ml}\frac{\S_{\frac{N+p+1}{N+1}}}{\S_{\frac{N+l+1}{N+1}}}\delta\left(\frac{l-s}{N+1}-1\right)\label{eq:K-defn}
\end{equation}
where the definitions for $\S_{y}$ and $\delta(X)$ can be found
in (\ref{eq:EandS}) and (\ref{eq:delta}).

In the soft momentum limit $k\to0$, each $G_{p}$ term splits into
the topological limit term $G^{(0)}_{p}$ and the soft correction
term $G^{(2)}_{p}$:
\begin{equation}
G_{p}=G^{(0)}_{p}+G^{(2)}_{p}k^{2}+O(k^{4})\,.\label{eq:G02}
\end{equation}
Then, we show in appendix \ref{subsec:Split-the-explicit} that the
explicit recursion relation (\ref{eq:G-recursion}) also splits into
two sets of recursion relations for $G^{(0)}_{p}$ and $G^{(2)}_{p}$,
respectively. We solve the split recursion relations in appendix \ref{subsec:Quadratic-coefficient-of}
to obtain closed form expressions for the partial sums of $G^{(0)}_{p}$
and $G^{(2)}_{p}$ (see eqs.\ (\ref{eq:Yq0}) and (\ref{eq:Yr2-1})).
This in turn yields closed form expressions for the $T^{(0)}_{1}$
and $T^{(2)}_{1}$ coefficients (see eqs.\ (\ref{eq:T10-cosphi})
and (\ref{eq:T12-closed-form-appendix})): 
\begin{align}
T^{(0)}_{1} & =-\sum^{\infty}_{p=1}G^{(0)}_{p}=2i\cos\left[\frac{\pi}{2(N+1)}\right]\,,\label{eq:T0-closed-form}\\
T^{(2)}_{1} & =-\sum^{\infty}_{p=1}G^{(2)}_{p}=-i\left[\frac{1}{i(N+1)}\right]^{\frac{2N}{N+1}}\pi^{3/2}\frac{\Gamma\left[\frac{N-1}{2(N+1)}\right]}{\Gamma\left(\frac{N}{N+1}\right)\Gamma\left[\frac{2N+1}{2(N+1)}\right]\Gamma\left[\frac{2N-1}{2(N+1)}\right]}\,,\label{eq:T2-closed-form}
\end{align}
where we have used eqs.\ (\ref{eq:T-split}) and (\ref{eq:TG}).

These two closed form expressions lead to a closed form expression
for the first Stokes constant in the soft limit
\begin{equation}
S_{1}(k)=e^{-2i\bar{\theta}(k)}T_{1}(k^{2})=e^{-2i\bar{\theta}(k)}\left[T^{(0)}_{1}+T^{(2)}_{1}k^{2}+O(k^{4})\right]\,.
\end{equation}
We have seen in eq.\ (\ref{eq:betaPropagatorExp}) that the particle
production coefficient  $\beta_{k}$ is equal to the 21 entry of a
matrix product expressed in terms of Stokes constants, and in eqs.\ (\ref{eq:SRot1})
to (\ref{eq:SRot3}) that all the Stokes constants are related to
the first Stokes constant via a discrete rotational symmetry. These
suggest that a closed form expression for $\beta_{k}$ in the soft
limit is obtainable. Deriving the closed form expression for $\beta_{k}$
to $O(k^{2})$ order is therefore the task of section \ref{subsec:Analytic-beta}.
\section{Soft corrections to Bogoliubov $\beta_{k}$\label{sec:Soft-corrections-to}}

Based on the framework established in sections \ref{sec:Discrete-symmetries}
and \ref{sec:The-first-Stokes}, we are now ready to compute soft
corrections to the Bogoliubov $\beta_{k}$ using eq.\ (\ref{eq:betaPropagatorExp}).
In section \ref{subsec:Analytic-beta}, we will show that the leading
order expansion of $\beta_{k}$ about the topological limit $k=0$
is quadratic in $k$, i.e.,
\begin{equation}
\beta_{k}=\beta_{k=0}+c_{1}k^{2}+O(k^{4})\,,\label{eq:betaQuad}
\end{equation}
with $\beta_{k=0}=-i\cot(\pi/(n+2))$ the topological limit result
\cite{Chung:2024pot} and $c_{1}$ a purely imaginary constant as
required by the time-reversal symmetry (cf.\ eq.\ (\ref{eq:betaImaginary})).
We first evaluate $c_{1}$ analytically via eq.\ (\ref{eq:betaPropagatorExp}),
utilizing $S_{1}(k)=\exp[-2i\bar{\theta}(k)]T_{1}(k^{2})$ where $T_{1}(k^{2})$
to leading order $O(k^{2})$ is given by eqs.\ (\ref{eq:T-split}),
(\ref{eq:T0-closed-form}) and (\ref{eq:T2-closed-form}). We then
compare this to a direct computation of $\beta_{k}$ using the Wronskian
projection (\ref{eq:betaReal}), obtained by numerically solving the
mode equation (\ref{eq:modeEqnChiReduced}) and find good agreement.

Surprisingly, for $n\ge4$, the non-analytic $k$ factors $\Delta_{j}(k)$
arising from branch cuts and the non-analytic residue $\bar{\theta}(k)$
in the Stokes constants explicitly cancel out. For $n=2$, we will
see from eq.\ (\ref{eq:betaPropagatorExp}) that $\beta_{k}$ only
receives contributions from $\Delta_{1}(k)$, which is analytic in
$k$.

\subsection{Analytic $\beta_{k}$\label{subsec:Analytic-beta}}

As we have shown in section \ref{subsec:Analytic-continuation-and},
propagating the WKB amplitudes on the lower half plane and utilizing
Stokes phenomena leads to the expression (\ref{eq:betaPropagatorExp})
where $\beta_{k}$ is written as the 21 entry of a product of propagators
across Stokes sectors and branch cut matrices. In this subsection,
we will further simplify the above expression, after which $\beta_{k}$
will be given as a product of matrices whose entries are the power
series $T_{1}(k^{2})$ with the arguments rotated. We will also use
the notations introduced in appendix \ref{sec:Propagator-rotation-details}.

We first consider the cases where $n\geq4$, i.e., $N\geq2$. Let
$j$ be an integer in the range $1\leq j\leq N$. As a first step,
note that the branch cut matrices $\mathscr{B}_{N+j}(k)$ defined
in eq.\ (\ref{eq:branchj}) can be rewritten as a product of rotational
representation matrices $B$ and $\mathscr{R}_{N+j}(k)$ defined in
eq.\ (\ref{eq:BRj}) as
\begin{equation}
\mathscr{B}_{N+j}(k)=-i\mathscr{R}[\Delta_{N+j}(k)]B\,.\label{eq:BjkAsRB}
\end{equation}
Next, we rotate the propagators $U_{N+2+j}(k)$ using eqs.\ (\ref{eq:SRot3})
and (\ref{eq:RUProperty})
\begin{equation}
U_{N+2+j}(k)=U_{+}[S_{N+2+j}(k)]=U_{+}[e^{4i\tilde{\Phi}_{j}}\tilde{S}^{(N+2+j)}_{1}(k)]=\mathscr{R}[\tilde{\Phi}_{j}]U_{+}[\tilde{S}^{(N+2+j)}_{1}(k)]\mathscr{R}^{-1}[\tilde{\Phi}_{j}]\,,\label{eq:UjkAsRURI}
\end{equation}
where $\tilde{S}^{(m)}_{1}(k)\equiv S_{1}(\gamma^{-N(m-1)}k)$ with
$1\leq m\leq2N+2$ denotes the first Stokes constant with the argument
$k$ rotated. From the decomposition (\ref{eq:betaSTrans}) and the
form of the non-analytic residue (\ref{eq:thetaBar}), we see that
\begin{equation}
\tilde{S}^{(m)}_{1}(k)=e^{-2i\bar{\theta}(\gamma^{-N(m-1)}k)}T_{1}(\gamma^{-2N(m-1)}k^{2})=e^{-2i(-1)^{m-1}\bar{\theta}(k)}T_{1}(\gamma^{-2N(m-1)}k^{2})\,.
\end{equation}
Combining eqs.\ (\ref{eq:BjkAsRB}) and (\ref{eq:UjkAsRURI}), we
find
\begin{align}
U_{N+2+j}\mathscr{B}_{N+j} & =-i\mathscr{R}[\Delta_{N+j}]U_{+}[e^{-4i(\Delta_{N+j}-\tilde{\Phi}_{j})}\tilde{S}^{(N+2+j)}_{1}]B\,,
\end{align}
where we have used eq.\ (\ref{eq:RUProperty}). Subsequently, we
may simplify the propagator $U(+\infty,-\infty)$ using the properties
$B\mathscr{R}[x]B=\mathscr{R}[-x]$ and $\mathscr{R}[x]U_{\pm}[S]\mathscr{R}^{-1}[x]=U_{\pm}[e^{\pm4ix}S]$
to find
\begin{align}
U_{0}(k,\eta_{+\infty},\eta_{-\infty}) & =\left[\prod^{N}_{j=1}(U_{N+2+j}\mathscr{B}_{N+j})\right]U_{N+2}\nonumber \\
 & =(-1)^{N}i^{N}\mathscr{R}[-\tilde{\Delta}_{N}]\prod^{N}_{j=1}\left(U_{+}[e^{4i(-1)^{j}\tilde{\Delta}_{N}}\tilde{S}^{(N+2+j)}_{1}]B\right)U_{+}[e^{4i\tilde{\Delta}_{N}}\tilde{S}^{(N+2)}_{1}]\,,
\end{align}
where in the last line we have used $-\tilde{\Delta}_{j}-\tilde{\Phi}_{j}=-(-1)^{j}\tilde{\Delta}_{N}$
and where $\prod^{N}_{j=1}$ denotes the matrix product from \textit{right
to left} with $j$ ascending. The Bogoliubov $\beta_{k}$ is then
given by the 21 entry of the above product as
\begin{align}
\beta_{k} & =(-1)^{N}i^{N}e^{2i\tilde{\Delta}_{N}}\left[\prod^{N}_{j=1}\left(U_{+}[e^{4i(-1)^{j}\tilde{\Delta}_{N}}\tilde{S}^{(N+2+j)}_{1}]B\right)\right]_{21}\,.\label{eq:betaProduct1}
\end{align}
In appendix \ref{sec:Non-analyticity-cancellation}, we detail the
$k$ non-analyticity cancellation and simplify the above result to
\begin{equation}
\beta_{k}=(-1)^{N}i^{N}\left[\prod^{N}_{j=1}\left(U_{+}[\tilde{T}^{(N+2+j)}_{1}(k^{2})]B\right)\right]_{21},\label{eq:betaT1}
\end{equation}
where 
\begin{equation}
\tilde{T}^{(N+2+j)}_{1}(k^{2})\equiv T_{1}(\gamma^{-2N(N+1+j)}k^{2})\label{eq:T1Rot}
\end{equation}
are rotated power series. Now the Bogoliubov $\beta_{k}$ is written
as a finite product of $\tilde{T}^{(j)}_{1}(k^{2})$. Hence, the soft
corrections to the Bogoliubov $\beta_{k}$ for $n\geq4$ (i.e., $N\geq2$)
are analytic in $k$ and specifically, the leading order correction
is quadratic.

For $n=2$ (i.e., $N=1$), the derivations up to eq.\ (\ref{eq:betaProduct1})
are still valid, giving
\begin{equation}
\beta_{k}=-ie^{2i\tilde{\Delta}_{1}}\left[U_{+}[e^{-4i\tilde{\Delta}_{1}}\tilde{S}^{(4)}_{1}]B\right]_{21}=-ie^{2i\Delta_{1}}\,,
\end{equation}
which is independent of the first Stokes constant. For $N=1$, we
have  $\Delta_{1}=i\pi k^{2}/4$ as can be seen from eq.\ (\ref{eq:Deltaj-1}),
then
\begin{equation}
\beta_{k}=-ie^{-\frac{\pi k^{2}}{2}}\,,\label{eq:betan2}
\end{equation}
which is also analytic in $k$, and agrees with the solvable model
in the literature (see e.g.\ \cite{Kofman:1997yn}).

With the above, the soft corrections to Bogoliubov $\beta_{k}$ to
quadratic order for arbitrary $n\in\text{even}$ are given by
\begin{equation}
\beta_{k}(n)=-i\cot\left(\frac{\pi}{n+2}\right)+c_{1}(n)k^{2}+O(k^{4})\label{eq:betaexpansion}
\end{equation}
with $c_{1}(n)$ derived in appendix \ref{sec:Closed-form-expression}
using eq.\ (\ref{eq:betaT1}) as
\begin{equation}
\boxed{c_{1}(n)=i\frac{2^{\frac{4}{n+2}-1}}{\pi\left(\frac{n+2}{2}\right)^{2-\frac{4}{n+2}}}\cos\left(\frac{\pi}{n+2}\right)\frac{\Gamma\left(\frac{1}{n+2}\right)\left[\Gamma\left(\frac{2}{n+2}\right)\right]^{2}\Gamma\left(\frac{3}{n+2}\right)}{\Gamma\left(\frac{4}{n+2}\right)}}\,.\label{eq:c1n}
\end{equation}
This is the main mathematical result of this paper.

To verify eq.~(\ref{eq:c1n}), we compute $\beta_{k}$ numerically
for $n=4,6,8,$ and $10$ by solving the mode equation (\ref{eq:modeEqnChiReduced})
with the WKB boundary conditions (\ref{eq:chi12}) and using the Wronskian
projection (\ref{eq:betaReal}). The object $(\beta^{\text{numerical}}_{k}-\beta_{k=0})/k^{2}$
(with $\beta_{k=0}$ given by eq.~(\ref{eq:topo}) with $N_{\text{Stokes sectors}}=n+2$)
agrees with $c_{1}(n)$ to an accuracy of $O(10^{-5})$ in the region
$k\lesssim0.01$. It is noteworthy that the topological information
of $N_{\text{Stokes sectors}}=n+2$ is ubiquitous in eq.~(\ref{eq:c1n})
as well. If we restore the physical dispersion relationship, $\omega^{2}_{k}(\eta)=C(k)+A\eta^{n}$,
the Bogoliubov coefficient contains factors of $A$, as will be displayed
explicitly later in eqs.~(\ref{eq:approx2}) and (\ref{eq:explicit}).
In the next subsection, we will see how $c_{1}(n)$ controls the dominant
portion of the particles produced for the class of models of $\omega^{2}_{k}(\eta)=k^{2}+A\eta^{n}$.

\subsection{Semi-analytic approximate number density for physical applications\label{subsec:Numerical-results}}

In this section, we combine eq.~(\ref{eq:c1n}) and a numerically
motivated ansatz to obtain a closed form approximation to the physical
number density of particles produced in the dispersion relationship
scenario of eq.~(\ref{eq:dispersionRelation}) for $n\geq4$ and
\begin{equation}
C(k)=k^{2}.\label{eq:example}
\end{equation}
Focusing on $C(k)=k^{2}$ instead of $C(k)=k^{2}-\Lambda^{2}_{\text{IR}}$
allows us to demonstrate the impact of the $c_{1}(n)$ computation
most strongly since without the $c_{1}(n)k^{2}$ term, the rigorous
validity of the $\beta_{k}$ coefficient is at a vanishing phase space
point $k=0$ for $C(k)=k^{2}$.

\selectlanguage{english}%
As implied by the scaling in eq.~(\ref{eq:rescalek}), we can restore
the (under)bar notation and write the physical number density of particles
produced as
\begin{align}
n_{\chi} & =\frac{1}{a^{3}2\pi^{2}}\int^{\infty}_{k_{m}}dkk^{2}\left|\underline{\beta}_{\bar{k}(k)}\right|^{2}\label{eq:kbarisntk}
\end{align}
where $k_{m}$ is the IR cutoff on particle modes satisfying $C(k_{m})=0$
(i.e.~with eq.~(\ref{eq:example}), $k_{m}=0$) and 
\begin{equation}
\bar{k}(k)=\sqrt{\frac{C(k)}{A^{\frac{2}{n+2}}}}\label{eq:explicit}
\end{equation}
which will make the $c_{1}(n)\bar{k}^{2}$ in $\underline{\beta}_{\bar{k}(k)}$
non-perturbative in $A$. For any finite $k_{\text{max}}$ defining
an interval $[0,k_{\text{max}}]$ and $|\beta^{\text{part }}_{\bar{k}(k)}|$
definition in which $|\beta^{\text{part}}|\lesssim|\beta|$ in that
interval, a part of the total $n_{\chi}$ can be written to reside
in 
\begin{align}
n^{\text{part}}_{\chi} & \equiv\frac{1}{a^{3}2\pi^{2}}\int^{k_{\text{max}}}_{0}dkk^{2}\left|\beta^{\text{part}}_{\bar{k}(k)}\right|^{2}.\label{eq:nchipartdef}
\end{align}

\begin{figure}[H]
\centering{}\includegraphics[scale=0.3]{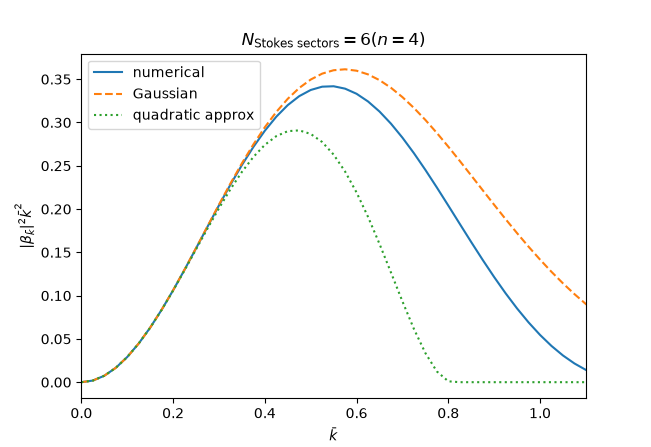}\includegraphics[scale=0.3]{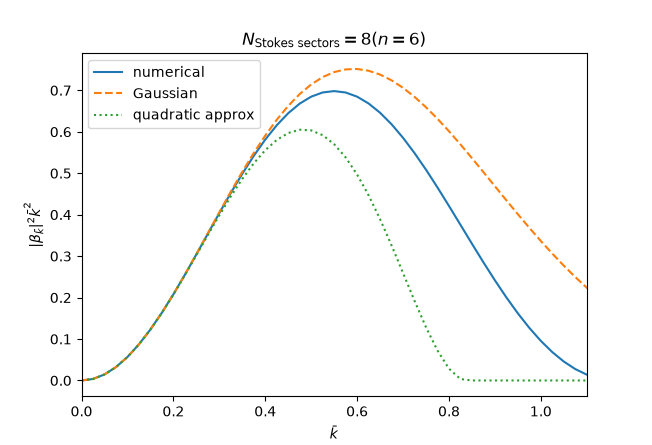}\includegraphics[bb=0bp 8.63448bp 439bp 313bp,scale=0.321]{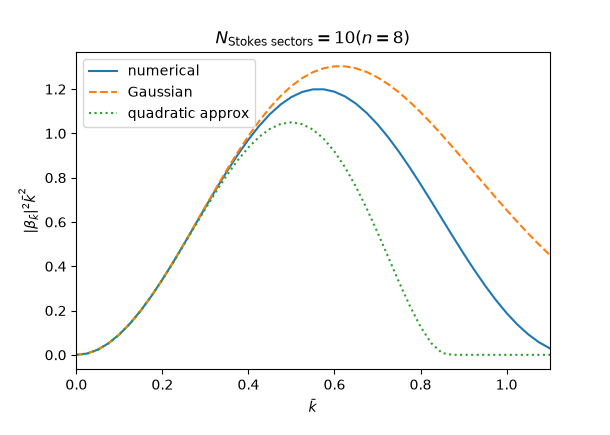}\caption{\label{fig:Comparison-of-different}Comparison of different approximations
for $n=4$, 6, 8 cases. The ``quadratic approx'' model is given
by eq.~(\ref{eq:approx2}) up to $\bar{k}=\bar{k}_{\text{max bd}}$
given in eq.~(\ref{eq:kmaxbd}) and underestimates the particle number
density. In contrast, the Gaussian approximation of eq.~(\ref{eq:gaussian})
overestimates compared to the true value.}
\end{figure}

As justified by numerical results of figure \ref{fig:Comparison-of-different}
for $n=4,6,8$, we can choose $\beta^{\text{part}}_{\bar{k}(k)}$
as
\begin{align}
\beta^{\text{part}}_{\bar{k}} & =\beta_{0}+c_{1}(n)\bar{k}^{2}\label{eq:c1approxdef}\\
 & =-i\cot\frac{\pi}{n+2}+i\left|c_{1}(n)\right|\frac{C(k)}{A^{\frac{2}{n+2}}}\label{eq:approx2}
\end{align}
in the $\bar{k}$ range $[0,\bar{k}(k_{\text{max}})]$ to obtain
\begin{equation}
n^{\text{part}}_{\chi}a^{3}2\pi^{2}=\cot^{2}\left(\frac{\pi}{n+2}\right)\left[\frac{k^{3}_{\text{max}}}{3}\right]-2\frac{|c_{1}(n)|}{A^{\frac{2}{n+2}}}\cot\left(\frac{\pi}{n+2}\right)\left[\frac{k^{5}_{\text{max}}}{5}\right]+\frac{|c_{1}(n)|^{2}}{A^{\frac{4}{n+2}}}\left[\frac{k^{7}_{\text{max}}}{7}\right]\label{eq:part}
\end{equation}
for a judiciously chosen $k_{\text{max}}$ that we will make explicit
next. Note the nontrivial $A$ dependence of the $c_{1}$ dependent
piece signaling the non-topological nature of these contributions.
Although eq.~(\ref{eq:approx2}) is nonperturbative in $A$ through
$\bar{k}$, that nonperturbativity cannot be used to make $\beta^{\text{part}}_{\bar{k}}$
large because we are expanding in small $\bar{k}$. On the other hand
the $\cot(\pi/(n+2))$ factor (i.e.~the topological factor) can be
made arbitrarily large for large $n$.

Let's choose $k_{\text{max}}$ for the ``quadratic approximation''
choice of eq.~(\ref{eq:approx2}). Because of eq.~(\ref{eq:explicit}),
the upper limit of the $\beta_{\bar{k}}$ index appearing in eq.~(\ref{eq:nchipartdef})
is 
\begin{equation}
\bar{k}_{\text{max}}=k_{\text{max}}A^{\frac{-1}{n+2}}
\end{equation}
and not $\bar{k}_{\text{max}}=k_{\text{max}}$. The quadratic behavior
makes $\beta^{\text{part}}_{\bar{k}}$ (eq.~(\ref{eq:approx2}))
vanish at finite $\bar{k}$ while explicit numerical computations
do not indicate this behavior (see figure \ref{fig:Comparison-of-different}).
That discrepancy is obviously expected from the fact that we are computing
a small $\bar{k}^{2}$ expansion in eq.~(\ref{eq:approx2}). Note
that one can analytically compute where $\beta^{\text{part}}_{\bar{k}}$
vanishes (see eq.~(\ref{eq:approx2})) as
\begin{equation}
\sqrt{\frac{\cot\left(\frac{\pi}{6}\right)}{|c_{1}(4)|}}\approx0.8<\bar{k}_{\text{max bd}}=\sqrt{\frac{\cot\left(\frac{\pi}{n+2}\right)}{|c_{1}(n)|}}<\sqrt{\frac{3}{2}}\approx1.2\label{eq:kmaxbd}
\end{equation}
between $n=4$ and $n=\infty$. Since the particle production spectrum
is expected to be dominantly one bump (attested by numerical computations),
we can naturally set $\bar{k}_{\text{max}}=\bar{k}_{\text{max bd}}$,
and find using eq.~(\ref{eq:part})
\begin{equation}
n^{\text{part}}_{\chi}=\frac{A^{\frac{3}{n+2}}}{a^{3}}\frac{4\cot^{7/2}\left(\frac{\pi}{n+2}\right)}{105\pi^{2}|c_{1}(n)|^{3/2}}
\end{equation}
where the non-perturbative nature in $A$ disappeared due to $k_{\text{max}}\propto A^{1/(n+2)}$.
As one can see in figure \ref{fig:Comparison-of-different}, $n^{\text{part}}_{\chi}$
underestimates the true spectrum by design. The integrated spectrum
numerically for $n=4,6,8$ is 
\begin{equation}
n^{\text{semi-anal}}_{\chi}\approx n^{\text{part}}_{\chi}\times\begin{cases}
1.6 & n=4\\
1.5 & n=6\\
1.5 & n=8
\end{cases}
\end{equation}
which is remarkably constant. Hence, we estimate
\begin{equation}
\boxed{n^{\text{semi-anal}}_{\chi}\approx\frac{A^{\frac{3}{n+2}}}{a^{3}}\frac{2\cot^{7/2}\left(\frac{\pi}{n+2}\right)}{35\pi^{2}|c_{1}(n)|^{3/2}}}\label{eq:quadraticresult}
\end{equation}
where $c_{1}(n)$ is given by eq.~(\ref{eq:c1n}).

As another approach to use $c_{1}(n)$ to obtain an approximate formula
for $n_{\chi}$, note we can appeal to the oscillatory nature of the
amplitude and stationary phase approximation intuition to consider
the ansatz
\begin{equation}
|\beta^{\text{Gaussian}}_{\bar{k}}|\equiv\cot\left(\frac{\pi}{n+2}\right)\exp\left(\frac{-|c_{1}(n)|}{\cot\left(\frac{\pi}{n+2}\right)}\bar{k}^{2}\right)\label{eq:gaussian}
\end{equation}
(matching eq.~(\ref{eq:approx2}) to leading $\bar{k}^{2}$ expansion)
which can be numerically checked to give an overestimate approximation
for different values of $n$ (see figure \ref{fig:Comparison-of-different}).
This will lead to the number density
\begin{equation}
\left(n^{\text{Gaussian}}_{\chi}\right)_{\Lambda_{\text{IR}}=0}=\frac{A^{\frac{3}{n+2}}}{a^{3}}\frac{\cot^{7/2}\left(\frac{\pi}{n+2}\right)}{16\sqrt{2}\pi^{3/2}|c_{1}(n)|^{3/2}}\label{eq:gaussianresult}
\end{equation}
which is an overestimate as seen in figure \ref{fig:Comparison-of-different}.
What is reassuring is that the power dependences of $n$ show robustness
since the dependences in eqs.~(\ref{eq:quadraticresult}) and (\ref{eq:gaussianresult})
agree. As an aside, we can express $A$ as
\begin{equation}
A=\frac{a^{n+2}_{i}\left[m^{2}_{\text{eff}}(\eta)\right]^{\frac{n+2}{2}}}{n^{n}}\left(\frac{\frac{1}{a_{i}}\partial_{\eta}m^{2}_{\text{eff}}(\eta)}{\left[m^{2}_{\text{eff}}(\eta)\right]^{\frac{3}{2}}}\right)^{n}
\end{equation}
where $m^{2}_{\text{eff}}(\eta)a^{2}_{i}\equiv A\eta^{n}$ (where
$a_{i}$ is a fiducial scale factor) to write the number density of
particles produced more physically suggestively as 
\begin{equation}
n^{\text{semi-anal}}_{\chi}\approx\frac{a^{3}_{i}}{a^{3}(\eta)}\frac{2\cot^{7/2}\left(\frac{\pi}{n+2}\right)}{35\pi^{2}|c_{1}(n)|^{3/2}}\frac{\left[m^{2}_{\text{eff}}(\eta)\right]^{\frac{3}{2}}}{n^{\frac{3n}{n+2}}}\left(\frac{\frac{1}{a_{i}}\partial_{\eta}m^{2}_{\text{eff}}(\eta)}{\left[m^{2}_{\text{eff}}(\eta)\right]^{\frac{3}{2}}}\right)^{\frac{3n}{n+2}}.\label{eq:suggestive}
\end{equation}

Eq.~(\ref{eq:suggestive}) shows that the particle production is
a direct consequence of time-translation symmetry breaking explicitly.
What is still counter-intuitive in this expression is that it looks
like there is a nontrivial time-dependence through the time-dependent
effective mass squared even though there is not. This is a non-generic
characteristic of the model defined by eq.~(\ref{eq:dispersionRelation}).
Another interesting item to note is that the topological factor of
$\cot(\pi/N_{\text{Stokes sectors}})=\cot\left(\pi/(n+2)\right)$
survives here even though that naively governs only the $k=0$ part
which does not have any phase space:
\begin{equation}
\frac{\cot^{7/2}\left(\frac{\pi}{n+2}\right)}{|c_{1}(n)|^{3/2}}=\left(\frac{\cot\left(\frac{\pi}{n+2}\right)}{|c_{1}(n)|}\right)^{3/2}\cot^{2}\left(\frac{\pi}{n+2}\right)
\end{equation}
where we noted in eq.~(\ref{eq:kmaxbd}) that $\cot(\pi/(n+2))/|c_{1}(n)|$
is insensitive to $n$ for large $n$.
\selectlanguage{american}%

\section{Conclusions}

By extending the exponentially improved asymptotic expansion techniques
of \cite{hinton1979stokesmultipliers,zhu1992stokesconstants} and
discrete symmetries that involve the complexification of the momentum
related variable $C(k)$ in addition to the conformal time $\eta$,
we have in eq.~(\ref{eq:approx2}) computed the leading soft momentum
contribution to the particle production amplitude in the situation
with a dispersion relationship in conformal time coordinates of $\omega^{2}_{k}(\eta)=C(k)+A\eta^{n\in\text{positive even}}$
as
\begin{equation}
\beta_{k}=-i\cot\frac{\pi}{n+2}+i\left|c_{1}(n)\right|\frac{C(k)}{A^{\frac{2}{n+2}}}+O(C^{2})\label{eq:finresult}
\end{equation}
where $c_{1}(n)$ is given by eq.~(\ref{eq:c1n}) and is purely imaginary.
The fact that this amplitude has no non-analyticity in $C$ is noteworthy given
that intermediate Schroedinger phase results have non-analytic $C$
dependence as seen in eqs.~(\ref{eq:betaProduct1}) and (\ref{eq:nonanalytic-factor}).
Indeed, the survival of non-analytic phase would be inconsistent with
the time-reversal symmetry in the system. Note also that this amplitude
is \foreignlanguage{english}{non-perturbative} in the coupling $A$.
Combining this analytic result with numerics, we established a semi-analytic
estimate of $n_{\chi}$ given by eq.~(\ref{eq:quadraticresult})
where the non-perturbative signal of $A^{-2/(n+2)}$ from $\beta_{k}$
disappeared due to the phase space volume having powers of $A^{1/(n+2)}$.
Nonetheless, unlike in standard saddle-point approximations where
$\beta_{k}$ and the number density are exponentially suppressed,
the amplitude as well as the resulting number density here has no
exponential suppression as evinced by $\cot[\pi/(n+2)]=\cot[\pi/N_{\text{Stokes sectors}}]$
which can be made arbitrarily large for large $n$.

There are several obvious future directions related to this paper.
In addition to the dark matter scenario given in \cite{Chung:2024pot},
it would be useful to apply eqs.~(\ref{eq:finresult}) and (\ref{eq:quadraticresult})
to other physical scenarios. For example, there may be enhanced symmetry
point type of isocurvature scenarios with nonrenormalizable operators
where this formula would naturally apply. Another natural application
of this paper would be to apply the results to a cosmologically viable
dark matter or isocurvature scenario with $C(k_{m})=0$ for $k_{m}>0$:
i.e. scenarios with tachyons. On a more technical side, there may
be ways to combine the present techniques with other techniques such
as the saddle-point approximation or other Stokes computational techniques
to derive a less numerically reliant large $k$ result leading to
the integrated number density result presumably approximated by eq.~(\ref{eq:quadraticresult}).
One may also be able to generalize the current results to higher spins.
It would also be interesting to see how one can apply the algorithmically
simplified Stokes phenomena exposing exponentially improved asymptotic
expansion method presented in this paper to other time-dependent cosmological
scenarios.

\acknowledgments
D.J.H.C. acknowledges the support by the U.S. Department of Energy, Office of Science, Office of High Energy Physics under Award Number DE-SC0017647.  N.S. and J.Z. acknowledge partial support from the Ray MacDonald Fund and Wisconsin Alumni Research Fund at the University of Wisconsin-Madison.

\appendix

\section{Analytically continuing the time coordinate\label{subsec:analytically-continue-time}}

In this appendix, we provide further details on analytically continuing
the time coordinate $\eta$ in the physical WKB basis $f^{\p}_{\pm}(k,\eta)$
to obtain the fictitious WKB basis $f_{\pm}(k,z)=\exp\left(\pm i\theta(k,z)\right)/\sqrt{2\omega_{k}(z)}$.
In particular, given the branch cuts specified in section \ref{subsec:Analytic-continuation-and}
(see above eq.\ (\ref{eq:Uj}) and figure \foreignlanguage{english}{\ref{fig:Branch-cut-structures}}),
we will express the complexified dispersion relationship $\omega_{k}(z)$
in terms of a sign factor determined by the topology of branch cuts
and a Taylor series globally defined in the asymptotic region. This
yields expression (\ref{eq:omega-continued-sj}) that is used to derive
the rotational symmetries in section \ref{sec:Discrete-symmetries}.
Based on the complexified $\omega_{k}(z)$, we will then specify the
branch structure for $\sqrt{\omega_{k}(z)}$, thereby completing the
analytic continuation of  the time coordinate in the WKB basis.

\selectlanguage{english}%

Consider $\omega_{k}(z)=\sqrt{k^{2}+z^{n}}$ that is globally defined
as a function of complex number $z$ (see figure \ref{fig:Branch-cut-structure.-1})
and analytic except along the branch cuts which we choose to be coincident
with the radially emanating anti-Stokes boundaries. Because of the
branch cuts, this original representation of $\omega_{k}(z)$ is naturally
discontinuous across each of the branch cuts on the first Riemann
sheet even though it is continuous within the disk of radius $|\psi_{j}|$
(the modulus of the branch point, recall that $\psi_{j}=k^{2/n}e^{(2j-1)i\pi/n}$):
a region which we call inner disk $D$. On the other hand, because
$\omega_{k}(z)$ naturally accesses the second Riemann sheet under
rotational algebraic manipulation $\omega_{k}(ze^{i\theta})$ in an
unintended manner for $|z|>|\psi_{j}|$ (for example in the asymptotic
annulus (AA)), working with only the original $\omega_{k}(z)$ representation
leads to an opaque construction of the rotational symmetry representation
that we build in section \ref{sec:Discrete-symmetries}.

To remedy this problem and make the symmetry representation construction
more transparent, we use a function $\underline{\omega}_{k}(z)$ which
cannot access the second Riemann sheet in $z$ under rotation and
represent the $\omega_{k}(z)$ staying strictly on the first Riemann
sheet as 
\begin{equation}
\omega_{k}(z)=\begin{cases}
\sqrt{k^{2}+z^{n}} & z\in D\\
s(z)\underline{\omega}_{k}(z) & |z|>|\psi_{j}|
\end{cases}\label{eq:sgf}
\end{equation}
where $\underline{\omega}_{k}(z)$ is defined as the Taylor expansion
\begin{equation}
\underline{\omega}_{k}(z)\equiv z^{n/2}\sum^{\infty}_{m=0}\begin{pmatrix}1/2\\
m
\end{pmatrix}\left(\frac{k^{2}}{z^{n}}\right)^{m}\label{eq:underomega}
\end{equation}
which is analytic in $z$ (since $n$ is even) and converges for $|z|\geq|\psi_{j}|$
but not for $|z|<|\psi_{j}|$. Since the disk $D$ defined by $|z|\leq|\psi_{j}|$
for $j\in\{1,...,n\}$ is therefore the natural boundary for $\underline{\omega}_{k}(z)$,
we define branch sectors $\mathfrak{B}_{j}$ as the region bounded
by two branch cuts and outside of the disk $D$. We will conventionally
put the $\mathscr{\mathfrak{B}}_{j=1}$ to include the positive real
axis and go counter-clockwise in assigning the indices. Because $\underline{\omega}_{k}(z)$
has no branch cuts in the AA nor for any $|z|>|\psi_{j}|$ (in contrast
with $\omega_{k}(z)$), its rotational transformation is manifestly
trivial. All the nontrivialities of the rotation are in the discrete
branch cut information which is now separated in $s(z)$. We will
determine $s(z)$ below by matching to the properties of $\omega_{k}(z)$.

The original function $\omega_{k}(z)$ is analytic on $\mathfrak{B}_{1}$
that includes positive real axis just like $\underline{\omega}_{k}(z)$
and their values coincide by construction. Therefore, $\underline{\omega}_{k}(z)$
is isomorphic to $\omega_{k}(z)$ in this region such that $s(z)$
of eq.\ (\ref{eq:sgf}) takes the value of unity in this region.
To find the function $s(z)$ in the other $\mathfrak{B}_{j}$ regions,
note that since 
\begin{equation}
\omega^{2}_{k}(z)=z^{n}\left(1+\frac{k^{2}}{z^{n}}\right)=\underline{\omega}^{2}_{k}(z)
\end{equation}
is analytic in all $|z|>|\psi_{j}|$, we find the identity 
\begin{equation}
\omega_{k}(z)=\pm\underline{\omega}_{k}(z)
\end{equation}
in any given branch sector $\mathscr{\mathfrak{B}}_{j}$.\footnote{It is a well-appreciated fact that the infinite series equation (\ref{eq:underomega})
collapses to two types of terms in the series when squared.} This says that $s(z)$ on any given $\mathfrak{B}_{j}$ is a constant
sign that only depends on $j$ since both $\omega_{k}(z)$ and $\underline{\omega}_{k}(z)$
are analytic in $\mathfrak{B}_{j}$. To find the dependence on $j$,
consider $\omega_{k}(z)$ on the two adjacent points $z_{+}\in\mathscr{\mathfrak{B}}_{j+1}$
and $z_{-}\in\mathfrak{B}_{j}$ on either side of a branch cut to
establish a relative minus sign:
\begin{equation}
\omega_{k}(z_{+})=-\omega_{k}(z_{-})
\end{equation}
which says that $s(z)$ flips signs every time $j$ in $\mathfrak{B}_{j}$
increments by unity. Hence, we have shown 
\begin{equation}
\omega_{k}(z)=\begin{cases}
\sqrt{k^{2}+z^{n}} & z\in D\\
(-1)^{q-1}\underline{\omega}_{k}(z) & z\in\mathfrak{B}_{q}
\end{cases}\label{eq:omega-continued}
\end{equation}
or equivalently $s(z)=(-1)^{q-1}$ for any $z\in\mathfrak{B}_{q}$.
By construction, we see now that $\omega_{k}(ze^{i\theta})$ cannot
access the second Riemann sheet in $z$ in the representation of the
right hand side of equation (\ref{eq:omega-continued}). Note that
if we want to be more formulaically explicit regarding the Stokes
sector embedding, then we can write $\mathscr{S}_{j}\subset\mathfrak{B}_{\mathcal{I}(j)}$
with 
\begin{align}
\mathcal{I}(j) & =\begin{cases}
j & \text{if }j\in\{1,...,n/2+1\}\\{}
[(j-2)\mod n]+1 & \text{if }j\in\{n/2+2,...,n+2\}
\end{cases}\label{eq:Ijdef}
\end{align}
Even more explicit is the form in terms of Stokes sector label instead:
\begin{equation}
\omega_{k}(z)=\begin{cases}
\sqrt{k^{2}+z^{n}} & z\in D\\
s_{j}\underline{\omega}_{k}(z) & z\in\mathscr{S}^{E}_{j}
\end{cases}\label{eq:omega-continued-sj}
\end{equation}
where
\begin{align}
s_{j} & \equiv\begin{cases}
(-1)^{j-1} & 1\leq j\leq n/2+1\\
(-1)^{j} & n/2+2\leq j\leq n+2
\end{cases}\label{eq:sj}
\end{align}
and where the superscript ``E'' in $\mathscr{S}^{E}_{j}$ corresponds
to the extension of the Stokes sector to cover the region between
the Stokes sector and the inner disk. Since we will be rotating between
Stokes sectors in section \ref{sec:Discrete-symmetries}, it will
be more convenient to use $s_{j}$ which is a function of $\mathscr{S}^{E}_{j}$
indices $j$ and not the branch sector indices $\mathfrak{B}_{j}$.

Next, we analytically continue the time coordinate in the square root
$\sqrt{\omega_{k}(z)}$ appearing in the denominator of the fictitious
WKB basis $f_{\pm}(k,z)$. \foreignlanguage{american}{Since we want
$f_{\pm}(k,z)$ to match the physical WKB basis $f^{{\rm phy}}_{\pm}(k,\eta)$
on the positive real axis, we require $\sqrt{\omega_{k}(z)}>0$ for
$z\in\mathbb{R}_{>0}$. The values of $\sqrt{\omega_{k}(z)}$ elsewhere
are then fixed by continuity. Inside the inner disk $D$, we have
\begin{equation}
\sqrt{\omega_{k}(z)}=+\left(k^{2}+z^{n}\right)^{1/4},\quad z\in D\,.\label{eq:sqrt-omega-D}
\end{equation}
Outside the $D$, using eq.\ (\ref{eq:omega-continued}) we get
\begin{equation}
\sqrt{\omega_{k}(z)}=\sqrt{(-1)^{q-1}\underline{\omega}_{k}(z)}=\sqrt{(-1)^{q-1}}\sqrt{\underline{\omega}_{k}(z)},\quad z\in\mathfrak{B}_{q}
\end{equation}
where the square root branches of $\sqrt{(-1)^{q-1}}$ and $\sqrt{\underline{\omega}_{k}(z)}$
are fixed by continuity}\footnote{\selectlanguage{american}%
One can show that if we retain eq.\ (\ref{eq:sqrt-omegaunderline-branch})
and choose $\sqrt{(-1)^{q-1}}=e^{\frac{i\pi}{2}(q-1)}$ instead, then
$\sqrt{\omega_{k}(z)}$ is discontinuous at $z=-k^{2/n}$ for odd
$N=n/2$.\selectlanguage{english}%
}\foreignlanguage{american}{
\begin{align}
\sqrt{(-1)^{q-1}} & =e^{-\frac{i\pi}{2}(q-1)}\\
\sqrt{\underline{\omega}_{k}(z)} & =|z|^{\frac{n}{4}}e^{\frac{in}{4}\arg z}\sum^{\infty}_{m=0}\begin{pmatrix}1/4\\
m
\end{pmatrix}\left(\frac{k^{2}}{z^{n}}\right)^{m},\quad|z|>k^{2/n}\,.\label{eq:sqrt-omegaunderline-branch}
\end{align}
Thus, the analytic continuation of $\sqrt{\omega_{k}(z)}$ is
\begin{equation}
\sqrt{\omega_{k}(z)}=\begin{cases}
+\left(k^{2}+z^{n}\right)^{1/4} & z\in D\\
e^{-\frac{i\pi}{2}(q-1)}\sqrt{\underline{\omega}_{k}(z)} & z\in\mathfrak{B}_{q}
\end{cases}\label{eq:sqrt-omega-continued}
\end{equation}
with $\sqrt{\underline{\omega}_{k}(z)}$ specified by eq.\ (\ref{eq:sqrt-omegaunderline-branch}).
Consequently, the analytical continuation $\sqrt{\omega_{k}(z)}$
matches the $\sqrt{\omega_{k}(\eta)}$ in the physical WKB basis.
To see this, note they match on the positive real axis by construction.
On the negative real axis, they match in the inner disk $D$ because
of eq.\ (\ref{eq:sqrt-omega-D}), and also match for $z\in\mathbb{R}_{<-k^{2/n}}$
because 
\begin{equation}
\sqrt{\omega_{k}(z)}|_{z\in\mathbb{R}_{<-k^{2/n}}}=e^{-\frac{i\pi N}{2}}|z|^{\frac{N}{2}}e^{\frac{i\pi N}{2}}\sum^{\infty}_{m=0}\begin{pmatrix}1/4\\
m
\end{pmatrix}\left(\frac{k^{2}}{z^{n}}\right)^{m}=|z|^{\frac{N}{2}}\sum^{\infty}_{m=0}\begin{pmatrix}1/4\\
m
\end{pmatrix}\left(\frac{k^{2}}{z^{n}}\right)^{m}>0\,.
\end{equation}
}

\selectlanguage{american}%
The analytic continuations of $\omega_{k}(z)$ and $\sqrt{\omega_{k}(z)}$
have two further implications. First, the fictitious WKB basis functions
defined by eq.\ (\ref{eq:WKBFic}),
\begin{equation}
f_{\pm}(k,z)=\frac{e^{\pm i\int^{z}_{0}dz'\,\omega_{k}(z')}}{\sqrt{2\omega_{k}(z)}}\,,
\end{equation}
are uniquely specified by the analytic continuations (\ref{eq:omega-continued})
and (\ref{eq:sqrt-omega-continued}). This analytic continuation matches
the fictitious and physical WKB bases throughout the real axis. This
means the Bogoliubov $\beta_{k}$ computed using the propagator expression
(\ref{eq:betaPropagatorExp}) of the fictitious WKB basis is equal
to that computed using the Wronskian projection (\ref{eq:betaReal})
of the physical WKB basis. Secondly, the analytic continuation of
$\sqrt{\omega_{k}(z)}$ implies that it gains a $(-i)$ factor when
crossing a branch cut counterclockwise. This is used in deriving eq.\ (\ref{eq:Fpm}),
the transformation of the fictitious WKB basis across branch cuts.\selectlanguage{american}%

\section{Complexifying the momentum variable\label{subsec:analytically-continue-momentum}}

In this appendix, we provide a different perspective for complexifying
the physical momentum variable $k\in\mathbb{R}_{>0}$ in the WKB phase
$\theta(k,z)=\int^{z}_{0}dz'\,\omega_{k}(z')$ to obtain $\theta(p,z)$,
where $p\in\mathbb{C}$ denotes the complexified momentum. The complexified
WKB phase $\theta(p,z)$ is used to derive the rotational symmetry
in section \ref{subsec:Rotational-symmetry} and to isolate the $k$
non-analyticities in section \ref{subsec:Non-analyticity-isolation}.
Throughout this appendix we assume $n\ge4$.

Our prescription for complexification is to split $\theta(k,z)$ into
terms that are analytic and non-analytic in $k$. We complexify the
analytic term by replacing $k$ with the complex momentum $p$, while
the non-analytic terms are continued by specifying the Riemann sheet
in the complex $p$-plane. As a generalization to section \ref{subsec:Non-analyticity-isolation},
for $z\in\mathscr{S}_{j}$, this decomposition can be achieved by
splitting the WKB phase at $\psi_{j}$, the endpoint of the $j$-th
branch cut on the $z$-plane, which intuitively captures the non-analytic
dependence on $k$. So we write
\begin{equation}
\theta(k,z)=\int^{\psi_{j}}_{0}dz'\omega_{k}(z')+\int^{z}_{\psi_{j}}dz'\omega_{k}(z')=\Delta_{j}(k)+\int^{z}_{\psi_{j}}dz'\omega_{k}(z')\label{eq:theta-split}
\end{equation}
where recall that $\Delta_{j}(k)=\int^{\psi_{j}}_{0}dz\omega_{k}(z)$
denotes the phase factor, which can be computed with a change of variable
\begin{equation}
\Delta_{j}(k)=\int^{\psi_{j}}_{0}dz\sqrt{k^{2}+z^{n}}=e^{\frac{2j-1}{n}i\pi}k^{\frac{n+2}{n}}d_{n},\quad d_{n}\equiv\frac{\Gamma\left(\frac{3}{2}\right)\Gamma\left(1+\frac{1}{n}\right)}{\Gamma\left(\frac{3}{2}+\frac{1}{n}\right)}\,,\label{eq:Deltaj-1}
\end{equation}
To further evaluate the second term in eq.\ (\ref{eq:theta-split}),
we use the Taylor series expression derived in eq.\ (\ref{eq:omega-continued-sj})
\begin{equation}
\omega_{k}(z)=s_{j}z^{n/2}\sum^{\infty}_{m=0}\begin{pmatrix}1/2\\
m
\end{pmatrix}\left(\frac{k^{2}}{z^{n}}\right)^{m},\quad z\in\mathscr{S}^{E}_{j}\,.
\end{equation}
This series converges uniformly for $z\in\mathscr{S}^{E}_{j}$, so
we can integrate it term by term to get
\begin{align}
\int^{z}_{\psi_{j}}dz'\omega_{k}(z') & =s_{j}\frac{z^{n/2+1}}{n/2+1}-s_{j}A(k,z)+s_{j}\Psi_{j}(k)\label{eq:theta-seg-2}
\end{align}
where
\begin{align}
A(k,z) & \equiv\sum^{\infty}_{m=1}\begin{pmatrix}1/2\\
m
\end{pmatrix}\frac{1}{mn-(n/2+1)}\frac{k^{2m}}{z^{mn-(n/2+1)}}\,,\label{eq:Akz}\\
\Psi_{j}(k) & \equiv\sum^{\infty}_{m=0}\begin{pmatrix}1/2\\
m
\end{pmatrix}\frac{1}{mn-(n/2+1)}\frac{k^{2m}}{\psi^{mn-(n/2+1)}_{j}}\,.\label{eq:Psij-0}
\end{align}
Here we note for $z\in\mathscr{S}^{E}_{j}$, $A(k,z)$ is analytic
in $k$. The $\Psi_{j}(k)$ term can be further simplified using $\psi_{j}=k^{2/n}e^{i(2j-1)\pi/n}$:
\begin{align}
\Psi_{j}(k) & =k^{\frac{n+2}{n}}\sum^{\infty}_{m=0}\begin{pmatrix}1/2\\
m
\end{pmatrix}\frac{e^{-i\frac{2j-1}{n}\pi\left(nm-n/2-1\right)}}{nm-(n/2+1)}\,.\label{eq:Psij}
\end{align}
which implies $\Psi_{j}(k)$ is proportional to $k^{(n+2)/n}$ as
the series in $m$ is convergent, making $\Psi_{j}(k)$ non-analytic
in $k$. Combining eqs.\ (\ref{eq:theta-split}) and (\ref{eq:theta-seg-2})
we obtain the decomposition
\begin{equation}
\theta(k,z)=\underset{\text{analytic in }k}{\underbrace{s_{j}\frac{z^{n/2+1}}{n/2+1}-s_{j}A(k,z)}}+\underset{\text{non-analytic in }k}{\underbrace{\Delta_{j}(k)+s_{j}\Psi_{j}(k)}}\,,\quad k\in\mathbb{R}_{\geq0},\quad z\in\mathscr{S}^{E}_{j}\,.\label{eq:theta-k-z-split}
\end{equation}

The analytic term is then complexified by replacing $k$ with $p$
in the power series expression (\ref{eq:Akz}) 
\begin{equation}
s_{j}A(k,z)\to s_{j}A(p,z)
\end{equation}
with
\begin{equation}
A(p,z)\equiv\sum^{\infty}_{m=1}\begin{pmatrix}1/2\\
m
\end{pmatrix}\frac{1}{mn-(n/2+1)}\frac{p^{2m}}{z^{mn-(n/2+1)}},\quad p\in\mathbb{C}\,.\label{eq:A-pz}
\end{equation}
We next consider the non-analytic terms $\Delta_{j}(k)$ and $\Psi_{j}(k)$.
Both can be expressed in terms of the branch point $\psi_{j}$. Since
integration contour for $\Delta_{j}(k)$ lies inside the inner disk
$D$, we may Taylor expand $\omega_{k}(z')$ in powers of $z'^{n}/k^{2}$
and integrate term by term:
\begin{align}
\Delta_{j}(k) & =\int^{\psi_{j}}_{0}dz'k\sum^{\infty}_{m=0}\begin{pmatrix}1/2\\
m
\end{pmatrix}\left(\frac{z'^{n}}{k^{2}}\right)^{m}\nonumber \\
 & =[k\psi_{j}(k)]\sum^{\infty}_{m=0}\begin{pmatrix}1/2\\
m
\end{pmatrix}\frac{1}{nm+1}\frac{[\psi_{j}(k)]^{nm}}{k^{2m}}\,.
\end{align}
On the other hand, eq.\ (\ref{eq:Psij-0}) can be rewritten as
\begin{equation}
\Psi_{j}(k)=[\psi_{j}(k)]^{n/2+1}\sum^{\infty}_{m=0}\begin{pmatrix}1/2\\
m
\end{pmatrix}\frac{1}{nm-(n/2+1)}\frac{k^{2m}}{[\psi_{j}(k)]^{nm}}\,.
\end{equation}
Since $\psi_{j}\propto k^{2/n}$ implies $\psi^{nm}_{j}\propto k^{2m}$,
the factors $\psi^{nm}_{j}$ are analytic in $k$. It is therefore
sufficient to obtain the complexifications $\Delta_{j}(p)$ and $\Psi_{j}(p)$
by specifying the analytic continuation of the branch point $\psi_{j}(p)$.
By definition, $\psi_{j}(p)$ is the complexification of the $j$-th
zero of $\omega_{k}(z)$, which should satisfy $p^{2}+[\psi_{j}(p)]^{n}=0$.
Specifically, we choose
\begin{equation}
\psi_{j}(p)=e^{i\frac{2j-1}{n}\pi}|p|^{\frac{2}{n}}e^{i\frac{2}{n}\arg p},\quad\arg p\in(-2\pi,0]\,.
\end{equation}
Here we have the freedom to choose the branch cut for $\psi_{j}(p)$
as long as the cut is not on the positive real momentum axis, then
$\psi_{j}(p=k)=e^{i(2j-1)\pi/n}k^{2/n}$ is guaranteed. Since we are
interested in rotating $k$ to $(\gamma^{-n/2})^{j}k$ with $1\leq j\leq n+1$
as in section \ref{subsec:Rotational-symmetry}, we stipulate $\arg p\in(-2\pi,0]$
to fix the branch cut for $\psi_{j}(p)$. This in turn fixes the branch
cut for $\theta(p,z)$, giving the complexified WKB phase
\begin{equation}
\theta(p,z)=s_{j}\frac{z^{n/2+1}}{n/2+1}-s_{j}A(p,z)+\Delta_{j}(p)+s_{j}\Psi_{j}(p),\ \arg p\in(-2\pi,0],\ z\in\mathscr{S}_{j}\,.\label{eq:theta-continued}
\end{equation}
with the analytic term $A(p,z)$ given by eq.\ (\ref{eq:A-pz}) and
the non-analytic terms given by
\begin{align}
\Delta_{j}(p) & =\psi_{j}(p)\sum^{\infty}_{m=0}\begin{pmatrix}1/2\\
m
\end{pmatrix}\frac{1}{nm+1}\frac{[\psi_{j}(p)]^{nm}}{p^{2m-1}}\,,\\
\Psi_{j}(p) & =[\psi_{j}(p)]^{n/2+1}\sum^{\infty}_{m=0}\begin{pmatrix}1/2\\
m
\end{pmatrix}\frac{1}{nm-(n/2+1)}\frac{p^{2m}}{[\psi_{j}(p)]^{nm}}\,.
\end{align}
This is the main result of this appendix.

Strictly speaking, we should have included the $p$ dependence in
the Stokes sector $\mathscr{S}_{j}$, since varying $\arg p$ rotates
the zeros $\psi_{j}(p)$ and hence drags the branch cuts emanating
from them. Nevertheless, as shown in figure \ref{fig:Shown-is-our-1},
we may choose to keep the branch-cut segments fixed in the AA under
this continuation. The corresponding branch sectors, and therefore
the factors $s_{j}$, may then be kept fixed. We therefore suppress
the momentum dependence of the notation $\mathscr{S}_{j}$.
\section{Relation among Stokes constants\label{sec:Propagator-rotation-details}}

Here we show how to obtain the relation (\ref{eq:SRot1})-(\ref{eq:SRot3})
among the Stokes constants from the rotational symmetry representation
on the propagators (\ref{eq:propagatorRotation}) and the property
(\ref{eq:DeltaRotation}) of the phase factor $\Delta_{j}(k)$.

We start with the propagator $U_{1}(k)$ across the first Stokes sector,
and rotate counterclockwise to obtain all the other propagators $U_{j}(k)$.
First, we introduce the notations
\begin{equation}
U_{-}[S]\equiv\begin{pmatrix}1 & 0\\
S & 1
\end{pmatrix},\quad U_{+}[S]\equiv\begin{pmatrix}1 & S\\
0 & 1
\end{pmatrix},\quad\mathscr{R}[x]\equiv\begin{pmatrix}e^{2ix} & 0\\
0 & e^{-2ix}
\end{pmatrix}\,.
\end{equation}
It follows immediately that
\begin{equation}
\mathscr{R}^{-1}[x]=\mathscr{R}[-x],\quad\mathscr{R}[x]\mathscr{R}[y]=\mathscr{R}[x+y]\,,\quad\mathscr{R}[x]U_{\pm}[S]\mathscr{R}^{-1}[x]=U_{\pm}[e^{\pm4ix}S]\,.\label{eq:RUProperty}
\end{equation}
On the upper half plane, the rotational symmetry representation (\ref{eq:propagatorRotation})
reads
\begin{equation}
U_{-}\left[S_{j+1}(k)\right]=\mathscr{R}\left[\Delta_{j}(k)\right]U_{-}\left[S_{j}(\gamma^{-N}k)\right]\mathscr{R}^{-1}\left[\Delta_{j}(k)\right],\quad1\leq j\leq N,
\end{equation}
Iterating the above relation and using eqs.\ (\ref{eq:DeltaRotation})
and (\ref{eq:RUProperty}), for $1\leq j\leq N$ we obtain
\begin{align}
U_{-}\left[S_{j+1}(k)\right] & =\mathscr{R}\left[\Delta_{j}(k)\right]\mathscr{R}\left[\Delta_{j-1}(\gamma^{-N}k)\right]U_{-}\left[S_{j-1}(\gamma^{-2N}k)\right]\mathscr{R}^{-1}\left[\Delta_{j-1}(\gamma^{-N}k)\right]\mathscr{R}^{-1}\left[\Delta_{j}(k)\right]\nonumber \\
 & =\mathscr{R}[\tilde{\Delta}_{j}(k)]U_{-}\left[S_{1}(\gamma^{-Nj}k)\right]\mathscr{R}^{-1}[\tilde{\Delta}_{j}(k)]\nonumber \\
 & =U_{-}\left[e^{-4i\tilde{\Delta}_{j}(k)}S_{1}(\gamma^{-Nj}k)\right]
\end{align}
with $\tilde{\Delta}_{j}(k)\equiv\sum^{j}_{r=1}(-1)^{j-r}\Delta_{r}(k)$
an alternating sum. This leads to eq.\ (\ref{eq:SRot1}). For $j=N+1$,
as the common boundary of the $(N+1)$-th Stokes sector and the $(N+2)$-th
Stokes sector is on the negative real axis, the rotation from $U_{N+1}(k)$
to $U_{N+2}(k)$ crosses no branch cuts and is given by eq.\ (\ref{eq:propagatorRotation})
\begin{equation}
U_{+}\left[S_{N+2}(k)\right]=BU_{-}\left[S_{N+1}(\gamma^{-N}k)\right]B\,,
\end{equation}
which yields eq.\ (\ref{eq:SRot2}). On the lower half plane, the
rotational symmetry representation (\ref{eq:propagatorRotation})
implies
\begin{equation}
U_{+}\left[S_{j+1}(k)\right]=\mathscr{R}\left[\Delta_{j-1}(k)\right]U_{+}\left[S_{j}(\gamma^{-N}k)\right]\mathscr{R}^{-1}\left[\Delta_{j-1}(k)\right],\quad N+2\leq j\leq2N+1.
\end{equation}
Repeating the derivation for the upper half plane, we find 
\begin{align}
U_{+}\left[S_{N+2+l}(k)\right] & =\mathscr{R}[-\tilde{\Delta}_{l}(k)]U_{+}\left[S_{N+2}(\gamma^{-Nl}k)\right]\mathscr{R}^{-1}[-\tilde{\Delta}_{l}(k)]\nonumber \\
 & =U_{+}\left[e^{-4i\tilde{\Delta}_{l}(k)}S_{N+2}(\gamma^{-Nl}k)\right].\label{eq:UPN2l}
\end{align}
 With eqs.\ (\ref{eq:SRot1}) and (\ref{eq:SRot2}) the $(N+2)$-th
Stokes constant can be written as
\begin{equation}
S_{N+2}(k)=e^{4i\tilde{\Delta}_{N}}S_{1}(\gamma^{-N(N+1)}k),
\end{equation}
which with eq.\ (\ref{eq:UPN2l}) proves the relation eq.\ (\ref{eq:SRot3}).
Hence, through eqs.\ (\ref{eq:SRot1})-(\ref{eq:SRot3}) all the
Stokes constants are expressed in terms of the first Stokes constant
with argument rotated and certain $k$-dependent phase factors.
\section{Remnant symmetry\label{sec:Remnant-symmetry}}

In this appendix, we will show explicitly that for a small but nonzero
$k$, the first Stokes constant $S_{1}(k)$ \textit{cannot} be fixed
merely by the rotational symmetry (\ref{eq:SRot1}-\ref{eq:SRot3}),
the time-reversal (\ref{eq:betaImaginary}) and the single-valuedness
of the exact mode $\chi_{k}(z)$. Specifically, we will prove this
by constructing a nontrivial remnant transformation on the Stokes
constants that is compatible with the symmetries and the single-valuedness.

We begin by considering a $2\pi$ propagation of the WKB amplitude
vector $V_{k}(z)$ along a closed contour around the origin, $z\to e^{2\pi i}z$.
As the branch cuts have been specified in section \ref{subsec:Analytic-continuation-and},
both the exact mode $\chi_{k}(z)$ and the WKB basis vector $F_{k}(z)$
are single-valued
\begin{equation}
\chi_{k}(e^{2\pi i}z)=\chi_{k}(z)\,,\quad F_{k}(e^{2\pi i}z)=F_{k}(z)\,.
\end{equation}
Given the decomposition $\chi_{k}(z)=F^{T}_{k}(z)V_{k}(z)$, we find
that the WKB amplitude vector must also be single-valued 
\begin{equation}
V_{k}(e^{2\pi i}z)=V_{k}(z)\,.
\end{equation}
On the other hand, the WKB amplitude vector propagation along this
contour is governed by a sequence of Stokes phenomena and branch cut
factors, giving the single-valuedness constraint
\begin{equation}
U_{2N+2}\mathscr{B}_{2N}U_{2N+1}\cdots\mathscr{B}_{N+1}U_{N+2}U_{N+1}\mathscr{B}_{N}\cdots U_{2}\mathscr{B}_{1}U_{1}=\mathbb{I}_{2\times2}\,.\label{eq:singleValue}
\end{equation}
We now show that the nontrivial remnant transformation on the Stokes
constants
\begin{equation}
S_{j}(k)\to\bar{S}_{j}(k)=e^{4if(\gamma^{-N(j-1)}k)}S_{j}(k),\quad1\leq j\leq2N+2\label{eq:SjTrans}
\end{equation}
\begin{equation}
f(k)=i\sum_{p\in\text{odd}}d_{p}(k^{\frac{N+1}{N}})^{p},\quad d_{p}\in\mathbb{R}\label{eq:fk}
\end{equation}
with at least one nonzero $d_{p}$ is compatible with the rotational
symmetry, the time-reversal and the single-valuedness constraint (\ref{eq:singleValue}).

The transformations (\ref{eq:SjTrans}) on $S_{j}(k)$ are naturally
induced from the shift $S_{1}(k)\to\bar{S}_{1}(k)=e^{4if(k)}S_{1}(k)$
and the rotational symmetry representations (\ref{eq:SRot1}-\ref{eq:SRot3}).
This derivation utilizes the alternating parity property $f(\gamma^{-N}k)=-f(k)$
inherited from definition (\ref{eq:fk}). Furthermore, because the
phase factors $e^{4if(\gamma^{-N(j-1)}k)}$ are real, the transformed
Bogoliubov $\beta_{k}$ remains purely imaginary, preserving time-reversal
symmetry. Thus, the remnant transformation (\ref{eq:SjTrans}) is
manifestly compatible with these symmetries.

To verify compatibility with the single-valuedness constraint, we
apply the operator identity $\mathscr{R}[x]U_{\pm}[S]\mathscr{R}^{-1}[x]=U_{\pm}[e^{\pm4ix}S]$
to express the transformed Stokes matrices in the conjugated form
\begin{equation}
\bar{U}_{j}=\begin{cases}
\mathscr{R}\big[(-1)^{j}f(k)\big]U_{-}\big[S_{j}\big]\mathscr{R}^{-1}\big[(-1)^{j}f(k)\big]\,, & 1\leq j\leq N+1\,,\\[1ex]
\mathscr{R}\big[(-1)^{j-1}f(k)\big]U_{+}\big[S_{j}\big]\mathscr{R}^{-1}\big[(-1)^{j-1}f(k)\big]\,, & N+2\leq j\leq2N+2\,.
\end{cases}
\end{equation}
Note for $1\leq j\leq N$
\begin{align}
\bar{U}_{j+1}\mathscr{B}_{j}\bar{U}_{j}= & \mathscr{R}[(-1)^{j+1}f(k)]U_{-}[S_{j+1}]\mathscr{R}^{-1}[(-1)^{j+1}f(k)]\mathscr{B}_{j}\mathscr{R}[(-1)^{j}f(k)]\\
 & \times U_{-}[S_{j}]\mathscr{R}^{-1}[(-1)^{j}f(k)]
\end{align}
in which 
\begin{equation}
\mathscr{R}^{-1}[(-1)^{j+1}f(k)]\mathscr{B}_{j}\mathscr{R}[(-1)^{j}f(k)]=\mathscr{B}_{j}\,.
\end{equation}
where we have used $\mathscr{B}[\Delta]=-i\mathscr{R}[\Delta]B$ and
$B\mathscr{R}[x]=\mathscr{R}[-x]B$. Hence, the propagation across
the upper half plane is
\begin{equation}
\bar{U}_{N+1}\mathscr{B}_{N}\cdots\bar{U}_{2}\mathscr{B}_{1}\bar{U}_{1}=\mathscr{R}[(-1)^{N+1}f(k)]U_{N+1}\mathscr{B}_{N}\cdots U_{2}\mathscr{B}_{1}U_{1}\mathscr{R}^{-1}[-f(k)]
\end{equation}
and the propagation across the lower half plane is similarly given
by
\begin{align}
\bar{U}_{2N+2}\mathscr{B}_{2N}\bar{U}_{2N+1}\cdots\mathscr{B}_{N+1}\bar{U}_{N+2}= & \mathscr{R}[(-1)^{2N+1}f(k)]U_{2N+2}\mathscr{B}_{2N}U_{2N+1}\times\cdots\nonumber \\
 & \times\mathscr{B}_{N+1}U_{N+2}\mathscr{R}^{-1}[(-1)^{N+1}f(k)]
\end{align}
which leads to
\begin{align}
 & \bar{U}_{2N+2}\mathscr{B}_{2N}\bar{U}_{2N+1}\cdots\mathscr{B}_{N+1}\bar{U}_{N+2}\bar{U}_{N+1}\mathscr{B}_{N}\cdots\bar{U}_{2}\mathscr{B}_{1}\bar{U}_{1}\nonumber \\
= & \mathscr{R}[(-1)^{2N+1}f(k)]U_{2N+2}\mathscr{B}_{2N}U_{2N+1}\cdots\mathscr{B}_{N+1}U_{N+2}U_{N+1}\mathscr{B}_{N}\cdots U_{2}\mathscr{B}_{1}U_{1}\mathscr{R}^{-1}[-f(k)]\nonumber \\
= & \mathscr{R}[-f(k)]\mathbb{I}_{2\times2}\mathscr{R}^{-1}[-f(k)]\nonumber \\
= & \mathbb{I}_{2\times2}
\end{align}
Hence, the transformed propagator of a $2\pi$ rotation reduces to
identity, meaning that it is compatible with the single-valuedness
constraint. Consequently, the combined constraints of rotational symmetry,
time-reversal symmetry, and single-valuedness are insufficient to
determine the first Stokes constant.
\section{Non-analytic residue\label{sec:Non-analytic--factor}}

In this appendix we derive the non-analytic residue expression (\ref{eq:thetaBar}),
which reads
\begin{equation}
\bar{\theta}(k)=\Delta_{1}(k)+\Psi_{1}(k)=\frac{\Delta_{1}(k)e^{-\frac{i\pi}{n}}}{\cos\left(\frac{\pi}{n}\right)}\,.
\end{equation}

First, plugging $\psi_{1}=k^{2/n}e^{i\pi/n}$, $\Delta_{1}(k)=k^{(n+2)/n}e^{i\pi/n}d_{n}$
(cf.\ eq.\ (\ref{eq:Deltaj-1})) and the definition (\ref{eq:Psij})
for $\Psi_{1}(k)$ into $\bar{\theta}(k)$ yields
\begin{equation}
\bar{\theta}(k)=k^{\frac{n+2}{n}}e^{\frac{i\pi}{n}}d_{n}+k^{\frac{n+2}{n}}e^{i\pi\frac{n+2}{2n}}\sum^{\infty}_{m=0}\begin{pmatrix}1/2\\
m
\end{pmatrix}\frac{(-1)^{m}}{mn-\left(\frac{n}{2}+1\right)}\,.
\end{equation}
with
\begin{equation}
d_{n}=\frac{\Gamma\left(\frac{3}{2}\right)\Gamma\left(1+\frac{1}{n}\right)}{\Gamma\left(\frac{3}{2}+\frac{1}{n}\right)}\,.
\end{equation}
Let $\nu\equiv-(1/2+1/n)$ which satisfies $-1<\nu<0$ for $n\geq4$.
The series in $\bar{\theta}(k)$ can be written as
\begin{align}
\sum^{\infty}_{m=0}\begin{pmatrix}1/2\\
m
\end{pmatrix}\frac{(-1)^{m}}{mn-\left(\frac{n}{2}+1\right)} & =\frac{1}{n}\sum^{\infty}_{m=0}\begin{pmatrix}1/2\\
m
\end{pmatrix}\frac{(-1)^{m}}{m+\nu}\\
 & =\frac{1}{n}\frac{\Gamma\left(\frac{3}{2}\right)\Gamma(\nu)}{\Gamma\left(\nu+\frac{3}{2}\right)}
\end{align}
As a result, 
\begin{equation}
\bar{\theta}(k)=k^{\frac{n+2}{n}}e^{\frac{i\pi}{n}}d_{n}\left[1+e^{i\pi\left(\frac{n+2}{2n}-\frac{1}{n}\right)}\frac{\Gamma\left(-\frac{1}{2}-\frac{1}{n}\right)\Gamma\left(\frac{3}{2}+\frac{1}{n}\right)}{n\Gamma\left(1-\frac{1}{n}\right)\Gamma\left(1+\frac{1}{n}\right)}\right]\,.\label{eq:thetaBarC8}
\end{equation}
Noting $n\Gamma(1+1/n)=\Gamma(1/n)$ and $\Gamma(z)\Gamma(1-z)=\pi/\sin(\pi z)$,
we find that eq.\ (\ref{eq:thetaBarC8}) transforms to 
\begin{equation}
\bar{\theta}(k)=k^{\frac{n+2}{n}}d_{n}e^{\frac{i\pi}{n}}\left(1-i\frac{\sin(\pi/n)}{\cos(\pi/n)}\right)=\frac{e^{-\frac{i\pi}{n}}}{\cos\left(\frac{\pi}{n}\right)}\Delta_{1}(k)\,.
\end{equation}

\section{The generalized exponential integral\label{sec:The-generalized-exponential}}

In this appendix, we derive the integral formulae eqs.~(\ref{eq:geiPlus})
and (\ref{eq:geiMinus}) using the Stokes phenomena of the generalized
exponential integral
\begin{equation}
E_{p}(z)\equiv z^{p-1}\int^{\infty}_{z}dt\frac{e^{-t}}{t^{p}},\qquad z\neq0.
\end{equation}
The generalized exponential integral admits the following asymptotic
expansions~\cite{olver1994thegeneralized} 
\begin{equation}
E_{p}(z)\sim\frac{e^{-z}}{z}\sum^{\infty}_{s=0}\frac{(-1)^{s}(p)_{s}}{z^{s}},\quad\left|\arg z\right|<\frac{3\pi}{2}\,,\label{eq:Epz-ordinary-asymptotic-series}
\end{equation}
\begin{equation}
E_{p}(z)\sim\frac{e^{-z}}{z}\sum^{\infty}_{s=0}\frac{(-1)^{s}(p)_{s}}{z^{s}}+2\pi i\frac{e^{-ip\pi}}{\Gamma(p)}z^{p-1},\quad\frac{\pi}{2}<\arg z<\frac{7\pi}{2}\,,\label{eq:Epz-ordinary-asymptotic-series-2}
\end{equation}
where $(p)_{s}\equiv\Gamma(p+s)/\Gamma(p)$. Interestingly, in the
overlapping sector $\pi/2<\arg z<3\pi/2$, the same analytic function
$E_{p}(z)$ has two different asymptotic expansions (\ref{eq:Epz-ordinary-asymptotic-series})
and (\ref{eq:Epz-ordinary-asymptotic-series-2}). This difference
in the asymptotic expansions could be resolved using the Stokes phenomena,
which says that when the argument of $E_{p}(z)$ rotates counterclockwise
and crosses the negative real axis from above, the integral acquires
a Stokes term in its asymptotic expansion \cite{olver1991uniform}
\begin{equation}
2\pi i\frac{e^{-ip\pi}}{\Gamma(p)}z^{p-1}\,.
\end{equation}
After turning on this Stokes term at the Stokes-line of $e^{-z}$
at $\arg z=\pi$, this term remains exponentially suppressed compared
to the ordinary asymptotic series (\ref{eq:Epz-ordinary-asymptotic-series})
in the sector $\pi<\arg z<3\pi/2$, but becomes unsuppressed when
$z$ crosses the anti-Stokes line of $e^{-z}$ at $\arg z=3\pi/2$.
Therefore, in the overlapping sector $\pi/2<\arg z<3\pi/2$, the two
asymptotic expansions are not accurate enough to detect the exponentially
suppressed Stokes term. This means only when one invokes a more accurate
asymptotic expansion scheme, the exponentially improved asymptotic
expansion, one can obtain the Stokes term in the expansion (see e.g.$\,$\cite{olver1991uniform})
as
\begin{equation}
E_{p}(z)\dot{\sim}\frac{e^{-z}}{z}\sum^{\infty}_{s=0}\frac{(-1)^{s}(p)_{s}}{z^{s}}+\Theta_{\mathbb{R}}\left(\arg z-\pi\right)2\pi i\frac{e^{-ip\pi}}{\Gamma(p)}z^{p-1},\quad-\frac{3\pi}{2}<\arg z<\frac{7\pi}{2}\,,
\end{equation}
where $\Theta_{\mathbb{R}}(x)$ denotes a step-function that jumps
at $x=0$ with $\Theta_{\mathbb{R}}(0)=1/2$ and recall that $\dot{\sim}$
denotes exponentially improved asymptotic expansion.

With the above results, we evaluate the exponentially improved asymptotic
expansions of the integrals 
\begin{equation}
I^{\pm}_{p}\left(z\right)\equiv\int^{z}_{\infty}dw\frac{e^{\pm\frac{2i}{n/2+1}w^{n/2+1}}}{w^{p}}
\end{equation}
via changes of variables (here the lower limit $\infty$ is taken
on the positive real axis).

For $I^{+}_{p}(z)$, we introduce the variable $\bar{\zeta}\equiv-2iw^{N+1}/(N+1)$.
The upper integration limit maps to $-\zeta=-2iz^{N+1}/(N+1)=-\bar{R}e^{i\arg\zeta}$,
where $\bar{R}\equiv2R^{N+1}/(N+1)$ with $R=|z|$ is a large positive
number and $\arg\zeta\leq\pi$. Consequently, we obtain 
\begin{align}
I^{+}_{p}(z) & =\frac{1}{2i}\left(\frac{2i}{N+1}\right)^{\frac{N+p}{N+1}}(-1)^{1-\frac{N+p}{N+1}}\int^{-\zeta}_{-\infty}d\bar{\zeta}\frac{e^{-\bar{\zeta}}}{\bar{\zeta}^{\frac{N+p}{N+1}}}\nonumber \\
 & =\frac{1}{2i}\left(\frac{2i}{N+1}\right)^{\frac{N+p}{N+1}}(-1)^{1-\frac{N+p}{N+1}}\left[-\frac{E_{\frac{N+p}{N+1}}\left(-\zeta\right)}{(-\zeta)^{\frac{N+p}{N+1}-1}}\right]\nonumber \\
 & \dot{\sim}e^{\frac{2i}{N+1}z^{N+1}}\sum^{\infty}_{s=0}\frac{\left(N+1\right)^{s}}{(2i)^{s+1}}\frac{\left(\frac{N+p}{N+1}\right)_{s}}{z^{(N+1)s+N+p}}.
\end{align}

Applying the same change of variables to $I^{-}_{p}(z)$, we obtain
\begin{align}
I^{-}_{p}(z) & =\frac{1}{2i}\left(\frac{2i}{N+1}\right)^{\frac{N+p}{N+1}}\int^{\zeta}_{\infty}d\bar{\zeta}\frac{e^{-\bar{\zeta}}}{\bar{\zeta}^{\frac{N+p}{N+1}}}\nonumber \\
 & =\frac{1}{2i}\left(\frac{2i}{N+1}\right)^{\frac{N+p}{N+1}}\left[-\frac{E_{\frac{N+p}{N+1}}\left(\zeta\right)}{\zeta^{\frac{N+p}{N+1}-1}}\right].
\end{align}
The term involving $\zeta$ now acquires a Stokes contribution, as
the integration path from $\infty$ to $\zeta$ crosses the negative
real axis. Thus, the exponentially improved asymptotic expansion is
\begin{equation}
I^{-}_{p}(z)\dot{\sim}e^{-\frac{2i}{N+1}z^{N+1}}\sum^{\infty}_{s=0}\frac{\left(N+1\right)^{s}}{(2i)^{s+1}}\frac{(-1)^{s+1}\left(\frac{N+p}{N+1}\right)_{s}}{z^{(N+1)s+N+p}}-\Theta(z)\frac{\pi e^{-\frac{i\pi}{2}\frac{N+p}{N+1}}}{\Gamma\left(\frac{N+p}{N+1}\right)\left(\frac{N+1}{2}\right)^{\frac{N+p}{N+1}}}.
\end{equation}
We summarize the asymptotic behaviors of these integrals as follows:
\begin{align}
\int^{z}_{\infty}dw\frac{e^{\zeta(w)}}{w^{p}} & \dot{\sim}e^{\zeta(z)}\sum^{\infty}_{s=0}\frac{\E_{s,p}}{z^{(N+1)s+N+p}},\\
\int^{z}_{\infty}dw\frac{e^{-\zeta(w)}}{w^{p}} & \dot{\sim}e^{-\zeta(z)}\sum^{\infty}_{s=0}\frac{(-1)^{s+1}\E_{s,p}}{z^{(N+1)s+N+p}}-\Theta(z)\S_{\frac{N+p}{N+1}},
\end{align}
where the expansion coefficients and the Stokes constant are defined
by 
\begin{equation}
\E_{s,p}\equiv\frac{(N+1)^{s}}{(2i)^{s+1}}\frac{\Gamma\left(\frac{N+p}{N+1}+s\right)}{\Gamma\left(\frac{N+p}{N+1}\right)},\qquad\S_{\frac{N+p}{N+1}}\equiv\frac{\pi e^{-\frac{i\pi}{2}\frac{N+p}{N+1}}}{\Gamma\left(\frac{N+p}{N+1}\right)\left(\frac{N+1}{2}\right)^{\frac{N+p}{N+1}}}\,,\label{eq:EandS}
\end{equation}
and recall that $\zeta(z)\equiv2iz^{N+1}/(N+1).$

\begin{figure}[H]
\centering
\begin{tikzpicture}[>=Stealth, thick, line cap=round]

    \begin{scope}[local bounding box=ZPlane]
        \draw[->] (-2,0) -- (2.5,0) node[right] {Re $w$};
        \draw[->] (0,-2) -- (0,2.5) node[above] {Im $w$};
        
        \node[draw, anchor=south east] at (2.5, 2.5) {$w$};

        \coordinate (O) at (0,0);
        \coordinate (z0) at (1.8, 0);
        \path (O) -- (45:1.8) coordinate (z1);

        \draw[dashed, thin, blue!50!black] (O) -- (22.5:2.2); 

        \draw[blue!70!black, midarrow] (z0) arc[start angle=0, end angle=45, radius=1.8];
        \draw[blue!70!black] (O) -- (z1);

        \fill[blue!70!black] (z0) circle (1.5pt) node[below=2pt, black] {$z_0$};
        \fill[blue!70!black] (z1) circle (1.5pt) node[above left, black] {$z_1$};
        \node[below left] at (O) {$0$};
    \end{scope}

    \begin{scope}[xshift=10cm, yshift=3cm, local bounding box=ZetaPlane]
        \draw[->] (-2.5,0) -- (2.5,0) node[right] {Re $\bar{\zeta}$};
        \draw[->] (0,-2.5) -- (0,2.5) node[above] {Im $\bar{\zeta}$};
        
        \draw[dashed, blue!50!black] (-2.2, 0.2) -- (0, 0.2); 

        \node[draw, anchor=south east] at (2.5, 2.5) {$\bar{\zeta}$};

        \coordinate (zeta0) at (0, 1.8);
        \coordinate (zeta1) at (0, -1.8);
        \draw[blue!70!black, midarrow] (zeta0) arc[start angle=90, end angle=270, radius=1.8];

        \fill[blue!70!black] (zeta0) circle (1.5pt) node[right, black] {$\zeta_0$};
        \fill[blue!70!black] (zeta1) circle (1.5pt) node[right, black] {$\zeta_1$};
    \end{scope}

    \begin{scope}[xshift=10cm, yshift=-3.5cm, local bounding box=MinusZetaPlane]
        \draw[->] (-2.5,0) -- (2.5,0) node[right] {Re $\bar{\zeta}$};
        \draw[->] (0,-2.5) -- (0,2.5) node[above] {Im $\bar{\zeta}$};
        
        \draw[dashed, blue!50!black] (-2.2, 0.2) -- (0, 0.2);

        \node[draw, anchor=south east] at (2.5, 2.5) {$\bar{\zeta}$}; 

        \coordinate (mzeta0) at (0, -1.8); 
        \coordinate (mzeta1) at (0, 1.8); 
        \draw[blue!70!black, midarrow] (mzeta0) arc[start angle=-90, end angle=90, radius=1.8];

        \fill[blue!70!black] (mzeta0) circle (1.5pt) node[left, black] {$-\zeta_0$};
        \fill[blue!70!black] (mzeta1) circle (1.5pt) node[left, black] {$-\zeta_1$};
    \end{scope}

    
    \draw[->, shorten >=0.2cm, shorten <=0.2cm] 
        ($(ZPlane.east) + (0, 1.5)$) -- (ZetaPlane.west) 
        node[midway, above, sloped, yshift=0.1cm] {\scriptsize $\displaystyle \bar{\zeta} \equiv \frac{2i}{N+1}w^{N+1}$};

    \draw[->, shorten >=0.2cm, shorten <=0.2cm] 
        ($(ZPlane.east) + (0, -1.5)$) -- (MinusZetaPlane.west) 
        node[midway, below, sloped, yshift=-0.1cm] {\scriptsize $\displaystyle \bar{\zeta} \equiv -\frac{2i}{N+1}w^{N+1}$};

\end{tikzpicture}\caption{Integration paths in the complex $w$-plane and the mapped $\bar{\zeta}$-plane.
The path for $I^{-}_{p}$ (top right) crosses the Stokes line (dashed),
resulting in the associated Stokes phenomena, whereas the path for
$I^{+}_{p}$ (bottom right) does not cross the Stokes line.\label{fig:stokes_paths}}
\end{figure}
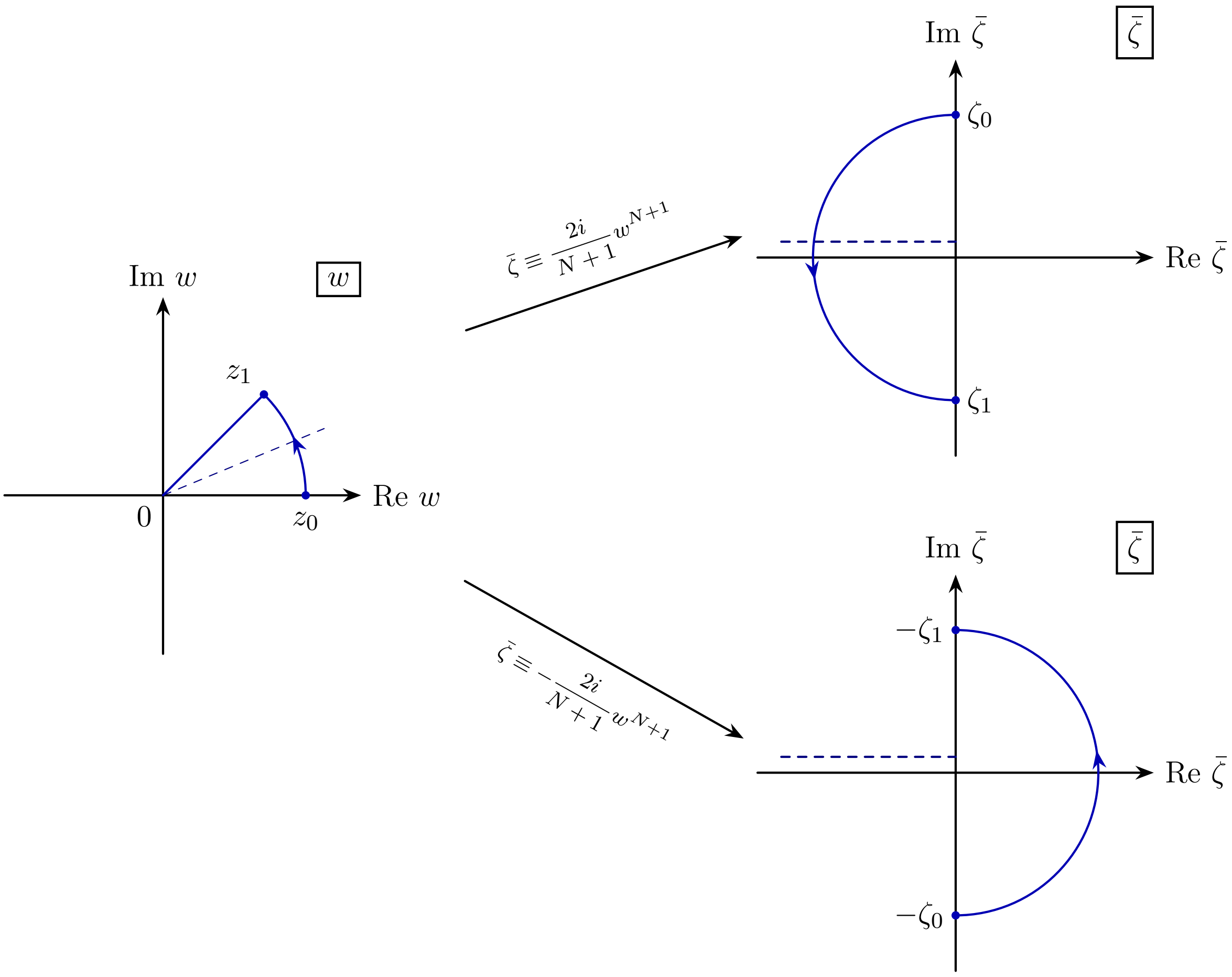

The asymmetry between the integrals $I^{+}_{p}$ and $I^{-}_{p}$
arises from the distinct topologies of their integration paths in
the complex $\zeta$-plane. As illustrated in figure \ref{fig:stokes_paths},
the mapping to the $\zeta$-variable reveals that the path for $I^{-}_{p}$
intersects the Stokes line (the negative real axis), thereby enforcing
the inclusion of the discontinuous Stokes term. In contrast, the path
for $I^{+}_{p}$ does not encounter the Stokes line, thus avoids the
Stokes term.

\section{Linear maps constructed from amplitude ODE\label{sec:Linear-maps-from}}

In this appendix, we derive explicit representations for the linear
operators defined in eqs. (\ref{eq:polyTimes})-(\ref{eq:ordinaryInt}).
While the abstract definitions suffice for formal developments of
the solvable technique, solving the recursion relation to obtain the
first Stokes constant requires the explicit action of these operators
on the space of power series.

We begin with the linear maps induced by the generalized exponential
integrals. Rewriting the constituent integrals acting on $z^{-p}$
yields 
\begin{align}
\int^{z}_{\infty}dw\frac{e^{+\zeta(w)}}{w^{p}} & \dot{\sim}e^{\zeta(z)}\sum^{\infty}_{m=0}\frac{\E_{\frac{m-p-N}{N+1},p}}{z^{m}}\delta\left(\frac{m-p-N}{N+1}\right)\,,\\
\int^{z}_{\infty}dw\frac{e^{-\zeta(w)}}{w^{p}} & \dot{\sim}e^{-\zeta(z)}\sum^{\infty}_{m=0}\frac{\E_{\frac{m-p-N}{N+1},p}}{z^{m}}(-1)^{\frac{m-p-N}{N+1}+1}\delta\left(\frac{m-p-N}{N+1}\right)-\Theta(z)\S_{\frac{N+p}{N+1}}\,,
\end{align}
where $\delta(X)$ enforces the integer constraint arising from the
dummy index shift
\begin{equation}
\delta(X)\equiv\begin{cases}
1 & X\in\mathbb{Z}_{\geq0}\\
0 & \text{otherwise.}
\end{cases}\label{eq:delta}
\end{equation}
Consequently, the linear maps $\mathcal{E}_{\pm}$ and the Stokes
functional $\mathcal{S}$ act on $f(z)=\sum^{\infty}_{m=1}f_{m}z^{-m}\in\mathcal{P}(z)$
as 
\begin{align}
\mathcal{E}_{+}\left[f(z)\right] & =\sum^{\infty}_{m=1}\frac{1}{z^{m+N}}\sum^{m-1}_{p=0}f_{p+1}\E_{\frac{m-p-1}{N+1},p+1}\delta\left(\frac{m-p-1}{N+1}\right)\,,\label{eq:repEp}\\
\mathcal{E}_{-}\left[f(z)\right] & =\sum^{\infty}_{m=1}\frac{1}{z^{m+N}}\sum^{m-1}_{p=0}f_{p+1}(-1)^{\frac{m-p-1}{N+1}+1}\E_{\frac{m-p-1}{N+1},p+1}\delta\left(\frac{m-p-1}{N+1}\right)\,,\label{eq:repEm}\\
\mathcal{S}\left[f(z)\right] & =-\sum^{\infty}_{p=1}f_{p}\S_{\frac{N+p}{N+1}}\,.\label{eq:repStokes}
\end{align}
Next, we consider multiplication by a fixed power series $B(z)=\sum^{\infty}_{m=1}B_{m}z^{-m}\in\mathcal{P}(z)$.
The associated linear map $\mathcal{B}$ acting on $f(z)$ is given
by
\begin{equation}
\mathcal{B}\left[f(z)\right]=B(z)f(z)=\sum^{\infty}_{p=2}\left(\sum_{m+l=p}\frac{B_{m}f_{l}}{z^{p}}\right)\,.\label{eq:repB}
\end{equation}
Finally, for the ordinary integral operator acting on $g(z)=\sum^{\infty}_{m=2}g_{m}z^{-m}\in\mathcal{P}_{\geq2}(z)$,
we obtain 
\begin{equation}
\mathcal{I}\left[g(z)\right]\sim\sum^{\infty}_{p=2}\frac{-g_{p}}{p-1}\frac{1}{z^{p-1}}=-\sum^{\infty}_{p=1}\frac{g_{p+1}}{p}\frac{1}{z^{p}}\in\mathcal{P}(z)\,.\label{eq:repI}
\end{equation}

\section{Justifying $\boldsymbol{\mathcal{V}_{k}}$\label{sec:Detailed-derivation-of}}

In this appendix we provide a rigorous justification for the form
(\ref{eq:Vk}) of $\mathcal{V}_{k}$, the functional space of the
exponentially improved asymptotic expansions of the WKB amplitudes.

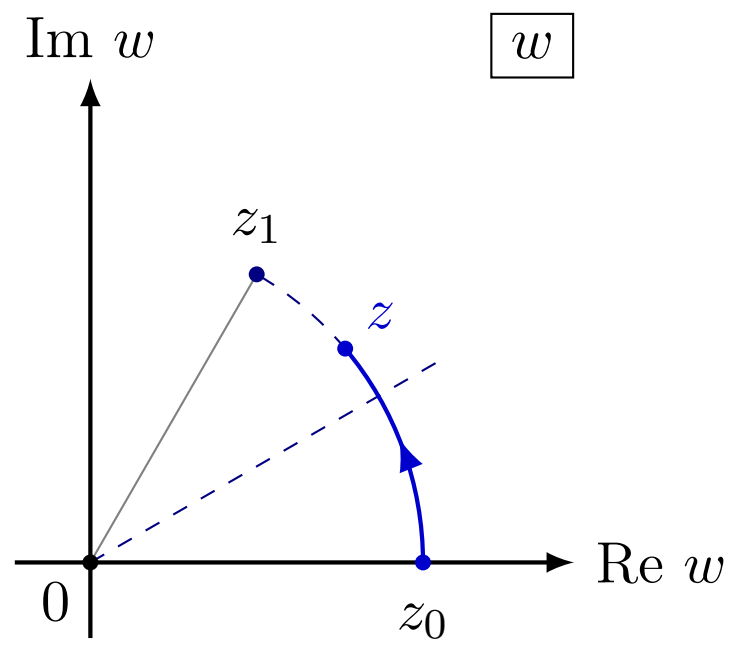
\begin{figure}[H]
\centering
\begin{tikzpicture}
\tikzset{midarrow/.style={postaction={decorate, decoration={
markings, mark=at position 0.55 with {\arrow[scale=1.2]{latex}}
}}}}
\begin{scope}[local bounding box=WPlane, >=latex]
\def\Rrad{2.2} 
\def\Zangle{40} 
\def\EndAngle{60} 
\coordinate (O) at (0,0);
\coordinate (z0) at (\Rrad, 0);
\coordinate (z_point) at (\Zangle:\Rrad);
\coordinate (z1) at (\EndAngle:\Rrad);

\draw[->, thick, black] (-0.5,0) -- (\Rrad+1,0) node[right, black] {Re $w$};
\draw[->, thick, black] (0,-0.5) -- (0,\Rrad+1) node[above, black] {Im $w$};

\node[draw, anchor=south east, fill=white] at (\Rrad+1, \Rrad+1) {$w$};

\draw[dashed, thin, blue!50!black] (O) -- (\EndAngle/2 :\Rrad+0.5);
\draw[thin, gray] (O) -- (z1);

\draw[blue!80!black, thick, midarrow] (z0) arc[start angle=0, end angle=\Zangle, radius=\Rrad];
\draw[blue!50!black, dashed] (z_point) arc[start angle=\Zangle, end angle=\EndAngle, radius=\Rrad];

\fill[black] (O) circle (1.5pt) node[below left] {$0$};
\fill[blue!80!black] (z0) circle (1.5pt) node[below=4pt, black] {$z_0$};
\fill[blue!80!black] (z_point) circle (1.5pt) node[above right] {$z$};
\fill[blue!50!black] (z1) circle (1.5pt) node[above=2pt, black] {$z_1$};
\end{scope}
\end{tikzpicture}\caption{The integration contour from $z_{0}$ to $z$ in the complex $w$-plane
used in the definition of the path-ordered exponential.\label{fig:The-integration-contour}}
\end{figure}

In the following, we let $z_{0}=R$ and $z_{1}=R\exp\left(\pi i/(N+1)\right)$
with real $R\to\infty$ be boundary points of the first Stokes sector.
We start with the propagator version of the amplitude ODE, which is
given by eqs.\ (\ref{eq:propagation-complex})-(\ref{eq:M0Complex})
as
\begin{equation}
V_{k}(z)=U_{0}(k,z,z_{0})V_{k}(z_{0}),\quad V_{k}(z_{0})=\begin{pmatrix}1\\
0
\end{pmatrix}
\end{equation}
\begin{equation}
U_{0}(k,z,z_{0})=\mathbb{P}\left[\mathbb{I}_{2\times2}\exp\left(\int^{z}_{\infty}dw\,M_{0}(k,w)\right)\right]\,.
\end{equation}
where $z$ denotes a point on the $w$-integration arc from $z_{0}$
to $z_{1}$, as shown in figure \ref{fig:The-integration-contour}.

The propagator has a formal expansion
\begin{align}
U_{0}(k,z,z_{0})= & \mathbb{P}\left[\mathbb{I}_{2\times2}\exp\left(\int^{z}_{\infty}dwM_{0}(k,w)\right)\right]\nonumber \\
= & \mathbb{I}_{2\times2}+\int^{z}_{\infty}dw_{1}M_{0}(w_{1})\nonumber \\
 & +\int^{z}_{\infty}dw_{1}M_{0}(w_{1})\int^{w_{1}}_{\infty}dw_{2}M_{0}(w_{2})\nonumber \\
 & +\int^{z}_{\infty}dw_{1}M_{0}(w_{1})\int^{w_{1}}_{\infty}dw_{2}M_{0}(w_{2})\int^{w_{2}}_{\infty}dw_{3}M_{0}(w_{3})\nonumber \\
 & +\cdots\label{eq:propagatorExpansion}
\end{align}
where due to the path-ordering in the integrations we require $\arg z_{0}\leq\cdots\leq\arg w_{3}\leq\arg w_{2}\leq\arg w_{1}\leq\arg z$
and we have suppressed the $k$ dependence to avoid notational clutter.
Using the definition (\ref{eq:Bpm}) for $B_{\pm}(z)$ we find
\begin{equation}
M_{0}(w)=\begin{pmatrix}0 & e^{\zeta(w)}e^{2i\bar{\theta}(k)}B_{-}(w)\\
e^{-\zeta(w)}e^{-2i\bar{\theta}(k)}B_{+}(w) & 0
\end{pmatrix}\,.
\end{equation}
Since $B_{\pm}(z)\in\mathcal{P}(z)$, the functional space of the
exponentially improved asymptotic expansions of the first few terms
in the formal expansion eq.\ (\ref{eq:propagatorExpansion}) is determined
by eqs.\ (\ref{eq:automorPlus}) and (\ref{eq:automorMinus}) 
\begin{equation}
\left[\int^{z}_{\infty}dw_{1}M_{0}(w_{1})\right]_{{\rm e.a.e.}}\in\begin{pmatrix}0 & e^{\zeta(z)}\mathcal{P}(z)\\
e^{-\zeta(z)}\mathcal{P}(z)+\Theta(z)\left[\mathcal{P}(z)+\mathbb{C}\right] & 0
\end{pmatrix}\label{eq:propagatorAE1st}
\end{equation}
\begin{equation}
\left[\int^{z}_{\infty}dw_{1}M_{0}(w_{1})\int^{w_{1}}_{\infty}dw_{2}M_{0}(w_{2})\right]_{{\rm e.a.e.}}\in\begin{pmatrix}\mathcal{P}(z)+e^{\zeta(z)}\Theta(z)\mathcal{P}(z) & 0\\
0 & \mathcal{P}(z)
\end{pmatrix}\label{eq:propagatorAE2nd}
\end{equation}
\begin{equation}
\left[\int^{z}_{\infty}dw_{1}M_{0}(w_{1})\int^{w_{1}}_{\infty}dw_{2}M_{0}(w_{2})\int^{w_{2}}_{\infty}dw_{3}M_{0}(w_{3})\right]_{{\rm e.a.e.}}\in\begin{pmatrix}0 & e^{\zeta(z)}\mathcal{P}(z)\\
e^{-\zeta(z)}\mathcal{P}(z)+\Theta(z)\left[\mathcal{P}(z)+\mathbb{C}\right] & 0
\end{pmatrix}
\end{equation}
Therefore, we conclude that the exponentially improved asymptotic
expansions of all higher order terms are contained in the union of
the functional spaces given by $\mathbb{I}_{2\times2}$, that of the
first order term (\ref{eq:propagatorAE1st}), and that of the second
order term (\ref{eq:propagatorAE2nd}). Hence, collecting the functional
spaces of all order terms in the formal expansion (\ref{eq:propagatorExpansion}),
we obtain the functional space of the exponentially improved asymptotic
expansion of the propagator:
\begin{equation}
\left[U_{0}(z,z_{0})\right]_{{\rm e.a.e.}}\in\begin{pmatrix}1+\mathcal{P}(z)+e^{\zeta(z)}\Theta(z)\mathcal{P}(z) & e^{\zeta(z)}\mathcal{P}(z)\\
e^{-\zeta(z)}\mathcal{P}(z)+\Theta(z)\left[\mathcal{P}(z)+\mathbb{C}\right] & 1+\mathcal{P}(z)
\end{pmatrix}\,.
\end{equation}

The exponentially improved asymptotic expansion of the WKB amplitude
propagated from $V(z_{0})=\begin{pmatrix}1 & 0\end{pmatrix}^{T}$
is therefore
\begin{equation}
[V_{k}(z)]_{{\rm e.a.e.}}\in\begin{pmatrix}1+\mathcal{P}(z)+e^{\zeta(z)}\Theta(z)\mathcal{P}(z)\\
e^{-\zeta(z)}\mathcal{P}(z)+\Theta(z)\left[\mathcal{P}(z)+\mathbb{C}\right]
\end{pmatrix}=\mathcal{V}_{k}
\end{equation}
providing a rigorous justification for eq.\ (\ref{eq:Vk}).

\section{Non-analyticity cancellation details\label{sec:Non-analyticity-cancellation}}

In this appendix, we provide a detailed derivation from eq.\ (\ref{eq:betaProduct1})
to eq.\ (\ref{eq:betaT1}), showing that for $n\geq4$ the non-analytic
$k$ factors in $\beta_{k}$ all cancel out explicitly.

We start with eq.\ (\ref{eq:betaProduct1})
\begin{equation}
\beta_{k}=(-1)^{N}i^{N}e^{2i\tilde{\Delta}_{N}}\left[\prod^{N}_{j=1}\left(U_{+}[e^{4i(-1)^{j}\tilde{\Delta}_{N}}\tilde{S}^{(N+2+j)}_{1}]B\right)\right]_{21}\,.
\end{equation}
Using the first Stokes constant decomposition (\ref{eq:betaSTrans})
we find
\begin{equation}
\tilde{S}^{(N+2+j)}_{1}(k)=S_{1}(\gamma^{-N(N+j+1)}k)=e^{-2i(-1)^{N+j+1}\bar{\theta}(k)}\tilde{T}^{(N+2+j)}_{1}(k^{2})\,,
\end{equation}
where we have used $\bar{\theta}(\gamma^{-N(N+j+1)}k)=(-1)^{N+j+1}\bar{\theta}(k)$.
This leads to
\begin{equation}
\beta_{k}=(-1)^{N}i^{N}e^{2i\tilde{\Delta}_{N}}\left[\prod^{N}_{j=1}\left(U_{+}[e^{4i(-1)^{j}x}\tilde{T}^{(N+2+j)}_{1}]B\right)\right]_{21}\label{eq:betaH3}
\end{equation}
\begin{equation}
x\equiv\tilde{\Delta}_{N}(k)+\frac{(-1)^{N}}{2}\bar{\theta}(k)\,.
\end{equation}
We use matrix property eq.\ (\ref{eq:RUProperty}) and $B\mathscr{R}[x]B=\mathscr{R}[-x]$
to get 
\begin{equation}
U_{+}[e^{4i(-1)^{j}x}\tilde{T}^{(N+2+j)}_{1}]B=\mathscr{R}[(-1)^{j}x]U_{+}[\tilde{T}^{(N+2+j)}_{1}]B\mathscr{R}^{-1}[(-1)^{j-1}x]\,,
\end{equation}
which leads to a cancellation in adjacent rotation matrices
\begin{align*}
\prod^{N}_{j=1}\left(U_{+}[e^{4i(-1)^{j}x}\tilde{T}^{(N+2+j)}_{1}]B\right)= & \mathscr{R}[(-1)^{N}x]\prod^{N}_{j=1}\left(U_{+}[\tilde{T}^{(N+2+j)}_{1}]B\right)\mathscr{R}^{-1}[x]\,.
\end{align*}
Therefore, eq.\ (\ref{eq:betaH3}) becomes
\begin{align}
\beta_{k} & =(-1)^{N}i^{N}e^{2i\tilde{\Delta}_{N}}\left[\mathscr{R}[(-1)^{N}x]\prod^{N}_{j=1}\left(U_{+}[\tilde{T}^{(N+2+j)}_{1}]B\right)\mathscr{R}^{-1}[x]\right]_{21}\nonumber \\
 & =(-1)^{N}i^{N}e^{2i\left[\tilde{\Delta}_{N}-(1+(-1)^{N})x\right]}\left[\prod^{N}_{j=1}\left(U_{+}[\tilde{T}^{(N+2+j)}_{1}]B\right)\right]_{21},\label{eq:betaH7}
\end{align}
which contains a non-analytic factor $e^{2i\left[\tilde{\Delta}_{N}-(1+(-1)^{N})x\right]}$
and a finite product of power series analytic in $k$ to leading order.
Next, we show that the non-analytic phase factor is trivial. With
the definitions (\ref{eq:Deltaj-1}) and (\ref{eq:alternatingPhase})
we obtain
\begin{equation}
\tilde{\Delta}_{N}=k^{\frac{n+2}{n}}d_{n}\sum^{N}_{r=1}(-1)^{N-r}e^{\frac{2r-1}{n}i\pi}=-\frac{k^{\frac{n+2}{n}}d_{n}}{2}\frac{1+(-1)^{N}}{\cos\left(\frac{\pi}{n}\right)}\,.\label{eq:nonanalytic-factor}
\end{equation}
Also, using eq.\ (\ref{eq:thetaBar}) for the non-analytic residue
we find
\begin{align}
\tilde{\Delta}_{N}-(1+(-1)^{N})x & =-(-1)^{N}\tilde{\Delta}_{N}-\frac{1+(-1)^{N}}{2}\bar{\theta}(k)\nonumber \\
 & =\frac{k^{\frac{n+2}{n}}d_{n}}{2}\frac{1+(-1)^{N}}{\cos\left(\frac{\pi}{n}\right)}-\frac{1+(-1)^{N}}{2}\frac{k^{\frac{n+2}{n}}d_{n}}{\cos\left(\frac{\pi}{n}\right)}\nonumber \\
 & =0\,,
\end{align}
which shows the non-analyticity cancellation explicitly. Hence, eq.\ (\ref{eq:betaH7})
becomes
\begin{equation}
\beta_{k}=(-1)^{N}i^{N}\left[\prod^{N}_{j=1}\left(U_{+}[\tilde{T}^{(N+2+j)}_{1}]B\right)\right]_{21}\,,
\end{equation}
which is eq.\ (\ref{eq:betaT1}) we wanted to derive.

\newpage{}

\section{Closed form expression for $\boldsymbol{c_{1}}$\label{sec:Closed-form-expression}}

The goal of this appendix is to derive the following closed form expression
for the $c_{1}$ coefficient defined in the Bogoliubov $\beta_{k}$
expansion (\ref{eq:betaQuad})\footnote{The closed form expression for $c_{1}(n)$ was initially speculated
by OpenAI's ChatGPT 5.6 Sol Ultra, then independently derived by human
authors.}
\begin{equation}
c_{1}(n)=i\frac{\cos\left[\frac{\pi}{2(N+1)}\right]}{\sqrt{\pi}(N+1)^{2(1-\frac{1}{N+1})}}\frac{\g\left[\frac{1}{2(N+1)}\right]\g\left(\frac{1}{N+1}\right)\g\left[\frac{3}{2(N+1)}\right]}{\Gamma\left(\frac{1}{N+1}+\frac{1}{2}\right)}\,,\label{eq:c1-closed-form}
\end{equation}
where $N=n/2$. The sketch of this derivation is as follows.
\begin{itemize}
\item As a first step, in appendix \ref{subsec:Split-the-explicit}, we
will split the explicit recursion relation (\ref{eq:G-recursion})
for $G(z)$ into two sets of recursion relations, one for the topological
limit terms $G^{(0)}_{p}$ and the other for the soft correction terms
$G^{(2)}_{p}$, where we recall that $G^{(0)}_{p}$ and $G^{(2)}_{p}$
are defined through the expansion $G_{p}=G^{(0)}_{p}+G^{(2)}_{p}k^{2}+O(k^{4})$.
\item We then solve the explicit recursion relation for $G^{(0)}_{p}$ in
appendix \ref{subsec:Reproducing-the-Stokes}, which will yield the
closed form expression (\ref{eq:T0-closed-form}) for $T^{(0)}_{1}$
defined in eq.\ (\ref{eq:T-split}). This will be used to reproduce
the Stokes topological limit result $\beta_{k=0}=-i\cot\left[\pi/(2(N+1))\right]$
via the propagator product expression (\ref{eq:betaT1}).
\item In appendix \ref{subsec:Quadratic-coefficient-of}, we will solve
the soft correction term $G^{(2)}_{p}$ based on its recursion relation
(\ref{eq:G2-recursion}) and the solution (\ref{eq:G0-sol}) of $G^{(0)}_{p}$.
Summing over $G^{(2)}_{p}$ then yields the quadratic coefficient
(coefficient of $k^{2}$) $T^{(2)}_{1}$ in $T_{1}(k^{2})$ as $T^{(2)}_{1}=-\sum^{\infty}_{p=1}G^{(2)}_{p}$.
\item Finally, in appendix \ref{subsec:Quadratic-coefficient-of-1}, we
isolate the quadratic coefficient in the propagator product expression
for $\beta_{k}$. This will express the quadratic coefficient of $\beta_{k}$,
$c_{1}$, in terms of $T^{(0)}_{1}$, $T^{(2)}_{1}$ and trigonometric
functions of $N$. Simplifying it leads to the closed form expression
(\ref{eq:c1-closed-form}) for $c_{1}$.
\end{itemize}
We repeat the explicit recursion relation (\ref{eq:G-recursion})
for $G(z)$ and relevant definitions (\ref{eq:K-defn}) and (\ref{eq:EandS})
here for reader's convenience:
\begin{equation}
G_{p+1}=\begin{cases}
B^{(+)}_{p+1}\S_{\frac{N+p+1}{N+1}} & 0\leq p\leq N\\
B^{(+)}_{p+1}\S_{\frac{N+p+1}{N+1}}-\sum^{p}_{m=N+1}\sum^{m}_{l=N+1}\sum^{l-N-1}_{s=0}K_{p;m,l,s}G_{s+1} & p\geq N+1
\end{cases}\label{eq:G-recursion-appendix}
\end{equation}
\begin{equation}
K_{p;m,l,s}=\frac{B^{(+)}_{p-m+1}B^{(-)}_{m-l+1}}{ml}\frac{\S_{\frac{N+p+1}{N+1}}}{\S_{\frac{N+l+1}{N+1}}}\delta\left(\frac{l-s}{N+1}-1\right)
\end{equation}
\begin{equation}
\S_{y}=\frac{\pi e^{-\frac{i\pi}{2}y}}{\Gamma(y)\left(\frac{N+1}{2}\right)^{y}},\quad\delta(X)=\begin{cases}
1 & X\in\mathbb{Z}_{\geq0}\\
0 & \text{otherwise}
\end{cases}\label{eq:S-delta-appendix}
\end{equation}
where the coefficients $B^{(\pm)}_{m}$ are defined by eq.\ (\ref{eq:Bpm}).
Throughout this appendix, we will be using the following index notations:
\begin{equation}
\l\equiv N+1,\quad a\equiv\frac{N}{2(N+1)},\quad d\equiv\frac{N-1}{N+1},\quad\varphi\equiv\frac{\pi}{2(N+1)}\,.\label{eq:index-notation}
\end{equation}

\subsection{Splitting the explicit recursion relation\label{subsec:Split-the-explicit}}

To split the explicit recursion relation (\ref{eq:G-recursion-appendix})
in the soft momentum limit, we first expand the coefficients $B^{(\pm)}_{m}$
to leading order of $k^{2}$, which yields
\begin{equation}
B^{(\pm)}_{m}=b^{(0)}_{m}+b^{(2\pm)}_{m}k^{2}+O(k^{4})\label{eq:Bpm-expansion}
\end{equation}
\begin{equation}
b^{(0)}_{m}=\frac{N}{2}\delta_{m,1},\quad b^{(2\pm)}_{m}=\pm\frac{iN}{2(N-1)}\delta_{m,N}-\frac{N}{2}\delta_{m,2N+1}\label{eq:b02}
\end{equation}
where we have used eq.\ (\ref{eq:Bpm}) and where $\delta_{m,p}$
denotes the Kronecker delta. These expressions imply that the supports
of $b^{(0)}_{m}$ and $b^{(2\pm)}_{m}$ are
\begin{equation}
\text{supp}[b^{(0)}]=\{1\},\quad\text{supp}[b^{(2\pm)}]=\{N,2N+1\}\,,\label{eq:supp-b}
\end{equation}
respectively.

Plugging the soft momentum expansions (\ref{eq:Bpm-expansion}) for
$B^{(\pm)}_{m}$ and $G_{p}=G^{(0)}_{p}+G^{(2)}_{p}k^{2}+O(k^{4})$
into the explicit recursion relation (\ref{eq:G-recursion-appendix})
and collecting the constant terms gives the explicit recursion relation
for the topological limit term $G^{(0)}_{p}$:
\begin{equation}
G^{(0)}_{p+1}=\begin{cases}
b^{(0)}_{p+1}\S_{\frac{N+p+1}{N+1}} & 0\leq p\leq N\\
-\sum^{p}_{m=N+1}\sum^{m}_{l=N+1}\sum^{l-N-1}_{s=0}K^{(0)}_{p;m,l,s}G^{(0)}_{s+1} & p\geq N+1
\end{cases}\label{eq:G0-recursion}
\end{equation}
\begin{equation}
K^{(0)}_{p;m,l,s}=\frac{b^{(0)}_{p-m+1}b^{(0)}_{m-l+1}}{ml}\frac{\S_{\frac{N+p+1}{N+1}}}{\S_{\frac{N+l+1}{N+1}}}\delta\left(\frac{l-s}{N+1}-1\right)\,,\label{eq:K0}
\end{equation}
where in the second line of eq.\ (\ref{eq:G0-recursion}) we have
used the fact that $b^{(0)}_{p+1}=0$ for $p\geq N+1$. Collecting
the $k^{2}$ coefficients in eq.\ (\ref{eq:G-recursion-appendix})
instead yields the recursion relation for the soft correction term
$G^{(2)}_{p}$:

\begin{equation}
G^{(2)}_{p+1}=\begin{cases}
b^{(2+)}_{p+1}\S_{\frac{N+p+1}{N+1}} & 0\leq p\leq N\\
b^{(2+)}_{p+1}\S_{\frac{N+p+1}{N+1}}-\sum^{p}_{m=N+1}\sum^{m}_{l=N+1}\sum^{l-N-1}_{s=0}\left(K^{(2)}_{p;m,l,s}G^{(0)}_{s+1}+K^{(0)}_{p;m,l,s}G^{(2)}_{s+1}\right) & p\geq N+1
\end{cases}\label{eq:G2-recursion}
\end{equation}
\begin{equation}
K^{(2)}_{p;m,l,s}=\frac{\left(b^{(0)}_{p-m+1}b^{(2-)}_{m-l+1}+b^{(2+)}_{p-m+1}b^{(0)}_{m-l+1}\right)}{ml}\frac{\S_{\frac{N+p+1}{N+1}}}{\S_{\frac{N+l+1}{N+1}}}\delta\left(\frac{l-s}{N+1}-1\right)\,.\label{eq:K2}
\end{equation}
As a result, eqs.\ (\ref{eq:G0-recursion}) and (\ref{eq:G2-recursion})
constitute the split explicit recursion relations. We will solve them
in the next two appendices.

\subsection{Reproducing the Stokes topological limit\label{subsec:Reproducing-the-Stokes}}

In this appendix, we will reproduce the Stokes topological limit result
\begin{equation}
\beta_{k=0}=-i\cot\left[\frac{\pi}{2(N+1)}\right]\label{eq:beta-k=00003D0-appendix}
\end{equation}
 by solving the explicit recursion relation (\ref{eq:G0-recursion})
for $G^{(0)}_{p}$ and using the propagator product expression (\ref{eq:betaT1}).
In particular, we will first solve eq.\ (\ref{eq:G0-recursion})
to obtain the finite sum $\sum^{r}_{t=1}G^{(0)}_{t}$, then take the
$r\to\infty$ limit to get the closed form expression for $T^{(0)}_{1}$
defined in eq.\ (\ref{eq:T-split}):
\begin{equation}
T^{(0)}_{1}=-\lim_{r\to\infty}\sum^{r}_{m=1}G^{(0)}_{m}\,.\label{eq:T10-Gm0}
\end{equation}
Plugging this into eq.\ (\ref{eq:betaT1}) then yields eq.\ (\ref{eq:beta-k=00003D0-appendix}).
A key step in this derivation is to show that the sequence $G^{(0)}_{p}$
has a sparse support. This sparsity follows from the selection rules
imposed by $K^{(0)}_{p;m,l,s}$ in the recursion relation (\ref{eq:G0-recursion}).
Once this support has been determined explicitly, as in eq. (\ref{eq:supp-G0}),
the original recursion reduces to a recursion for a nonzero subsequence
and can be solved in closed form.

According to the first line of eq.\ (\ref{eq:G0-recursion}), the
only nonzero $G^{(0)}_{p}$ element for $1\leq p\leq N+1$ is
\begin{equation}
G^{(0)}_{1}=b^{(0)}_{1}\S_{1}=-\frac{i\pi N}{N+1}=-2i\pi a\,,
\end{equation}
where we have used eqs.\ (\ref{eq:S-delta-appendix}),\ (\ref{eq:index-notation})
and (\ref{eq:b02}). For $p\geq N+1$, the object $K^{(0)}_{p;m,l,s}$
given by eq.\ (\ref{eq:K0}) is nonzero if and only if $b^{(0)}_{p-m+1}$,
$b^{(0)}_{m-l+1}$ and $\delta\left((l-s)/(N+1)-1\right)$ are all
nonzero. Thus,
\begin{equation}
p-m+1\in\text{supp}[b^{(0)}],\quad m-l+1\in\text{supp}[b^{(0)}],\quad\frac{l-s}{N+1}-1\in\mathbb{Z}_{\geq0}\,,\label{eq:K0-nonzero}
\end{equation}
where the last condition follows from the definition of $\delta(X)$
in eq.\ (\ref{eq:S-delta-appendix}). Since $\text{supp}[b^{(0)}]=\{1\}$,
for a given $p\geq N+1$, the first two conditions require $m=l=p$,
and the third condition implies
\begin{equation}
p=s+(j+1)\l\,,\quad j\in\mathbb{Z}_{\geq0}\label{eq:ps-Np}
\end{equation}
with $\l$ defined in eq.\ (\ref{eq:index-notation}). The selection
rule imposed by $K^{(0)}_{p;m,l,s}$ can be summarized as
\begin{equation}
K^{(0)}_{p;m,l,s}\neq0\quad\text{iff}\quad m=l=p\text{\ and\ }p=s+(j+1)\l\,,\,j\in\mathbb{Z}_{\geq0}\,.\label{eq:K0-selection}
\end{equation}
This selection rule implies that $G^{(0)}_{p+1}$ can receive contributions
only from $G^{(0)}_{s+1}$ whose indices differ from $p+1$ by a positive
integer multiple of $\l$. Furthermore, since $G^{(0)}_{p}=(-2\pi ia)\delta_{p,1}$
for $1\leq p\leq\l$, iterating the recursion relation (\ref{eq:G0-recursion})
gives
\begin{equation}
\text{supp}[G^{(0)}]\subseteq\{1+r\l\mid r\in\mathbb{Z}_{\geq0}\}\,.\label{eq:supp-G0}
\end{equation}
Hence, the support of $G^{(0)}_{p}$ is sparse, with successive nonzero
coefficients separated by $\l$.

Given this sparse support, we may simplify the recursion relation
(\ref{eq:G0-recursion}) by considering the nonzero subsequence of
$G^{(0)}_{p}$:
\begin{equation}
g^{(0)}_{r}\equiv G^{(0)}_{1+r\l},\quad r\geq0\,.\label{eq:gr0}
\end{equation}
We note
\begin{equation}
T^{(0)}_{1}=-\sum^{\infty}_{m=1}G^{(0)}_{m}=-\sum^{\infty}_{m\in\text{supp}[G^{(0)}]}G^{(0)}_{m}=-\sum^{\infty}_{r=0}g^{(0)}_{r}\,.
\end{equation}
This motivates us to introduce the partial sum series
\begin{equation}
Y^{(0)}_{r}\equiv\sum^{r}_{t=0}g^{(0)}_{t},\quad r\geq0\label{eq:Yq0-defn}
\end{equation}
such that
\begin{equation}
T^{(0)}_{1}=-\lim_{r\to\infty}Y^{(0)}_{r}\,.
\end{equation}
Setting $p=r\l$ and $s=t\l$ in eq.\ (\ref{eq:G0-recursion}) then
leads to
\begin{equation}
g^{(0)}_{r}=\begin{cases}
-2\pi ia & r=0\\
-\frac{a^{2}}{r^{2}}\sum^{r-1}_{t=0}g^{(0)}_{t} & r\geq1
\end{cases}\,.\label{eq:g0-recursion}
\end{equation}
where we have used the conditions (\ref{eq:K0-nonzero}) to reduce
$K^{(0)}_{p;m,l,s}$. Since for $r\geq1$, $g^{(0)}_{r}=Y^{(0)}_{r}-Y^{(0)}_{r-1}$,
the recursion relation (\ref{eq:g0-recursion}) for $g^{(0)}_{r}$
translates to a recursion relation for $Y^{(0)}_{r}$:
\begin{equation}
Y^{(0)}_{r}=\begin{cases}
-2\pi ia & r=0\\
\left(1-\frac{a^{2}}{r^{2}}\right)Y^{(0)}_{r-1} & r\geq1
\end{cases}\,.
\end{equation}
Iterating this recursion relation yields
\begin{equation}
Y^{(0)}_{r}=\left[\prod^{r}_{j=1}\left(1-\frac{a^{2}}{j^{2}}\right)\right]Y^{(0)}_{0}=\frac{\Gamma(r+1+a)\Gamma(r+1-a)}{\Gamma(1+a)\Gamma(1-a)(r!)^{2}}(-2\pi ia)\,,\quad r\geq1\,,\label{eq:Yq0}
\end{equation}
where we have used the gamma function property $\prod^{r}_{j=1}(j+x)=\Gamma(r+x+1)/\Gamma(x+1)$
to derive
\begin{equation}
\prod^{r}_{j=1}\left(1-\frac{a^{2}}{j^{2}}\right)=\left(\prod^{r}_{j=1}\frac{j-a}{j}\right)\left(\prod^{r}_{j=1}\frac{j+a}{j}\right)=\frac{\Gamma(r+1+a)\Gamma(r+1-a)}{\Gamma(1+a)\Gamma(1-a)(r!)^{2}}\,.
\end{equation}
Given the second line of eq.\ (\ref{eq:g0-recursion}), the definition
(\ref{eq:Yq0-defn}) of $Y^{(0)}_{r}$, and the expression (\ref{eq:Yq0}),
we find
\begin{equation}
g^{(0)}_{r}=-\frac{a^{2}}{r^{2}}Y^{(0)}_{r-1}=(2\pi ia)\frac{a^{2}}{r^{2}}\frac{(1+a)_{r-1}(1-a)_{r-1}}{[(r-1)!]^{2}}\,,\quad r\geq1\,,
\end{equation}
where we have used the Pochhammer symbol $(x)_{s}=\Gamma(x+s)/\Gamma(x)$.
Since $g^{(0)}_{r}$ is the nonzero subsequence of $G^{(0)}_{p}$,
the solution to the explicit recursion relation (\ref{eq:G0-recursion})
is
\begin{equation}
G^{(0)}_{p}=\begin{cases}
-2\pi ia & p=1\\
2\pi ia\frac{a^{2}}{r^{2}}\frac{(1+a)_{r-1}(1-a)_{r-1}}{[(r-1)!]^{2}} & p=1+r\l,\ r\geq1\\
0 & \text{otherwise}
\end{cases}\,.\label{eq:G0-sol}
\end{equation}

The topological limit term in $T_{1}(k^{2})$ is then given by eq.\ (\ref{eq:T10-Gm0})
as
\begin{equation}
T^{(0)}_{1}=-\lim_{r\to\infty}Y^{(0)}_{r}=(2\pi ia)\lim_{r\to\infty}\frac{\Gamma(r+1+a)\Gamma(r+1-a)}{\Gamma(1+a)\Gamma(1-a)(r!)^{2}}=\frac{2\pi ia}{\Gamma(1+a)\Gamma(1-a)}\,.
\end{equation}
This can be further reduced using the gamma function identity $\Gamma(z)\Gamma(1-z)=\pi/\sin(\pi z)$:
\begin{equation}
T^{(0)}_{1}=2i\sin(\pi a)=2i\cos\varphi\,,\label{eq:T10-cosphi}
\end{equation}
where we have used eq.\ (\ref{eq:index-notation}).

We now compute the Bogoliubov $\beta_{k}$ using eq.\ (\ref{eq:betaT1}),
which at the topological limit $k=0$ reads
\begin{equation}
\beta_{k=0}=(-1)^{N}i^{N}\left[X^{N}_{0}\right]_{21}\,,\label{eq:beta-k=00003D0-21}
\end{equation}
where
\begin{equation}
X_{0}\equiv\begin{pmatrix}T^{(0)}_{1} & 1\\
1 & 0
\end{pmatrix}\,.\label{eq:X0-1}
\end{equation}
We compute $X^{N}_{0}$ by solving a recursion relation for $X^{r\geq0}_{0}$.
Since $X_{0}$ is symmetric, so is $X^{r}_{0}$. We therefore parameterize
$X^{r}_{0}$ as
\begin{equation}
X^{r}_{0}=\begin{pmatrix}\mathfrak{a}_{r} & \mathfrak{b}_{r}\\
\mathfrak{b}_{r} & \mathfrak{c}_{r}
\end{pmatrix}\,,\label{eq:X0r}
\end{equation}
then
\begin{equation}
X^{r+1}_{0}=\begin{pmatrix}\mathfrak{a}_{r} & \mathfrak{b}_{r}\\
\mathfrak{b}_{r} & \mathfrak{c}_{r}
\end{pmatrix}\begin{pmatrix}T^{(0)}_{1} & 1\\
1 & 0
\end{pmatrix}=\begin{pmatrix}\mathfrak{b}_{r}+T^{(0)}_{1}\mathfrak{a}_{r} & \mathfrak{a}_{r}\\
\mathfrak{c}_{r}+T^{(0)}_{1}\mathfrak{b}_{r} & \mathfrak{b}_{r}
\end{pmatrix}=\begin{pmatrix}\mathfrak{a}_{r+1} & \mathfrak{b}_{r+1}\\
\mathfrak{b}_{r+1} & \mathfrak{c}_{r+1}
\end{pmatrix}\,.
\end{equation}
This gives a set of recursion relations for the entries of $X^{r}_{0}$:
\begin{equation}
\begin{cases}
\mathfrak{a}_{r+1}=\mathfrak{b}_{r}+T^{(0)}_{1}\mathfrak{a}_{r}\\
\mathfrak{b}_{r+1}=\mathfrak{a}_{r}\\
\mathfrak{c}_{r+1}=\mathfrak{b}_{r}
\end{cases},\quad r\geq0\label{eq:abc-recursion}
\end{equation}
with initial conditions given by $(X_{0})^{0}=\mathbb{I}_{2\times2}$
and eq.\ (\ref{eq:X0-1}) as
\begin{equation}
\begin{cases}
\mathfrak{a}_{0}=1\\
\mathfrak{b}_{0}=0\\
\mathfrak{c}_{0}=1
\end{cases},\quad\begin{cases}
\mathfrak{a}_{1}=T^{(0)}_{1}\\
\mathfrak{b}_{1}=1\\
\mathfrak{c}_{1}=0
\end{cases}\,.\label{eq:abc-initial-condition}
\end{equation}
From eq.\ (\ref{eq:abc-recursion}) we obtain the recursion relation
for $\mathfrak{b}_{r}$ using eq.\ (\ref{eq:T10-cosphi}):
\begin{equation}
\mathfrak{b}_{r+1}=\mathfrak{b}_{r-1}+(2i\cos\varphi)\mathfrak{b}_{r},\quad r\geq1\,.\label{eq:mathfrak-c-recursion}
\end{equation}
This can be solved using the characteristic equation method. With
the initial condition (\ref{eq:abc-initial-condition}) we get
\begin{equation}
\mathfrak{b}_{r}=\frac{\lambda^{r}_{+}-\lambda^{r}_{-}}{\lambda_{+}-\lambda_{-}},\quad\lambda_{\pm}=ie^{\mp i\varphi},\quad r\geq0\,,
\end{equation}
where $\lambda_{\pm}$ denote the characteristic roots of the recursion
relation (\ref{eq:mathfrak-c-recursion}). This can be further simplified:
\begin{equation}
\mathfrak{b}_{r}=i^{r-1}\frac{e^{-ir\varphi}-e^{ir\varphi}}{e^{-i\varphi}-e^{i\varphi}}=i^{r-1}\frac{\sin(r\varphi)}{\sin\varphi},\quad r\geq0\,.\label{eq:br}
\end{equation}
As a result, the Bogoliubov $\beta_{k}$ at the topological limit
is obtained from eq.\ (\ref{eq:beta-k=00003D0-21}) as
\begin{equation}
\beta_{k=0}=(-1)^{N}i^{N}\mathfrak{b}_{N}=-i\frac{\sin(N\varphi)}{\sin\varphi}\,.
\end{equation}
Restoring $\varphi=\pi/(2(N+1))$ as defined in eq.\ (\ref{eq:index-notation}),
we express $\beta_{k=0}$ in terms of $N$ as
\begin{equation}
\beta_{k=0}=-i\cot\left[\frac{\pi}{2(N+1)}\right]\,.
\end{equation}
This reproduces the Stokes topological limit result in \cite{Chung:2024pot}.

\subsection{Quadratic coefficient of $T_{1}(k^{2})$\label{subsec:Quadratic-coefficient-of}}

This appendix is devoted to deriving the closed form expression (\ref{eq:T2-closed-form})
for $T^{(2)}_{1}$, the quadratic coefficient in the soft momentum
expansion (\ref{eq:T-split}) for $T_{1}(k^{2})$. Analogous to the
logic of appendix \ref{subsec:Reproducing-the-Stokes}, we will first
solve the explicit recursion relation (\ref{eq:G2-recursion}) for
$G^{(2)}_{p}$, then use the propagator product expression (\ref{eq:betaT1}).
Specifically, we will first solve eq.\ (\ref{eq:G2-recursion}) to
get the finite sum $\sum^{r}_{t=1}G^{(2)}_{t}$ then take the $r\to\infty$
limit to obtain the closed form expression for $T^{(2)}_{1}$ given
by
\begin{equation}
T^{(2)}_{1}=-\lim_{r\to\infty}\sum^{r}_{t=1}G^{(2)}_{t}\,.\label{eq:T12-limit-appendix}
\end{equation}
The key to solving $G^{(2)}_{p}$ is also similar to that in appendix
\ref{subsec:Reproducing-the-Stokes}. We will show that the sequence
$G^{(2)}_{p}$ has a sparse support where the sparsity follows from
the selection rules imposed by $K^{(2)}_{p;m,l,s}$. By determining
this support we will reduce the original recursion relation (\ref{eq:G2-recursion})
to that of a nonzero subsequence, which can be solved in closed form.
The main difference between the derivation here and that in the previous
appendix arises when taking $r\to\infty$ in eq.\ (\ref{eq:T12-limit-appendix}).
Because the recursion relation for the nonzero subsequence of $G^{(2)}_{p}$
is more involved than that of $G^{(0)}_{p}$, we will need to invoke
hypergeometric function properties to obtain the closed form as $r\to\infty$
in eq.\ (\ref{eq:T12-limit-appendix}).

We start with working out the support of $G^{(2)}_{p}$ by reducing
its recursion relation (\ref{eq:G2-recursion}). For $1\leq p+1\leq N+1$,
eq.\ (\ref{eq:G2-recursion}) gives
\begin{equation}
G^{(2)}_{p+1}=b^{(2+)}_{p+1}\S_{\frac{N+p+1}{N+1}}=\begin{cases}
0 & 1\leq p+1\leq N-1\\
b^{(2+)}_{N}\S_{\frac{2N}{N+1}} & p+1=N\\
0 & p+1=N+1
\end{cases}\,,\label{eq:G2-N}
\end{equation}
where we have used eq.\ (\ref{eq:b02}). For $p+1\geq N+2$, the
second line of eq.\ (\ref{eq:G2-recursion}) implies that $G^{(2)}_{p+1}$
is equal to a source term minus a propagator term:
\begin{equation}
G^{(2)}_{p+1}=\underset{\equiv\ \text{source term}}{\underbrace{b^{(2+)}_{p+1}\S_{\frac{N+p+1}{N+1}}-\sum^{p}_{m=N+1}\sum^{m}_{l=N+1}\sum^{l-N-1}_{s=0}K^{(2)}_{p;m,l,s}G^{(0)}_{s+1}}}-\underset{\equiv\ \text{propagator term}}{\underbrace{\sum^{p}_{m=N+1}\sum^{m}_{l=N+1}\sum^{l-N-1}_{s=0}K^{(0)}_{p;m,l,s}G^{(2)}_{s+1}}}\,.\label{eq:G2-source-propagator}
\end{equation}
In this expression, the support of the source term can be determined
without knowledge of $G^{(2)}_{s+1}$, whereas the support of the
propagator term must be determined using the support of the source
term and the selection rule (\ref{eq:K0-selection}) imposed by $K^{(0)}_{p;m,l,s}$.

We first consider the source term in the range $p\geq N+1$. According
to eq.\ (\ref{eq:supp-G0}), $G^{(0)}_{s+1}\neq0$ requires
\begin{equation}
s=t\l,\quad t\in\mathbb{Z}_{\geq0}\,,
\end{equation}
where $\l=N+1$ was defined in eq.\ (\ref{eq:index-notation}). Furthermore,
given eq.\ (\ref{eq:K2}), $K^{(2)}_{p;m,l,s}$ splits into two terms
\begin{equation}
K^{(2)}_{p;m,l,s}=K^{(2+)}_{p;m,l,s}+K^{(2-)}_{p;m,l,s}\label{eq:K2-split}
\end{equation}
\begin{align}
K^{(2+)}_{p;m,l,s} & \equiv\frac{b^{(2+)}_{p-m+1}b^{(0)}_{m-l+1}}{ml}\frac{\S_{\frac{N+p+1}{N+1}}}{\S_{\frac{N+l+1}{N+1}}}\delta\left(\frac{l-s}{N+1}-1\right)\label{eq:K2p}\\
K^{(2-)}_{p;m,l,s} & \equiv\frac{b^{(0)}_{p-m+1}b^{(2-)}_{m-l+1}}{ml}\frac{\S_{\frac{N+p+1}{N+1}}}{\S_{\frac{N+l+1}{N+1}}}\delta\left(\frac{l-s}{N+1}-1\right)\,.\label{eq:K2m}
\end{align}
Similar to eq.\ (\ref{eq:ps-Np}), the common delta factor $\delta\left((l-s)/\l-1\right)$
is nonzero when
\begin{equation}
l=s+(j+1)\l=(t+j+1)\l,\quad j\in\mathbb{Z}_{\geq0}\,.\label{eq:1-Np}
\end{equation}
Moreover, using the supports of $b^{(0)}$ and $b^{(2\pm)}$ given
in eq.\ (\ref{eq:b02}) we find that 
\begin{equation}
K^{(2+)}_{p;m,l,s}\neq0\quad\text{iff}\quad m=l,\quad p-l+1\in\{N,\ 2N+1\}\,,\label{eq:K2p-selection}
\end{equation}
\begin{equation}
K^{(2-)}_{p;m,l,s}\neq0\quad\text{iff}\quad m=p,\quad m-l+1\in\{N,\ 2N+1\}\,.\label{eq:K2m-selection}
\end{equation}
Both cases require
\begin{equation}
p+1\in\{l+N,\ l+2N+1\}\,.
\end{equation}
Using $l=(t+j+1)\l$ obtained in eq.\ (\ref{eq:1-Np}) and noting
$2N+1=N+\l$, we obtain
\begin{equation}
p+1\in\{N+(t+j+1)\l,\ N+(t+j+2)\l\}\,.
\end{equation}
Consequently, the $K^{(2)}G^{(0)}$ contribution has support
\begin{equation}
\text{supp}\left[\sum^{p}_{m=N+1}\sum^{m}_{l=N+1}\sum^{l-N-1}_{s=0}K^{(2)}_{p;m,l,s}G^{(0)}_{s+1}\right]\subseteq\{N+r\l\mid r\in\mathbb{Z}_{\geq1}\}\,,
\end{equation}
where $r\geq1$ because eq.\ (\ref{eq:G2-source-propagator}) is
only valid for $p+1\geq N+2$. The support of $b^{(2+)}_{p+1}$ given
by (\ref{eq:b02}) is a subset of the above support. Hence,
\begin{equation}
\text{supp}\left[\text{source term}\right]\subseteq\{N+r\l\mid r\in\mathbb{Z}_{\geq1}\}\,.\label{eq:supp-source}
\end{equation}

We next consider the propagator term in eq.\ (\ref{eq:G2-source-propagator}).
According to the selection rule (\ref{eq:K0-selection}), $K^{(0)}_{p;m,l,s}\neq0$
requires
\begin{equation}
m=l=p,\quad p=s+(j+1)\l,\quad j\in\mathbb{Z}_{\geq0}\,.
\end{equation}
Therefore, every nonzero contribution from $G^{(2)}_{s+1}$ to $G^{(2)}_{p+1}$
through the propagator term satisfies
\begin{equation}
(p+1)-(s+1)\equiv0\mod{\l}\,,\label{eq:ps-index-class}
\end{equation}
Furthermore, eq.\ (\ref{eq:G2-N}) shows that $G^{(2)}_{N}$ is the
only nonzero element of $G^{(2)}_{p+1}$ in the initial range $1\leq p+1\leq N+1$,
and eq.\ (\ref{eq:supp-source}) shows that the source term is supported
on the residue class $\{N+r\l\mid r\in\mathbb{Z}_{\geq1}\}$, then
we can show using eq. (\ref{eq:G2-source-propagator}) that
\begin{equation}
\text{supp}\left[G^{(2)}\right]\subseteq\{N+r\l\mid r\in\mathbb{Z}_{\geq0}\}\,.\label{eq:supp-G2}
\end{equation}
Specifically, we prove by induction on $r$  the statement $P(r)$:
for $1\leq p+1\leq N+r\l$, $G^{(2)}_{p+1}$ has support on $p+1\equiv N\mod{\l}$.
From eq.\ (\ref{eq:G2-N}) we see that $P(0)$ is true and $G^{(2)}_{N+1}=0$,
which together with eq.\ (\ref{eq:supp-source}) shows that $P(1)$
is also true. Suppose that $P(r)$ holds. We consider the index range
for $P(r+1)$ with $r\geq1$:
\begin{equation}
N+r\l+1\leq p+1\leq N+(r+1)\l\,.\label{eq:p-r-induction}
\end{equation}
We prove $P(r+1)$ by showing that for this index range, $G^{(2)}_{p+1}$
can be nonzero only if $p+1=N+(r+1)\l$. First, we note all the $G^{(2)}_{s+1}$
entering the propagator term in eq.\ (\ref{eq:G2-source-propagator})
with $p+1$ in the range (\ref{eq:p-r-induction}) must satisfy $1\leq s+1\leq N+r\l$,
therefore covered by the induction hypothesis $P(r)$. Thus, nonzero
$G^{(2)}_{s+1}$ terms in the propagator term satisfy $s+1\equiv N\mod{\l}$.
For $p+1$ in the range (\ref{eq:p-r-induction}), the selection rule
(\ref{eq:ps-index-class}) further implies that the propagator term
can be nonzero only when $p+1=N+(r+1)\l$; moreover, according to
eq.\ (\ref{eq:supp-source}), the source term in (\ref{eq:G2-source-propagator})
can be nonzero only at $p+1=N+(r+1)\l$. The recursion relation (\ref{eq:G2-source-propagator})
then implies that the statement $P(r+1)$ is true. Hence, $P(r)$
is true for any $r\in\mathbb{Z}_{\geq0}$ by induction, proving eq.\ (\ref{eq:supp-G2}).

Analogous to the recursion relation reduction in appendix \ref{subsec:Reproducing-the-Stokes},
the sparse support (\ref{eq:supp-G2}) motivates us to introduce a
nonzero subsequence of $G^{(2)}_{p}$:
\begin{equation}
g^{(2)}_{r}\equiv G^{(2)}_{N+r\l}\,,\quad r\geq0\,.\label{eq:gr2}
\end{equation}
It follows from eq.\ (\ref{eq:T12-limit-appendix}) that
\begin{equation}
T^{(2)}_{1}=-\sum^{\infty}_{m=1}G^{(2)}_{m}=-\sum^{\infty}_{m\in\text{supp}[G^{(2)}]}G^{(2)}_{m}=-\sum^{\infty}_{r=0}g^{(2)}_{r}\,.
\end{equation}
We therefore introduce the partial sum series
\begin{equation}
Y^{(2)}_{r}\equiv\sum^{r}_{t=0}g^{(2)}_{t},\quad r\geq0\label{eq:Yr2}
\end{equation}
such that 
\begin{equation}
T^{(2)}_{1}=-\lim_{r\to\infty}Y^{(2)}_{r}\,.\label{eq:T12-limit}
\end{equation}
Next, we translate the recursion relation (\ref{eq:G2-recursion})
of $G^{(2)}_{p}$ to that of $g^{(2)}_{r}$ to solve the partial sum
series $Y^{(2)}_{r}$, which will in turn yield the closed form expression
for $T^{(2)}_{1}$ by taking the $r\to\infty$ limit.

To start, we note the zeroth term $g^{(2)}_{0}$ is given by eq.\ (\ref{eq:G2-N})
as
\begin{equation}
g^{(2)}_{0}=G^{(2)}_{N}=b^{(2+)}_{N}\S_{\frac{2N}{N+1}}\,.\label{eq:g20}
\end{equation}
The remaining terms in $g^{(2)}_{r}$ satisfy $r\geq1$, which correspond
to $G^{(2)}_{p+1}$ terms with $p\geq2N$, which are covered by the
recursion relation (\ref{eq:G2-source-propagator}). To reduce eq.\ (\ref{eq:G2-source-propagator})
to a recursion relation of $g^{(2)}_{r}$, we set $p+1=N+r\l$, $s+1=1+t\l$
in the source term and $s+1=N+t\l$ in the propagator term, which
leads to
\begin{align}
g^{(2)}_{r} & =\underset{\text{source term}}{\underbrace{b^{(2+)}_{N+r\l}\S_{1+r+d}-\sum^{p_{r}}_{m=N+1}\sum^{m}_{l=N+1}\sum^{\left\lfloor \frac{l}{\l}\right\rfloor -1}_{t=0}K^{(2)}_{p_{r};m,l,t\l}\,g^{(0)}_{t}}}\nonumber \\
 & \quad-\underset{\text{propagator term}}{\underbrace{\sum^{p_{r}}_{m=N+1}\sum^{m}_{l=N+1}\sum^{\left\lfloor \frac{l-N+1}{\l}\right\rfloor -1}_{t=0}K^{(0)}_{p_{r};m,l,N-1+t\l}\,g^{(2)}_{t}}},\quad r\geq1\,,\label{eq:gr2-recursion-raw}
\end{align}
where we have used eqs.\ (\ref{eq:gr0}) and (\ref{eq:gr2}) and
\begin{equation}
p_{r}\equiv N+r\l-1\,.\label{eq:pr}
\end{equation}

To simplify eq.\ (\ref{eq:gr2-recursion-raw}), we first reduce the
propagator term using the selection rule (\ref{eq:K0-selection})
imposed by $K^{(0)}$, which gives
\begin{equation}
\text{propagator term}=\sum^{\left\lfloor \frac{p_{r}-N+1}{\l}\right\rfloor -1}_{t=0}K^{(0)}_{p_{r};p_{r},p_{r},N-1+t\l}\,g^{(2)}_{t}=\frac{a^{2}}{(r+d)^{2}}\sum^{r-1}_{t=0}g^{(2)}_{t}\,,\quad r\geq1\label{eq:propagator-int1}
\end{equation}
where we have used eqs.\ (\ref{eq:b02}) and (\ref{eq:K0}), and
where $a$ and $d$ are indices defined in eq.\ (\ref{eq:index-notation}).
To reduce the source term in eq.\ (\ref{eq:gr2-recursion-raw}),
we use eq.\ (\ref{eq:K2-split}) and the selection rules (\ref{eq:K2p-selection})
and (\ref{eq:K2m-selection}) to obtain
\begin{align}
\text{source term} & =b^{(2+)}_{N+r\l}\S_{1+r+d}-\sum^{p_{r}}_{l=N+1}\sum^{\left\lfloor \frac{l}{\l}\right\rfloor -1}_{t=0}\left(K^{(2+)}_{p_{r};l,l,t\l}+K^{(2-)}_{p_{r};p_{r},l,t\l}\right)g^{(0)}_{t}\nonumber \\
 & =b^{(2+)}_{N+r\l}\S_{1+r+d}-\sum^{p_{r}}_{l=\l}\left(\frac{b^{(2+)}_{p_{r}-l+1}}{l}+\frac{b^{(2-)}_{p_{r}-l+1}}{p_{r}}\right)\frac{N}{2l}\frac{\S_{1+\frac{p_{r}}{\l}}}{\S_{1+\frac{l}{\l}}}\sum^{\left\lfloor \frac{l}{\l}\right\rfloor -1}_{t=0}\delta\left(\frac{l}{\l}-t-1\right)g^{(0)}_{t}\nonumber \\
 & =b^{(2+)}_{N+r\l}\S_{1+r+d}-\sum^{\left\lfloor \frac{p_{r}}{\l}\right\rfloor }_{q=1}\left(\frac{b^{(2+)}_{p_{r}-q\l+1}}{q\l}+\frac{b^{(2-)}_{p_{r}-q\l+1}}{p_{r}}\right)\frac{N}{2q\l}\frac{\S_{1+\frac{p_{r}}{\l}}}{\S_{1+q}}Y^{(0)}_{q-1}\,,\ r\geq1\,,\label{eq:source-term-int1}
\end{align}
where in arriving at the last line we used $l=q\l$ with $q\in\mathbb{Z}_{\geq1}$
imposed by the delta factor and the partial sum $Y^{(0)}_{q-1}$ defined
in eq.\ (\ref{eq:Yq0-defn}). The definition (\ref{eq:pr}) of $p_{r}$
implies $\left\lfloor p_{r}/\l\right\rfloor =r$. This, with the index
notations (\ref{eq:index-notation}), further reduces eq.\ (\ref{eq:source-term-int1})
to
\begin{equation}
\text{source term}=b^{(2+)}_{N+r\l}\S_{1+r+d}-\frac{N}{2\l^{2}}\sum^{r}_{q=1}\left(\frac{b^{(2+)}_{N+(r-q)\l}}{q}+\frac{b^{(2-)}_{N+(r-q)\l}}{r+d}\right)\frac{1}{q}\frac{\S_{1+r+d}}{\S_{1+q}}Y^{(0)}_{q-1},\quad r\geq1\,.\label{eq:source-term-int2}
\end{equation}
For $r=1$, the above becomes
\begin{equation}
\text{source term}=b^{(2+)}_{2N+1}\S_{2+d}-\frac{N}{2\l^{2}}\left(b^{(2+)}_{N}+\frac{b^{(2-)}_{N}}{1+d}\right)\frac{\S_{2+d}}{\S_{2}}Y^{(0)}_{0},\quad r=1\,.\label{eq:source-term-int3-1}
\end{equation}
For $r\geq2$, the support of $b^{(2\pm)}$ given in eq.\ (\ref{eq:supp-b})
requires $q\in\{r,r-1\}$ in the summation in eq.\ (\ref{eq:source-term-int2}),
which gives
\begin{align}
\text{source term} & =-\frac{N}{2\l^{2}}\left(\frac{b^{(2+)}_{N}}{r}+\frac{b^{(2-)}_{N}}{r+d}\right)\frac{1}{r}\frac{\S_{1+r+d}}{\S_{1+r}}Y^{(0)}_{r-1}\nonumber \\
 & \quad-\frac{N}{2\l^{2}}\left(\frac{b^{(2+)}_{2N+1}}{r-1}+\frac{b^{(2-)}_{2N+1}}{r+d}\right)\frac{1}{r-1}\frac{\S_{1+r+d}}{\S_{r}}Y^{(0)}_{r-2},\quad r\geq2\,.\label{eq:source-term-int3-2}
\end{align}
where we have used $b^{(2+)}_{N+r\l}=0$ for $r\geq2$. Combining
the expression (\ref{eq:g20}) for the first term $g^{(2)}_{0}$,
the propagator term expression (\ref{eq:propagator-int1}), the source
term expressions (\ref{eq:source-term-int3-1}) and (\ref{eq:source-term-int3-2}),
and the expression (\ref{eq:b02}) for $b^{(0)}$ and $b^{(2\pm)}$,
we reduce the recursion relation (\ref{eq:gr2-recursion-raw}) for
$g^{(2)}_{r}$ to
\begin{equation}
g^{(2)}_{r}=\begin{cases}
\frac{ia}{d}\S_{1+d} & r=0\\
-\frac{N}{2}\S_{2+d}-\frac{1}{16}\frac{ia}{d}\S_{1+d}-\frac{\S_{2+d}}{\S_{2}}\frac{\pi a^{2}}{2\l} & r=1\\
J^{(2)}_{r}-\frac{a^{2}}{(r+d)^{2}}\sum^{r-1}_{t=0}g^{(2)}_{t} & r\geq2
\end{cases}\label{eq:g2-recursion}
\end{equation}
whered
\begin{equation}
J^{(2)}_{r}\equiv-\frac{ia^{2}}{\l r^{2}(r+d)}\frac{\S_{1+r+d}}{\S_{1+r}}Y^{(0)}_{r-1}+\frac{a^{2}(2r+d-1)}{(r-1)^{2}(r+d)}\frac{\S_{1+r+d}}{\S_{r}}Y^{(0)}_{r-2},\quad r\geq2\,.\label{eq:Jr2}
\end{equation}

Recall that we have defined in eq.\ (\ref{eq:Yr2}) the partial sum
$Y^{(2)}_{r}$ of $g^{(2)}_{t}$. The recursion relation (\ref{eq:g2-recursion})
then translates to
\begin{equation}
Y^{(2)}_{r}=\begin{cases}
g^{(2)}_{0}=\frac{ia}{d}\S_{1+d} & r=0\\
g^{(2)}_{0}+g^{(2)}_{1}=-\frac{N}{2}\S_{2+d}+\frac{15}{16}\frac{ia}{d}\S_{1+d}-\frac{\S_{2+d}}{\S_{2}}\frac{\pi a^{2}}{2\l} & r=1\\
\left[1-\frac{a^{2}}{(r+d)^{2}}\right]Y^{(2)}_{r-1}+J^{(2)}_{r} & r\geq2
\end{cases}\label{eq:Yr2-recursion}
\end{equation}
where we have used the property $g^{(2)}_{r}=Y^{(2)}_{r}-Y^{(2)}_{r-1}$
for $r\geq1$. Iterating the third line of eq.\ (\ref{eq:Yr2-recursion})
we obtain the expression
\begin{equation}
Y^{(2)}_{r}=\left(\prod^{r}_{j=2}\left[1-\frac{a^{2}}{(j+d)^{2}}\right]\right)Y^{(2)}_{1}+\sum^{r}_{q=2}\left(\prod^{r}_{j=q+1}\left[1-\frac{a^{2}}{(j+d)^{2}}\right]\right)J^{(2)}_{q},\quad r\geq2\,,\label{eq:Yr2-1}
\end{equation}
which is analogous to the product expression (\ref{eq:Yq0}) for $Y^{(0)}_{r}$
but with an additional source term. Note the product factor is expressible
in terms of gamma functions
\begin{align}
\prod^{r}_{j=q+1}\left[1-\frac{a^{2}}{(j+d)^{2}}\right] & =\left[\prod^{r}_{j=q+1}\left(\frac{j+d+a}{j+d}\right)\right]\left[\prod^{r}_{j=q+1}\left(\frac{j+d-a}{j+d}\right)\right]\nonumber \\
 & =\left[\frac{\Gamma(q+1+d)}{\Gamma(r+1+d)}\right]^{2}\frac{\Gamma(r+1+d+a)\Gamma(r+1+d-a)}{\Gamma(q+1+d+a)\Gamma(q+1+d-a)}\,.
\end{align}
In the $r\to\infty$ limit, we find
\begin{equation}
\lim_{r\to\infty}\prod^{r}_{j=q+1}\left[1-\frac{a^{2}}{(j+d)^{2}}\right]=\frac{[\Gamma(q+1+d)]^{2}}{\Gamma(q+1+d+a)\Gamma(q+1+d-a)}\,.
\end{equation}
The quadratic coefficient of $T_{1}(k^{2})$ is therefore given by
eqs.\ (\ref{eq:T12-limit}) and (\ref{eq:Yr2-1}) as
\begin{equation}
T^{(2)}_{1}=-\lim_{r\to\infty}Y^{(2)}_{r}=-\frac{Y^{(2)}_{1}}{(2+d)_{a}(2+d)_{-a}}-\sum^{\infty}_{r=2}\frac{J^{(2)}_{r}}{(r+1+d)_{a}(r+1+d)_{-a}}\,,\label{eq:T12-int-1}
\end{equation}
where we have used the Pochhammer symbol to make the expression more compact.
The summation term in the above equation can be expanded using the
definition (\ref{eq:Jr2}) for $J^{(2)}_{r}$ and the solution (\ref{eq:Yq0})
for $Y^{(0)}_{r}$:
\begin{equation}
\sum^{\infty}_{r=2}\frac{J^{(2)}_{r}}{(r+1+d)_{a}(r+1+d)_{-a}}=-\left(\frac{2}{i\l}\right)^{d}\frac{4\pi a^{3}}{\l}\left(\Sigma_{1}+2\Sigma_{2}\right)\label{eq:Jr2-Sigma}
\end{equation}
where we have used eqs.\ (\ref{eq:S-delta-appendix}) and (\ref{eq:index-notation}),
and defined 
\begin{align}
\Sigma_{1} & \equiv\sum^{\infty}_{r=2}\frac{1}{(r+1+d)_{a}(r+1+d)_{-a}(r+1)_{d}}\frac{1}{r^{2}(r+d)}\frac{(1+a)_{r-1}(1-a)_{r-1}}{[(r-1)!]^{2}}\,,\\
\Sigma_{2} & \equiv\sum^{\infty}_{r=2}\frac{1}{(r+1+d)_{a}(r+1+d)_{-a}(r+1)_{d}}\frac{(2r+d-1)}{r(r-1)^{2}(r+d)}\frac{(1+a)_{r-2}(1-a)_{r-2}}{[(r-2)!]^{2}}\,.
\end{align}

We next derive closed form expressions for $\Sigma_{1}$ and $\Sigma_{2}$,
which will yield a closed form expression for $T^{(2)}_{1}$ through
eq.\ (\ref{eq:T12-int-1}). The key to reducing $\Sigma_{1}$ and $\Sigma_{2}$
is to rewrite them in terms of the generalized hypergeometric functions
\begin{equation}
_{p}F_{q}\left(\begin{array}{c}
a_{1},\dots,a_{p}\\
b_{1},\dots,b_{q}
\end{array};z\right)\equiv\sum^{\infty}_{s=0}\frac{(a_{1})_{s}\cdots(a_{p})_{s}}{(b_{1})_{s}\cdots(b_{q})_{s}}\frac{z^{s}}{s!}\,.\label{eq:gen-hypergeo}
\end{equation}
We first shift the summation indices in $\Sigma_{1}$ and $\Sigma_{2}$
from $r$ to $s=r-2$, giving
\begin{align}
\Sigma_{1} & =\sum^{\infty}_{s=0}\frac{1}{(s+3+d)_{a}(s+3+d)_{-a}(s+3)_{d}}\frac{1}{(s+2)^{2}(s+2+d)}\frac{(1+a)_{s+1}(1-a)_{s+1}}{[(s+1)!]^{2}}\,,\\
\Sigma_{2} & =\sum^{\infty}_{s=0}\frac{1}{(s+3+d)_{a}(s+3+d)_{-a}(s+3)_{d}}\frac{(2s+d+3)}{(s+2)(s+1)^{2}(s+2+d)}\frac{(1+a)_{s}(1-a)_{s}}{(s!)^{2}}\,.
\end{align}
We then utilize the gamma function properties 
\begin{equation}
\Gamma(x+1)=x\Gamma(x),\quad s+x=x\frac{(x+1)_{s}}{(x)_{s}}\label{eq:gamma-id}
\end{equation}
to rewrite the summands in $\Sigma_{1}$ and $\Sigma_{2}$ in the
form of the summand in eq.\ (\ref{eq:gen-hypergeo}):
\begin{align}
\Sigma_{1} & =\frac{1-a^{2}}{2}\frac{\Gamma(2+d)}{\Gamma(3+d+a)\Gamma(3+d-a)}\sum^{\infty}_{s=0}\frac{(2+d)_{s}(2+a)_{s}(2-a)_{s}}{(3+d+a)_{s}(3+d-a)_{s}(3)_{s}}\,,\\
\Sigma_{2} & =\frac{2\Gamma(2)\Gamma(2+d)}{\Gamma(3+d+a)\Gamma(3+d-a)}\frac{\Gamma(\frac{d+5}{2})}{\Gamma(\frac{3+d}{2})}\sum^{\infty}_{s=0}\frac{(1+a)_{s}(1-a)_{s}(1)_{s}(\frac{d+5}{2})_{s}(2+d)_{s}}{(3+d+a)_{s}(3+d-a)_{s}(2)_{s}(\frac{3+d}{2})_{s}}\frac{1}{s!}\,.
\end{align}

Noting $s!=(1)_{s}$ and $d=4a-1$, we express the above in terms
of the generalized hypergeometric functions
\begin{equation}
\Sigma_{1}=\frac{(1-a^{2})\Gamma(4a+1)}{2\Gamma(5a+2)\Gamma(3a+2)}F_{4},\quad\Sigma_{2}=\frac{2(2a+1)\Gamma(4a+1)}{\Gamma(5a+2)\Gamma(3a+2)}F_{5}\label{eq:Sigma12-F45}
\end{equation}
\begin{align}
F_{4} & \equiv{}_{4}F_{3}\left(\begin{array}{c}
4a+1,\ 2+a,\ 2-a,\ 1\\
5a+2,\ 3a+2,\ 3
\end{array};z=1\right)\\
F_{5} & \equiv{}_{5}F_{4}\left(\begin{array}{c}
1+a,\ 1-a,\ 1,\ 2a+2,\ 4a+1\\
5a+2,\ 3a+2,\ 2,\ 2a+1
\end{array};z=1\right)\,.
\end{align}
where $F_{4}$ can be simplified using Mathematica
\begin{equation}
F_{4}=\frac{2(1+5a)}{(1-a)(1+a)(1-4a)}\left(-11-\frac{4}{a}+3a+\frac{(4^{1-4a})(5\pi)\g(5a)\g(2+3a)}{\g(\frac{1}{2}+a)[\g(1+2a)]^{2}\g(\frac{1}{2}+3a)}\right)\,,\label{eq:F4}
\end{equation}
whereas $F_{5}$ cannot be simplified so. To evaluate $F_{5}$, we
use Dixon's theorem (see e.g.\ \cite{bailey1964generalized})
\begin{equation}
\mathfrak{D}(A,B,C)\equiv{}_{3}F_{2}\left(\begin{array}{c}
A,\ B,\ C\\
1+A-B,\ 1+A-C
\end{array};z=1\right)=\frac{(1+A)_{-B}\left(1+\frac{A}{2}-C\right)_{-B}}{\left(1+\frac{A}{2}\right)_{-B}(1+A-C)_{-B}}
\end{equation}
valid for $\Re(1+A/2-B-C)>0$. Here $(A,B,C)$ are auxiliary complex
parameters, distinct from the quantities denoted by the same symbols
in the main text. We consider 
\begin{equation}
F_{5}=\sum^{\infty}_{s=0}u_{s},\quad\mathfrak{D}(4a+1,1+a,1-a)=\sum^{\infty}_{s=0}v_{s}\,,
\end{equation}
where 
\begin{equation}
\begin{aligned}u_{s} & \equiv\frac{(1+a)_{s}(1-a)_{s}(2a+2)_{s}(4a+1)_{s}}{(5a+2)_{s}(3a+2)_{s}(2)_{s}(2a+1)_{s}}\,,\\
v_{s} & \equiv\frac{(4a+1)_{s}(1+a)_{s}(1-a)_{s}}{(3a+1)_{s}(5a+1)_{s}\,s!}\,.
\end{aligned}
\end{equation}
Using eq.\ (\ref{eq:gamma-id}) we obtain 
\begin{equation}
\frac{u_{s}}{v_{s}}=\frac{(3a+1)(5a+1)}{2a+1}\frac{s+2a+1}{(s+3a+1)(s+5a+1)(s+1)},
\end{equation}
and 
\begin{equation}
\frac{v_{s+1}}{v_{s}}=\frac{(s+4a+1)(s+a+1)(s-a+1)}{(s+3a+1)(s+5a+1)(s+1)}.
\end{equation}
These ratios imply the identity\footnote{The identity (\ref{eq:usvs}) was initially noticed by ChatGPT 5.6
Sol Ultra, then verified by us.}
\begin{equation}
u_{s}=\frac{(3a+1)(5a+1)}{8a^{3}(2a+1)}\left[-(4a-1)v_{s}+(s+4a)v_{s}-(s+4a+1)v_{s+1}\right]\,,\label{eq:usvs}
\end{equation}
where the last two terms telescope. Summing the above from $s=0$
to $L$ then gives
\begin{equation}
\sum^{L}_{s=0}u_{s}=\frac{(3a+1)(5a+1)}{8a^{3}(2a+1)}\left[-(4a-1)\sum^{L}_{s=0}v_{s}+4a-(L+1+4a)v_{L+1}\right]
\end{equation}
Taking $L\to\infty$ yields
\begin{align}
F_{5} & =\frac{(3a+1)(5a+1)}{8a^{3}(2a+1)}\left[4a-(4a-1)\mathfrak{D}(4a+1,1+a,1-a)\right]\nonumber \\
 & =\frac{(3a+1)(5a+1)}{8a^{3}(2a+1)}\left[4a-(4a-1)\frac{(3a+\frac{1}{2})_{-a-1}(4a+2)_{-a-1}}{(2a+\frac{3}{2})_{-a-1}(5a+1)_{-a-1}}\right]\,,\label{eq:F5}
\end{align}
where we have used $\lim_{L\to\infty}(L+1+4a)v_{L+1}=0$.

We use eqs.\ (\ref{eq:Jr2-Sigma}) and (\ref{eq:Sigma12-F45}) to
turn eq.\ (\ref{eq:T12-int-1}) into
\begin{align}
T^{(2)}_{1} & =-\frac{Y^{(2)}_{1}}{(2+d)_{a}(2+d)_{-a}}\nonumber \\
 & \quad\quad+\left(\frac{2}{i\l}\right)^{d}\frac{4\pi a^{3}}{\l}\left(\frac{(1-a^{2})\Gamma(4a+1)}{2\Gamma(5a+2)\Gamma(3a+2)}F_{4}+\frac{4(2a+1)\Gamma(4a+1)}{\Gamma(5a+2)\Gamma(3a+2)}F_{5}\right)\,.
\end{align}
Then, using the second line of eq.\ (\ref{eq:Yr2-recursion}) for
$Y^{(2)}_{1}$ and eqs.\ (\ref{eq:F4}) and (\ref{eq:F5}) for the
closed form expressions for $F_{4}$ and $F_{5}$, we reduce and express
$T^{(2)}_{1}$ in a closed form:
\begin{equation}
T^{(2)}_{1}=-i\left(\frac{1}{i\l}\right)^{4a}\pi^{3/2}\frac{\g(2a-\frac{1}{2})}{\g(2a)\g(a+\frac{1}{2})\g(3a-\frac{1}{2})}\,.\label{eq:T12-closed-form-appendix}
\end{equation}

\subsection{Quadratic coefficient of $\beta_{k}$\label{subsec:Quadratic-coefficient-of-1}}

In this appendix, we compute the closed form expression of $c_{1}$,
the quadratic coefficient of $\beta_{k}$ defined in eq.\ (\ref{eq:betaQuad}),
using eq.\ (\ref{eq:betaT1}) with the closed form expressions (\ref{eq:T10-cosphi})
and (\ref{eq:T12-closed-form-appendix}) for $T^{(0)}_{1}$ and $T^{(2)}_{1}$.

We start with expanding eq.\ (\ref{eq:betaT1}) to $O(k^{2})$ using
the splitting (\ref{eq:T-split}) for $T_{1}(k^{2})$:
\begin{equation}
\beta_{k}=(-1)^{N}i^{N}\left[\prod^{N}_{j=1}\begin{pmatrix}T^{(0)}_{1}+\gamma^{-2N(N+1+j)}T^{(2)}_{1}k^{2} & 1\\
1 & 0
\end{pmatrix}\right]_{21}+O(k^{4})\label{eq:beta-quad-T12}
\end{equation}
where we have used eq.\ (\ref{eq:T1Rot}). Given eq.\ (\ref{eq:index-notation})
we note $\gamma^{-2N(N+1+j)}=e^{4ij\varphi}$, then we rewrite the
matrix term in eq.\ (\ref{eq:beta-quad-T12}) as
\begin{equation}
X_{j}(k^{2})\equiv X_{0}+e^{4ij\varphi}T^{(2)}_{1}\mathscr{E}k^{2},\quad\mathscr{E}\equiv\begin{pmatrix}1 & 0\\
0 & 0
\end{pmatrix}\label{eq:Xj}
\end{equation}
where $X_{0}$ is defined in eq.\ (\ref{eq:X0-1}). Expanding eq.\ (\ref{eq:beta-quad-T12})
with eq.\ (\ref{eq:Xj}) then yields $c_{1}$:
\begin{align}
c_{1} & =(-1)^{N}i^{N}\left[\prod^{N}_{j=1}\left(X_{0}+e^{4ij\varphi}T^{(2)}_{1}\mathscr{E}k^{2}\right)\right]_{21,k^{2}\text{ coeff}}\nonumber \\
 & =(-1)^{N}i^{N}T^{(2)}_{1}\sum^{N}_{j=1}e^{4ij\varphi}[X^{N-j}_{0}]_{21}[\mathscr{E}]_{11}[X^{j-1}_{0}]_{11}\,.\label{eq:c1-int1}
\end{align}
We use eqs.\ (\ref{eq:X0r}) and (\ref{eq:br}) to get
\begin{equation}
[X^{r}_{0}]_{21}=[X^{r-1}_{0}]_{11}=\mathfrak{b}_{r}=i^{r-1}\frac{\sin(r\varphi)}{\sin\varphi},\quad r\geq1\,.
\end{equation}
This turns eq.\ (\ref{eq:c1-int1}) into
\begin{equation}
c_{1}=-\frac{T^{(2)}_{1}}{\sin^{2}\varphi}Q_{N},\quad Q_{N}\equiv\sum^{N}_{j=1}e^{4ij\varphi}\sin[(N-j)\varphi]\sin(j\varphi)\,.\label{eq:c1-int2}
\end{equation}
To reduce $Q_{N}$, we note an identity $N\varphi=\pi/2-\varphi$
from the definition (\ref{eq:index-notation}) of $\varphi$, then
\begin{equation}
\sin[(N-j)\varphi]\sin(j\varphi)=\frac{1}{2}\left[\sin[(2j+1)\varphi]-\sin\varphi\right]\,.
\end{equation}
Plugging this into $Q_{N}$ and using $e^{\pm i\varphi}=\cos\varphi\pm i\sin\varphi$
gives
\begin{equation}
Q_{N}=\frac{e^{i\varphi}}{4i}\sum^{N}_{j=1}e^{6ij\varphi}-\frac{e^{-i\varphi}}{4i}\sum^{N}_{j=1}e^{2ij\varphi}-\frac{\sin\varphi}{2}\sum^{N}_{j=1}e^{4ij\varphi}=-e^{-2i\varphi}\frac{\cos(2\varphi)}{2\sin(3\varphi)}\,.
\end{equation}
Hence, plugging $Q_{N}$ and the closed form expression (\ref{eq:T12-closed-form-appendix})
for $T^{(2)}_{1}$ into $c_{1}$ given by eq.\ (\ref{eq:c1-int2}),
we get the closed form expression
\begin{equation}
c_{1}=i\frac{\pi^{3/2}}{\l^{2\left(1-\frac{1}{\l}\right)}}\frac{\g(2a-\frac{1}{2})}{\g(2a)\g(a+\frac{1}{2})\g(3a-\frac{1}{2})}\frac{\cos(2\varphi)}{2\sin^{2}(\varphi)\sin(3\varphi)}\,,\label{eq:c1-int3}
\end{equation}
where we have used $\varphi=\pi/(2\l)$ and $a=N/(2\l)$ to compute
the phase factor. Here we remark that $c_{1}$ is purely imaginary,
explicitly confirming the time-reversal symmetry result (\ref{eq:betaImaginary}).
Further simplifying eq.\ (\ref{eq:c1-int3}) using the gamma function
identity $\Gamma(z)\Gamma(1-z)=\pi/\sin(\pi z)$ and restoring the
index notations $a=N/(2(N+1))$, $\varphi=\pi/(2(N+1))$ and $\l=N+1$
yields the closed form expression
\begin{equation}
c_{1}(n)=i\frac{\cos\left[\frac{\pi}{2(N+1)}\right]}{\sqrt{\pi}(N+1)^{2(1-\frac{1}{N+1})}}\frac{\g\left[\frac{1}{2(N+1)}\right]\g\left(\frac{1}{N+1}\right)\g\left[\frac{3}{2(N+1)}\right]}{\Gamma\left(\frac{1}{N+1}+\frac{1}{2}\right)}\,.
\end{equation}
This is eq.\ (\ref{eq:c1-closed-form}), the main result of this
appendix.\bibliographystyle{JHEP}
\bibliography{merged-references}

\end{document}